\documentclass{aa}

\usepackage{graphicx} 
\usepackage[varg]{txfonts} 
\usepackage{amsmath} 
\usepackage{amssymb} 
\usepackage{natbib} 
\usepackage{subfig} 
\usepackage{float} 
\usepackage{lscape} 
\usepackage{booktabs} 
\usepackage{color} 
\usepackage{xfrac} 
\usepackage{multirow,bigdelim} 
\usepackage[multidot]{grffile} 
\usepackage{multicol} 
\usepackage{ragged2e} 
\usepackage{siunitx} 
\usepackage{tabulary} 

\usepackage{rotating} 
\usepackage[colorlinks=true]{hyperref} 
\usepackage{color} 
\hypersetup{colorlinks, citecolor=blue, linkcolor=blue, urlcolor=blue} 

\definecolor{purple}{rgb}{0.75,0.0,0.75} 

\usepackage{tabularx} 
\newcolumntype{X}{>{\raggedright\arraybackslash}X} 
\newcolumntype{Y}{>{\centering\arraybackslash}X} 
\newcolumntype{Z}{>{\raggedleft\arraybackslash}X} 

\renewcommand{\thefootnote}{\fnsymbol{footnote}} 

\makeatletter 
\newcommand\footnoteref[1]{\protected@xdef\@thefnmark{\ref{#1}}\@footnotemark}
\makeatother

\newcommand{\hersc}{\textit{Herschel}}
\newcommand{\planck}{\textit{Planck}}
\newcommand{\spitz}{\textit{Spitzer}}
\newcommand{\micron}{{\textmu m}}
\newcommand{\montage}{Montage}

\newcommand{\caapr}{CAAPR}

\begin{document}

\title{DustPedia: Multiwavelength photometry and imagery of 875 nearby galaxies in 42 ultraviolet--microwave bands
\thanks{The DustPedia photometry and imagery are available from IRSA at \url{https://irsa.ipac.caltech.edu/data/Herschel/DustPedia/overview.html}; photometry tables are also available at CDS VizieR at \url{http://cdsweb.u-strasbg.fr/cgi-bin/qcat?J/A+A/}.}}

\author{
C.\,J.\,R.\,Clark\thanks{E-mail: \tt \href{mailto:cclark@stsci.edu}{mailto:cclark@stsci.edu}}\inst{1}
\and
S.\,Verstocken\inst{2}
\and
S.\,Bianchi\inst{3}
\and
J.\,Fritz\inst{4}
\and
S.\,Viaene\inst{2,5}
\and
M.\,W.\,L.\,Smith\inst{1}
\and
M.\,Baes\inst{2}
\and
V.\,Casasola\inst{3}
\and
L.\,P.\,Cassara\inst{6}
\and
J.\,I.\,Davies\inst{1}
\and
I.\,De\,Looze\inst{2,7}
\and
P.\,De\,Vis\inst{8}
\and
R.\,Evans\inst{1}
\and
M.\,Galametz\inst{9}
\and
A.\,P.\,Jones\inst{8}
\and
S.\,Lianou\inst{10}
\and
S.\,Madden\inst{10}
\and
A.\,V.\,Mosenkov\inst{2,11,12}
\and
M.\,Xilouris\inst{6}
}

\institute{
School of Physics \& Astronomy, Cardiff University, Queen's Buildings, The Parade, Cardiff, CF24 3AA, UK
\and
Sterrenkundig Observatorium, Universiteit Gent, Krijgslaan 281 S9, B-9000 Gent, Belgium
\and
INAF, Osservatorio Astrofisico di Arcetri, Largo E. Fermi 5,I-50125, Florence, Italy
\and
Instituto de Radioastronom\'ia y Astrof\'isica, UNAM, Campus Morelia, AP 3-72, CP 58089 Michoac\'an, M\'exico
\and
Centre for Astrophysics Research, University of Hertfordshire, College Lane, Hatfield, AL10 9AB, UK
\and
National Observatory of Athens, Institute for Astronomy, Astrophysics, Space Applications and Remote Sensing, Ioannou Metaxa and Vasileos Pavlou GR-15236, Athens, Greece
\and
Department of Physics \& Astronomy, University College London, Gower Street, London, WC1E 6BT, UK
\and
Institut d'Astrophysique  Spatiale, UMR 8617, CNRS, Univerit\'e Paris Sud, Univerit\'e Paris-Saclay, Univerit\'e Paris Sud, Orsay, F-91405, France
\and
European Southern Observatory, Karl-Schwarzchild-Str, D-85748, Garching, Germany
\and
CEA/DSM/IRFU/Service d’Astrophysique, Astrophysique des Interactions Multi-eschelles UMR 7158, CEA, Saclay, Orme des Merisiers batiment 709, 91191 Gif-sur-Yvette, France
\and
St. Petersburg State University, Universitetskij Pr. 28, 198504, St. Petersburg, Stary Peterhof, Russia
\and
Central Astronomical Observatory of RAS, Pulkovskoye Chaussee 65/1, 196140, St. Petersburg, Russia
}

\date{Received 22 June 2017; accepted 14 August 2017}

\abstract
{}{The DustPedia project is capitalising on the legacy of the \hersc\ Space Observatory, using cutting-edge modelling techniques to study dust in the 875 DustPedia galaxies -- representing the vast majority of extended galaxies within 3000\,km\,s$^{-1}$ that were observed by \hersc. This work requires a database of multiwavelength imagery and photometry that greatly exceeds the scope (in terms of wavelength coverage and number of galaxies) of any previous local-Universe survey.}
{We constructed a database containing our own custom \hersc\ reductions, along with standardised archival observations from GALEX, SDSS, DSS, 2MASS, WISE, \spitz, and \planck. Using these data, we performed consistent aperture-matched photometry, which we combined with external supplementary photometry from IRAS and \planck.}
{We present our multiwavelength imagery and photometry across 42 UV--microwave bands for the 875 DustPedia galaxies. Our aperture-matched photometry, combined with the external supplementary photometry, represents a total of 21,857 photometric measurements. A typical DustPedia galaxy has multiwavelength photometry spanning 25 bands. We also present the Comprehensive \& Adaptable Aperture Photometry Routine (\caapr), the pipeline we developed to carry out our aperture-matched photometry. \caapr\ is designed to produce consistent photometry for the enormous range of galaxy and observation types in our data. In particular, \caapr\ is able to determine robust cross-compatible uncertainties, thanks to a novel method for reliably extrapolating the aperture noise for observations that cover a very limited amount of background. Our rich database of imagery and photometry is being made available to the community.}{}

\keywords{galaxies: photometry -- 
galaxies: general -- 
techniques: photometric -- 
ISM: dust -- 
catalogues  -- 
surveys}

\titlerunning{DustPedia photometry \& imagery}

\authorrunning{C.J.R. Clark et al.}

\maketitle 

\setcounter{footnote}{0}
\renewcommand{\thefootnote}{\textsuperscript{\arabic{footnote}}}

\section{Introduction} \label{Section:Introduction}

Over the past 10--15 years dramatic progress has been made in the study of cosmic dust as a window into the nature and evolution of galaxies. This advancement has been primarily driven by the wealth of excellent data provided by far-infrared (FIR) and submillimetre (submm) observatories such as \hersc\ \citep{Pilbratt2010D}, \planck\ \citep{Planck2011I}, \spitz\ \citep{Werner2004B}, the James Clerk Maxwell Telescope (JCMT, and more recently the Atacama Large Millimetre/submillimetre Array (ALMA). \hersc\ in particular was especially well suited for the study of nearby galaxies; its rapid mapping abilities enabled it to observe a sizeable portion of the galaxies in the local universe, with a combination of sensitivity, resolution, and broad wavelength coverage that remains unmatched.

The DustPedia project\footnote{DustPedia is funded by the European Union, as a European Research Council (ERC) 7\textsuperscript{th} Framework Program (FP7) project (PI Jon Davies, proposal 606824): \url{http://dustpedia.com/}.} \citep{Davies2017A} is working towards a definitive understanding of dust in the local Universe, by capitalising on the legacy of \hersc. The DustPedia sample consists of every galaxy that was observed by \hersc\ that lies within a velocity of 3000\,km\,s$^{-1}$ (corresponding to 41\,Mpc distance, assuming ${\rm H_{0} = 73.24\,km\,s^{-1}\,Mpc^{-1}}$; \citealp{Riess2016B}), and has D25\,\textgreater\,1\arcmin (D25 being the major axis isophote at which optical surface brightness falls beneath 25 mag arcsec$^{-2}$); these criteria were evaluated using the HyperLEDA database\footnote{\url{http://leda.univ-lyon1.fr/}}\textsuperscript{,}\footnote{The HyperLEDA database is continually updated to reflect newly-incorporated data; therefore note that the DustPedia sample is derived from HyperLEDA queries performed in January 2015.} \citep{Makarov2014A}. Additionally, the DustPedia sample only includes galaxies that have a detected stellar component; WISE observations at 3.4\,\micron\ are the deepest all-sky data sensitive to the stellar component of galaxies, and hence provide the most consistent way of implementing this stellar detection requirement. Therefore, the DustPedia sample only includes galaxies brighter than the WISE 3.4\,\micron\ all-sky 5\,$\sigma$ sensitivity limit of $19.91\,{\rm m_{AB}}$ (although this requreiment has little impact upon the final sample, as \textless\,1\% of candidate HyperLEDA galaxies fail to meet the brightness limit). For full sample details see \citet{Davies2017A}. Note that the DustPedia sample excludes the Local Group galaxies of Andromeda, Triangulum, and the Magellanic Clouds, as working with such exceptionally-extended systems would entail fundamentally different data processing and analysis.

DustPedia is combining cutting-edge methods for studying dust: physically-motivated dust modelling with The Heterogeneous Evolution Model for Interstellar Solids (THEMIS; \citealp{Jones2016A,Ysard2016A,Jones2017A}); hierarchical Bayesian Spectral Energy Distribution (SED) fitting with HiERarchical Bayesian Inference for dust Emission (HerBIE; Galliano et al.,\,in prep.); and 3-dimensional radiative transfer modelling and fitting with Stellar Kinematics Including Radiative Transfer (SKIRT; \citealp{Baes2011H,Camps2015A}) and FitSKIRT \citep{DeGeyter2013A}. 

The study of nearby galaxies now involves increasingly-extensive multiwavelength datasets, and DustPedia is no exception to this. The tools used in modern extragalactic astronomy require that data from across broad swathes of the spectrum can be directly compared, despite the very different natures of the data employed. This is especially true for the study of dust, where radiative transfer and SED modelling tools -- such as HerBIE, SKIRT, MAGPHYS \citep{DaCunha2008}, MOCASSIN \citep{Ercolano2003B,Ercolano2005A}, CIGALE \citep{Burgarella2005F,Noll2009B}, and GRASIL \citep{Silva1998C} -- critically require that the flux densities being used represent self-consistent measurements. For example, in energy-balance SED-fitting, it is vital that the stellar population sampled by ultraviolet (UV), optical, and near-infrared (NIR) data points is the same stellar population heating the dust observed at mid infrared (MIR) to submm wavelengths.

Moreover, in the realm of multiwavelength photometry, it is often overlooked that measuring flux densities in a consistent and accurate manner is invariably much easier than measuring the {\it uncertainties} on those flux densities in a similarly consistent and accurate manner. However, with the growing prevalence of Bayesian techniques in astronomy (see review in \citealp{Loredo2013A}, and references therein), it has never been more important that photometric uncertainties be robust and directly-comparable.

Therefore, in order to best exploit DustPedia's unique combination of advanced tools for the study of dust in galaxies, the cornerstone of the project is the DustPedia database\footnote{\url{http://dustpedia.astro.noa.gr}} -- which contains standardised multiwavelength imagery of all 875 galaxies in our sample\footnote{Subsequent to the publication of \citet{Davies2017A}, DustPedia source SDSSJ140459.26+533838.8 was found to be a shredded portion of NGC\,5474; so whereas \citet{Davies2017A} refer to the DustPedia sample containing 876 galaxies, we here only deal with the revised sample of 875.}, including both custom reductions and archival data, spanning over five orders of magnitude in wavelength from UV to microwave. Furthermore, the DustPedia database contains the results of consistent multiwavelength aperture-matched photometry, conducted using the Comprehensive \& Adaptable Aperture Photometry Routine (\caapr) -- along with supplementary photometry for additional instruments for which aperture-matched photometry was impractical. In total, we provide data for 42 UV--microwave bands. These data are already being employed in various accepted and in-preparation DustPedia works, including \citet{Casasola2017A}, Lianou et al.\,(in prep.), Cassara et al.\,(in prep.), Mosenkov et al.\,(in prep.), Evans et al.\,(in prep.), Verstocken et al.\,(in prep.), and Nersesian et al.\,(in prep.). Ultimately, the DustPedia database will also contain the results of the SED fitting and radiative transfer modelling performed on these data, for each galaxy.

In this paper, we present the imagery and photometry that form the centrepiece of the DustPedia database. The multiwavelength imagery -- both our custom \hersc\ reductions, and the standardised UV--microwave ancillary data -- are presented in Section~\ref{Section:Imagery}. The functionality of the \caapr\ pipeline, which we developed to produce our consistent multiwavelength aperture-matched photometry, is described in Section~\ref{Section:Pipeline}. The flagging we carry out for our photometry is explained in Section~\ref{Section:Photometry_Flagging}. Supplementary photometry from additional observatories is laid out in Section~\ref{Section:Supplementary_Photometry}. In Section~\ref{Section:Validation} we detail the various tests we carried out to validate the quality of our photometry. The contents and format of the DustPedia data products are specified in Section~\ref{Section:Data_Products}, including information on distance measures for each galaxy. 

For the sake of brevity and readability, we refer to `flux densities' as `fluxes' throughout the rest of this work. We adopt a Hubble constant of ${\rm H_{0} = 73.24\,km\,s^{-1}\,Mpc^{-1}}$ \citep{Riess2016B}.

\begin{figure*}
\begin{center}
\includegraphics[width=0.975\textwidth]{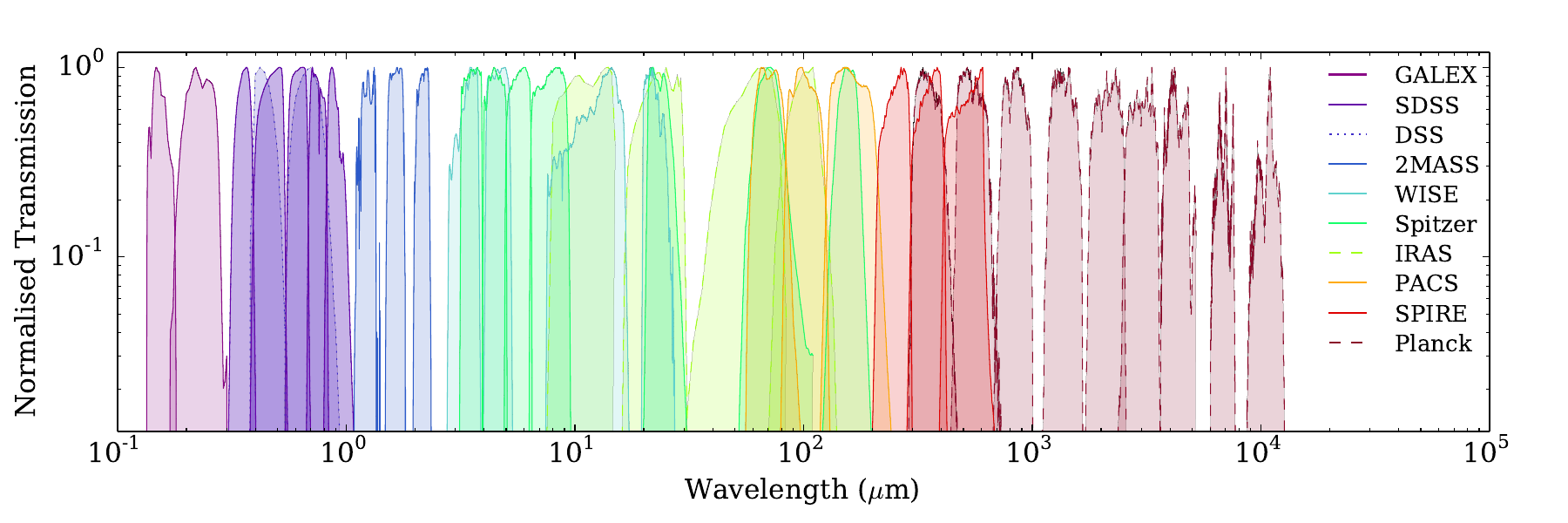}
\caption{Illustration of the spectral coverage provided by the DustPedia database, showing filter response functions of all bands for which we present data. As can be seen, the data we employ effectively provides complete sampling of over five orders of magnitude in wavelength. Response functions of the bands for which we present both imagery and aperture-matched photometry are traced with solid lines. Bands for which we present supplementary external photometry (see Section~\ref{Section:Supplementary_Photometry}) are traced with dashed lines. Bands for which we present imagery only (see Section~\ref{Subsubsection:DSS_Imagery}) are traced with dotted lines.}
\label{Fig:Transmission_Curves}
\end{center}
\end{figure*}

\section{Multiwavelength Imagery} \label{Section:Imagery}

For the DustPedia galaxies, we produced dedicated reductions of the \hersc-SPIRE and \hersc-PACS maps, along with standardised preparations of archive imagery for the ancillary data.

All maps were produced in Flexible Image Transport System (FITS; \citealp{Wells1981A, Hanisch2001F}) format, with standardised file names and headers. The file name indicates the target galaxy, facility, and band in question, taking the form: \texttt{[galaxy]\_[facility]\_[band].fits}; in cases where an error map was also available, it was stored in a separate FITS file, with a file name of the form: \texttt{[galaxy]\_[facility]\_[band]\_Error.fits}. As an example, the \hersc-SPIRE 250\,\micron\ map of DustPedia galaxy NGC\,0891 has the file name: \texttt{NGC0891\_SPIRE\_250.fits}. All galaxies are referred to by the name under which they are listed in the HyperLEDA database.

The FITS headers contain the standard World Coordinate System (WCS; \citealp{Greisen2002B}) fields, along with fields that provide a range of additional information: target galaxy, telescope, filter effective wavelength, filter name, instrument, map units, and the name of the database the original data was acquired from (if this is not the same for all observations in a given band). If there is an error map available, then the header additionally indicates whether the file contains the image map or the error map. 

All maps are in units of Jy\,pix$^{-1}$ (with the exception of the DSS maps, which are left in their native non-linear units of `photographic densities'), and hence can trivially be converted into units of AB magnitudes, erg\,s$^{-1}$, etc. For users interested in working with the native data units of the maps, conversion details are provided in Appendix~\ref{AppendixSection:Map_Unit_Conversions}.

We do not apply any colour-corrections (to account for the fact that the data from different facilities are calibrated assuming different reference spectra), because such corrections will depend upon the underlying SED of each source. Throughout this section, we state the default reference spectrum assumed for each instrument.

Throughout the rest of this section, we describe how the imagery was produced, providing details for each instrument from which we use data. The basic parameters of each band are listed in Table~\ref{Table:Band_Parameters}, whilst their spectral coverage is illustrated in Figure~\ref{Fig:Transmission_Curves}.

\subsection{Herschel Imagery} \label{Subsection:Herschel_Imagery}

Full details of our \hersc\ data reduction process are provided in \citet{Davies2017A}; here we provide only a brief summary, for the sake of completeness. For both Spectral and Photometric Imaging REceiver (SPIRE; \citealp{Griffin2010D}) and Photodetector Array Camera and Spectrometer (PACS; \citealp{Poglitsch2010B}) data, the \hersc\ Science Archive (HSA\footnote{\url{http://www.cosmos.esa.int/web/herschel/science-archive}}) was queried to find all \hersc\ photometer observations that provided coverage of each DustPedia galaxy. 

\subsubsection{\textit{Herschel}-SPIRE} \label{Subsubsection:Herschel-SPIRE_Reduction}

In the first stage of our SPIRE reduction, raw Level-0 data (acquired from the HSA) was processed up to Level-1 (calibrated pointed timelines) using {\sc HIPE}\,v13\footnote{At the time of processing, {\sc HIPE}\,v13 was the current release of the \hersc\ Interactive Processing Environment \citep{Ott2010B}. At the time of publication, the current release is {\sc HIPE}\,v15.0.1, which nonetheless uses the same photometric calibrations as {\sc HIPE}\,v13.}. The Level-1 data was processed up to Level-2 (final maps) using the Bright Galaxy Adaptive Element ({\sc BriGAdE}; \citealp{MWLSmithC}) pipeline. Observations that overlapped were combined into single contiguous maps\footnote{With an exception for observations of the Virgo Cluster, where the sheer quantity of data makes combining all available data computationally prohibitive -- instead, we only make use of the deep data taken by HeViCS (the Herschel Virgo Cluster Survey; \citealp{Davies2010A}).}. The final timelines were refined with the {\sc HIPE}\,v13 destriper; the improved processing provided by {\sc BriGAdE}, and the fact that we combine all overlapping observations, allows the destriper to operate more effectively than it would otherwise. Final SPIRE maps were produced using the {\sc HIPE}\,v13 na\"ive map-maker, with pixel sizes of 6, 8, and 12 arcseconds at 250, 350, and 500\,\micron\ respectively ($\approx\sfrac{1}{3}$ each band's FWHM). The photometric calibration of our maps used the beam area values provided in {\sc HIPE}\,v13; specifically, 469.7, 831.7, and 1793.5 arcsec$^{2}$ at 250, 350, and 500\,\micron. If a target galaxy was located in a reduced map larger than one degree in RA or dec, then a cutout of $1^{\circ} \times 1^{\circ}$, centred on the target, was produced as the final map; if the reduced map was smaller than one degree, then it was used in its entirety as the final map. In total, 844 (96\%) of the DustPedia galaxies have SPIRE coverage. SPIRE maps are calibrated assuming a $\nu^{-1}$ reference spectrum. 

\subsubsection{\textit{Herschel}-PACS} \label{Subsubsection:Herschel-PACS_Reduction}

For our PACS reduction, Level-0 raw data was processed up to Level-1 timelines using {\sc HIPE}\,v13. The Level-1 data was then processed into Level-2 maps using the standalone {\sc Scanamorphous}\,v24 pipeline \citep{Roussel2012B,Roussel2013A}. The {\sc Scanamorphous} processing took account of whether the PACS data in question was taken in scan map, mini map, or parallel mode. Observations taken in different observing modes were not combined; when data from more than one mode was available, the superior set of observations was identified, based upon depth, coverage sky area, and scan speed. In the case of some target galaxies, the PACS data was found to be unusable. The final PACS maps have pixel sizes of 2, 3, and 4 arcseconds at 70, 100, and 160\,\micron\ respectively ($\approx\sfrac{1}{3}$ the FWHM, in keeping with the SPIRE maps). For data from very large PACS fields, the final maps have a diameter of $0.35^{\circ}$, centred upon the target galaxies; this is smaller than for the SPIRE maps, due to the additional computational complexity of reducing PACS data. In total, 771 (88\%) of the DustPedia galaxies have usable PACS coverage. PACS maps are calibrated assuming a $\nu^{-1}$ reference spectrum.

\begin{table*}
\begin{center}
\footnotesize
\caption{Details of each band for which we have data.}
\label{Table:Band_Parameters}
\begin{tabular}{lrlSSllSl}
\toprule \toprule
\multicolumn{1}{c}{Facility} &
\multicolumn{1}{c}{Effective } &
\multicolumn{1}{c}{Band} &
\multicolumn{1}{c}{Imagery} &
\multicolumn{1}{c}{Photometry} &
\multicolumn{1}{c}{Pixel} &
\multicolumn{1}{c}{Resolution} &
\multicolumn{2}{c}{Calibration} \\
\multicolumn{1}{c}{} &
\multicolumn{1}{c}{Wavelength} &
\multicolumn{1}{c}{Name} &
\multicolumn{1}{c}{Present} &
\multicolumn{1}{c}{Present} &
\multicolumn{1}{c}{Width} &
\multicolumn{1}{c}{FWHM} &
\multicolumn{2}{c}{Uncertainty} \\
\cmidrule(lr){6-6}
\cmidrule(lr){7-7}
\cmidrule(lr){8-9}
\multicolumn{1}{c}{} &
\multicolumn{1}{c}{} &
\multicolumn{1}{c}{} &
\multicolumn{1}{c}{} &
\multicolumn{1}{c}{} &
\multicolumn{1}{c}{(\arcsec)} &
\multicolumn{1}{c}{(\arcsec)} &
\multicolumn{2}{c}{(\%)} \\
\midrule
GALEX & 153\,nm & FUV & 797 & 794 & 3.2 & 4.3 & 4.5 & \rdelim\}{2}{13pt}[$a$] \\
GALEX & 227\,nm & NUV & 832 & 830 & 3.2 & 5.3 & 2.7 & \\
SDSS & 353\,nm & {\it u} & 655 & 655 & 0.45 & 1.3 & 1.3 & \rdelim\}{5}{13pt}[$b$] \\
SDSS & 475\,nm & {\it g} & 655 & 655 & 0.45 & 1.3 & 0.8 & \\
SDSS & 622\,nm & {\it r} & 655 & 655 & 0.45 & 1.3 & 0.8 & \\
SDSS & 763\,nm & {\it i} & 655 & 655 & 0.45 & 1.3 & 0.7 & \\
SDSS &  905\,nm& {\it z} & 655 & 655 & 0.45 & 1.3 & 0.8 & \\
DSS1 & 450\,nm & {\it B} & 794 & \textendash & 1--1.7 & 1.9 (1.5--3.0) & \textendash & \\
DSS1 & 660\,nm & {\it R} & 794 & \textendash & 1--1.7 & 1.9 (1.5--3.0) & \textendash & \\
DSS2 & 450\,nm & {\it B} & 861 & \textendash & 1--1.7 & 1.9 (1.5--3.0) & \textendash & \\
DSS2 & 660\,nm & {\it R} & 861 & \textendash & 1--1.7 & 1.9 (1.5--3.0) & \textendash & \\
2MASS & 1.24\,\micron\ & {\it J} & 875 & 875 & 1 & 2.0 & 1.7 & \rdelim\}{3}{13pt}[$c$] \\
2MASS & 1.66\,\micron\ & {\it H} & 875 & 875 & 1 & 2.0 & 1.9 & \\
2MASS & 2.16\,\micron\ & {\it K$_{S}$} & 875 & 875 & 1 & 2.0 & 1.9 & \\
WISE & 3.4\,\micron\ & (W1) & 875 & 875 & 1.375 & 6.1 & 2.9 & \rdelim\}{4}{13pt}[$d$] \\
WISE & 4.6\,\micron\ & (W2) & 875 & 875 & 1.375 & 6.4 & 3.4 & \\
WISE & 12\,\micron\ & (W3) & 875 & 875 & 1.375 & 6.5 & 4.6 & \\
WISE & 22\,\micron\ & (W4) & 875 & 875 & 1.375 & 12 & 5.6 & \\
\spitz\ & 3.6\,\micron\ & (IRAC-1) & 644 & 623 & 0.75 & 1.66 & 3 & \rdelim\}{4}{13pt}[$e$] \\
\spitz\ & 4.5\,\micron\ & (IRAC-2) & 804 & 777 & 0.75 & 1.72 & 3 & \\
\spitz\ & 5.8\,\micron\ & (IRAC-3) & 392 & 374 & 0.6 & 1.88 & 3 & \\
\spitz\ & 8.0\,\micron\ & (IRAC-4) & 411 & 392 & 0.6 & 1.98 & 3 & \\
\spitz\ & 24\,\micron\ & (MIPS-1) & 491 & 477 & 2.4--2.6 & 6 & 5 & \rdelim\}{3}{13pt}[$f$] \\
\spitz\ & 70\,\micron\ & (MIPS-2) & 198 & 177 & 4 & 18 &10 & \\
\spitz\ & 160\,\micron\ & (MIPS-3) & 184 & 171 & 8 & 38 & 12 & \\
PACS & 70\,\micron\ & \textendash & 255 & 244 & 2 & 9 (5.8--12.2) & 7 & \rdelim\}{3}{13pt}[$g$] \\
PACS & 100\,\micron\ & \textendash & 716 & 701 & 3 & 10 (6.9--12.7) & 7 & \\
PACS & 160\,\micron\ & \textendash & 771 & 753 & 4 & 13 (12.1--15.7) & 7 & \\
SPIRE & 250\,\micron\ & (PSW) & 844 & 844 & 6 & 18 & 5.5 & \rdelim\}{3}{13pt}[$h$] \\
SPIRE & 350\,\micron\ & (PMW) & 844 & 844 & 8 & 25 & 5.5 & \\
SPIRE & 500\,\micron\ & (PLW) & 844 & 844 & 12 & 36 & 5.5 & \\
\midrule
\planck\ & 350\,\micron\ & (857\,GHz) & 875 & 394 & 102.1 & 278 & 6.4 & \rdelim\}{6}{13pt}[$i$] \\
\planck\ & 550\,\micron\ & (545\,GHz) & 875 & 279 & 102.1 & 290 & 6.1 & \\
\planck\ & 850\,\micron\ & (353\,GHz) & 875 & 197 & 102.1 & 296 & 0.78 & \\
\planck\ & 1.38\,mm & (217\,GHz) & 875 & 97 & 102.1 & 301 & 0.16 & \\
\planck\ & 2.10\,mm & (143\,GHz) & 875 & 29 & 102.1 & 438 & 0.07 & \\
\planck\ & 3.00\,mm & (100\,GHz) & 875 & 19 & 205.7 & 581 & 0.09 & \\
\planck\ & 4.26\,mm & (70\,GHz) & 875 & 11 & 205.7 & 799 & 0.20 & \rdelim\}{3}{13pt}[$j$] \\
\planck\ & 6.81\,mm & (44\,GHz) & 875 & 18 & 205.7 & 1630 & 0.26 & \\
\planck\ & 10.60\,mm & (30\,GHz) & 875 & 35 & 205.7 & 1940 & 0.35 & \\
\midrule
IRAS & 12\,\micron\ & \textendash & \textendash & 598 & \textendash & 270 & 20 & \rdelim\}{4}{13pt}[$k$] \\
IRAS & 25\,\micron\ & \textendash & \textendash & 578 & \textendash & 276 & 20 & \\
IRAS & 60\,\micron\ & \textendash & \textendash & 675 & \textendash & 282 & 20 & \\
IRAS & 100\,\micron\ & \textendash & \textendash & 682 & \textendash & 300 & 20 & \\
\bottomrule
\end{tabular}
\end{center}
\tablefoot{For FUV--$K_{S}$ bands, we refer to each band by its listed `Band Name'; otherwise we refer to bands by wavelength. The `Imagery Present' and `Photometry Present' columns give the number of galaxies in each band for which we present imagery and photometry (there are some targets for which we present imagery, but where photometry was not possible; see Section~\ref{Subsection:Aperture_Photometry}). For supplementary photometry, we give the number of galaxies in each band for which photometry was available from IRAS SCANPI and \planck\ CCS2 (see Section~\ref{Section:Supplementary_Photometry}). References for calibration uncertainties are provided below. For bands with significantly varying resolutions, a typical value is given, followed by the full range in parentheses.}
\footnotesize
\justify
{\bf References.} $^{a}$ \citet{Morrissey2007B}; $^{b}$ SDSS DR12 Data Release Supplement: \url{https://www.sdss3.org/dr12/scope.php}; $^{c}$ \citet{Cohen2003E}; $^{d}$ WISE All-Sky Release Explanatory Supplement: \url{http://wise2.ipac.caltech.edu/docs/release/allsky/expsup/sec4_4h.html}; $^{e}$ IRAC Instrument Handbook \url{https://irsa.ipac.caltech.edu/data/SPITZER/docs/irac/iracinstrumenthandbook/17/\#_Toc410728305}; $^{f}$ MIPS Instrument Handbook: \url{https://irsa.ipac.caltech.edu/data/SPITZER/docs/mips/mipsinstrumenthandbook/42/\#_Toc288032317}; $^{g}$ PACS Instrument \& Calibration Wiki: \url{http://herschel.esac.esa.int/twiki/bin/view/Public/PacsCalibrationWeb}; $^{h}$ SPIRE Instrument \& Calibration Wiki:\ \url{http://herschel.esac.esa.int/twiki/bin/view/Public/SpireCalibrationWeb}; $^{i}$ \citet{Planck2015VIII}; $^{j}$ \citet{Planck2015V}; $^{k}$ \citet{Sanders2003A}, \citet{Miville-Deschenes2005A}.
\end{table*}

\subsection{Ancillary Imagery} \label{Subsection:Ancillary_Imagery}

Our ancillary imaging data consists of observations from 8 facilities; the GALaxy Evolution eXplorer (GALEX; \citealp{Morrissey2007B}), the Sloan Digital Sky Survey (SDSS; \citealp{York2000B,Eisenstein2011B}), the Digitized Sky Survey (DSS), the 2 Micron All-Sky Survey (2MASS; \citealp{Skrutskie2006A}), the Wide-field Infrared Survey Explorer (WISE; \citealp{Wright2010F}), the \spitz\ Space Telescope, \citep{Werner2004B}, and \planck\ \citep{Planck2011I}. The basic parameters of each band are listed in Table~\ref{Table:Band_Parameters}. The following sections will detail the specifics of how the data for each facility was processed; however the same general procedure was employed throughout.

The various facilities from which we take our ancillary data were chosen on the basis of observation quality, data availability, and wavelength coverage. All of the facilities we employ provide coverage for the vast majority of the DustPedia galaxies (\textgreater\,75\%). Whilst data with improved resolution and/or sensitivity is available in certain wavelength ranges, coverage is only ever available for a small minority of our targets. The only potentially `borderline' case for inclusion was the {\sc UKirt} Infrared Deep Sky Survey (UKIDSS; \citealp{Lawrence2007E}). Whilst UKIDSS data is higher-resolution and deeper than 2MASS, it still only provides coverage for a minority (40\%) of the DustPedia galaxies, and questions remain regarding possible photometric inconsistencies when compared to the Visible and Infrared Survey Telescope for Astronomy (VISTA), 2MASS, and SDSS surveys (see \citealp{Driver2016A}). As such, we opted not to include UKIDSS data at present; alternative facilities operating in the UKIDSS wavelength range (such as VISTA) only provide coverage for \textless\,15\%\ of our target galaxies. Nonetheless, we do not rule out incorporating UKIDSS, or other datasets, in to the DustPedia database at some point in the future.

In general, the ancillary imagery takes the form of 0.5\textdegree\,$\times$\,0.5\textdegree\ cutouts (oriented East-North), centred on their respective target galaxies. However, for galaxies sufficiently extended that they would span more than 20\%\ of the diameter of such a cutout (ie, those with D25\,\textgreater\,6\arcmin), larger cutouts of 1\textdegree\,$\times$\,1\textdegree\ were used (although less than 10\%\ of the DustPedia galaxies are this extended); this ensures that all cutouts can contain enough sky to allow for proper background measurement, aperture noise estimation, etc, even for extremely extended sources. 

For a particular target galaxy, observed in a given band, by a given facility, all available imagery covering the region of interest was identified. The data was retrieved from the official online archive of the facility in question, and then mosaiced to produce the cutout map.

The tool employed to carry out the mosaicing was \montage\footnote{\url{http://montage.ipac.caltech.edu/ }}. \montage\ was used to re-grid the images to a common pixel projection, match their background levels, and co-add them; when the observations being co-added were of different depths, the contribution of each image to the final co-addition was weighted appropriately (by considering exposure time information, or using error maps, etc). 

\begin{figure}
\begin{center}
\includegraphics[width=0.475\textwidth]{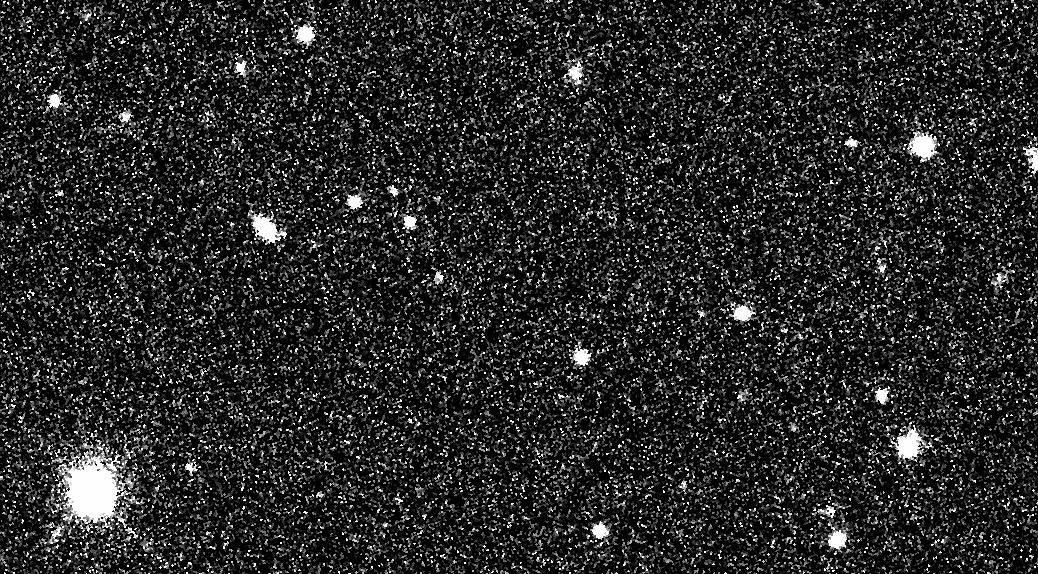}
\includegraphics[width=0.01\textwidth]{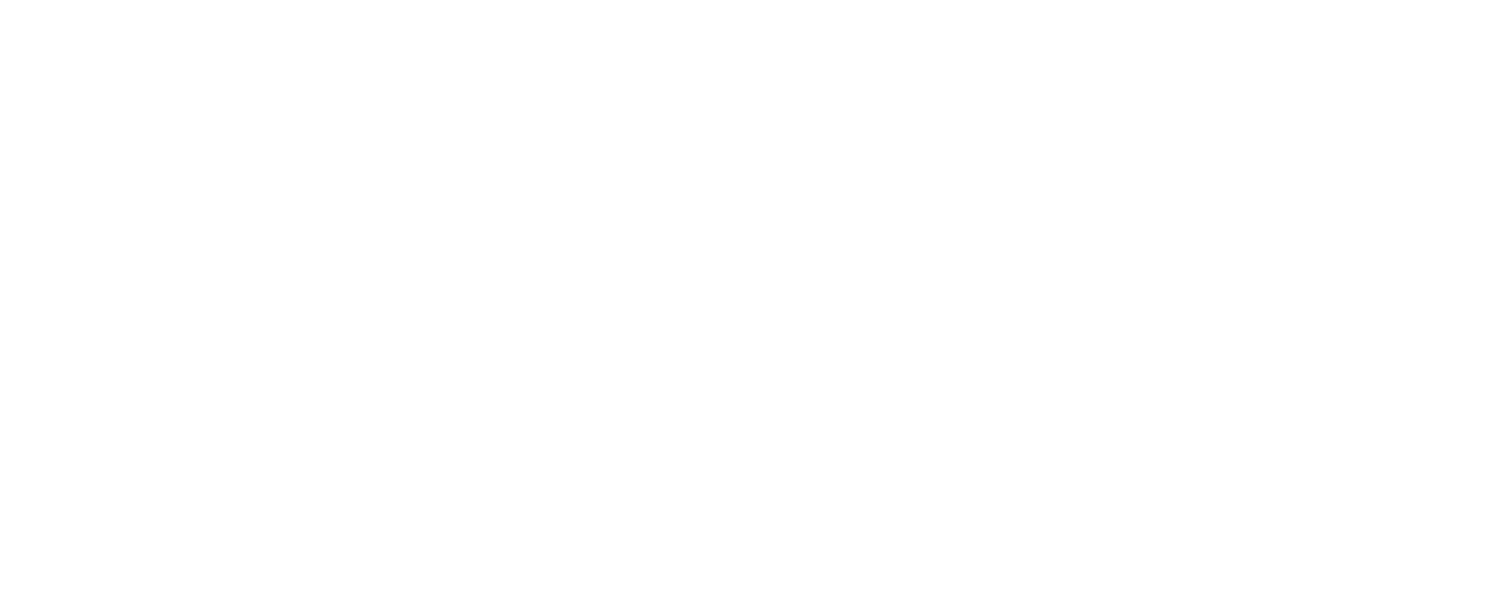}
\includegraphics[width=0.475\textwidth]{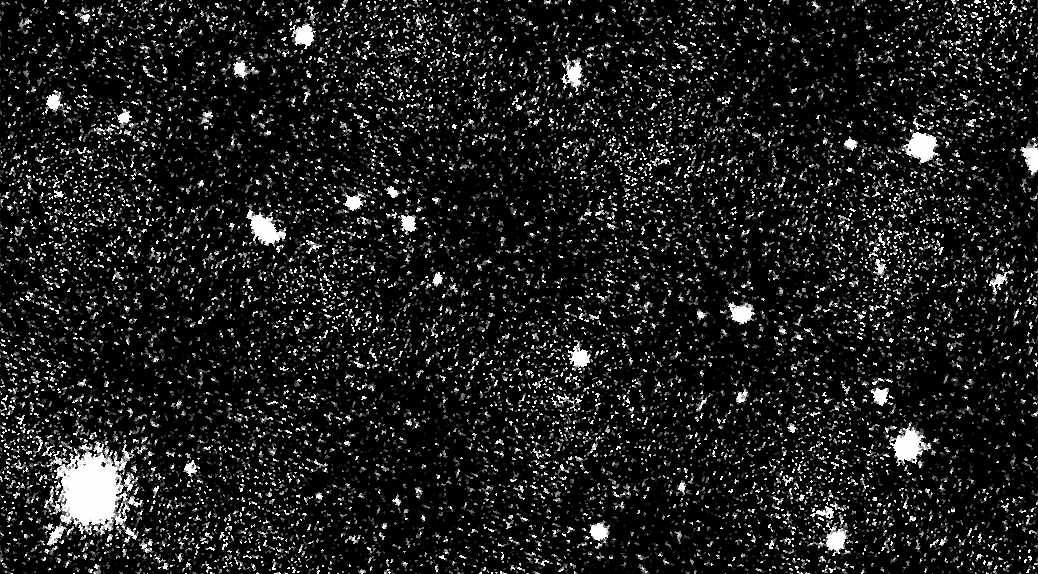}
\caption{Illustration of the moir\'e patterns which can arise when reprojecting an image to a different orientation, using the example of an SDSS {\it r}-band image, shown before ({\it upper}) and after ({\it lower}) reprojection. The RMS pixel noise at the `peaks' of the pattern is $\sim$\,50\%\ greater than in the `troughs'.}
\label{Fig:Moire_Example}
\end{center}
\end{figure}

Mosaicing via \montage\ has been found by previous authors to give rise to no consequential degradation of image quality \citep{Blanton2011A}. However, when reprojecting images to a different coordinate grid rotation (via either \montage\ or other tools), moir\'e patterns can appear, particularly if the noise budget is dominated by Poisson statistics. As well as being a conspicuous artefact, moir\'e patterns also have an effect upon map noise characteristics at smaller scales. An illustration of a reprojection-induced moir\'e pattern is shown in Figure~\ref{Fig:Moire_Example}. If necessary, it is possible to prevent the appearance of moir\'e patterns by slightly ($\sim$\,10\%) enlarging the pixel size of the reprojected maps.This was necessary for our GALEX and SDSS data (the specifics are discussed in their respective sections below); all photometry and other analysis was performed on maps that had undergone the pixel enlargement, hence preventing moir\'e patterns from affecting our results.

Note that the background-matching carried out by \montage\ is not a background {\it subtraction}. Rather, \montage\ adjusts the level of each map, so that it matches, as closely as possible, those maps it overlaps with. 
This background matching does not scale the maps to any absolute physical zero-point -- it merely minimises the difference between overlapping maps. This keeps the scaling of the map units as close as possible to that of the input data (whilst still providing good mosaicing), permitting users to convert the maps back to their native data units (for instance, in order to consider pixel Poisson statistics for resolved analyses) as per Table~\ref{AppendixTable:Map_Unit_Conversions}. As such, local background subtraction (eg, with a background annulus) must still be carried out when performing photometry on the resulting maps. Moreover, users will find local background subtraction unavoidable when working with much of the data, given the necessity of accounting for sky brightness, Galactic cirrus, etc.

\subsubsection{GALEX} \label{Subsubsection:GALEX_Imagery} 

UV observations made by the GALaxy Evolution eXplorer (GALEX; \citealp{Morrissey2007B}) were acquired from the GR6/7 data release \citep{LBianchi2014H}, hosted at the Mikulski Archive for Space Telescopes (MAST\footnote{\url{http://galex.stsci.edu/GR6/}}).

GALEX coverage is available for the vast majority of the DustPeda \hersc\ galaxies; 797 (91\%) in FUV, and 832 (95\%) in NUV. Coverage is more extensive in the NUV due to the fact that GALEX's FUV camera failed in June 2009, whilst the NUV camera continued operating until the satellite was decommissioned.

The {\sc GalexView} utility provided by MAST was used to identify all GALEX tiles in the vicinity of the DustPedia \hersc\ galaxies. These tiles were retrieved and co-added in line with the process detailed in Section~\ref{Subsection:Ancillary_Imagery}, but with some additional considerations specific to handling GALEX data, described here.

The final GALEX cutouts have pixel sizes of 3.2\arcsec\ in both FUV and NUV, in contrast to the standard GALEX pixel sizes of 1.5\arcsec; there are two reasons for this. In shallow GALEX tiles, particularly in the FUV, an extremely large fraction of the pixels have a value of zero -- pixels where no photons were detected during the exposure. Indeed, for the extremely shallow GALEX `all-sky' survey, over 99\%\ of the pixels in a FUV tile can be expected to be zero-value, and most of the pixels that {\it do} contain flux have a pixel value that represents only a single photon. This presents an issue when co-adding GALEX tiles. In order to perform the co-addition, it is necessary to re-sample all of the input images to a common pixel grid. But when re-sampling shallow GALEX tiles, this has the effect of spreading the flux in single-photon pixels over several reprojected pixels. Not only does this manifest as a very obvious aliasing artefact when inspected visually, but spreading the flux from a single photon over an extended area of sky is also physically dubious. Furthermore, in reprojected GALEX tiles with smaller numbers of zero-value pixels, noticeable moir\'e patterns often appear, introducing poorly-behaved noise on some scales. We avoided both of these artefacts by reprojecting the tiles to larger pixel sizes when performing the co-addition. Whilst this means that our GALEX maps are no longer Nyquist sampled, we will primarily be using them in concert with much lower-resolution data (WISE, \spitz, \hersc, etc), rendering this of less concern.

Care needs to be taken to correctly handle zero-value pixels in GALEX tiles. A zero-value pixel can either represent that a pixel lies outside of the coverage region for the observation, or it can represent a pixel that lies inside the coverage region but where no photons were detected during the exposure. We used the GALEX response maps corresponding to each tile to determine the exact region observed; only the zero-value pixels within the coverage area were included in mosaics.

Coverage in the outer portions of GALEX tiles tend to be of very low quality, due to aperture vignetting \citep{Morrissey2007B}. To prevent this leading to degradation of the final cutouts, we follow WiggleZ \citep{Drinkwater2010A} and GAMA \citep{Driver2016A} in masking all pixels outside the central 35\arcmin\ of each tile, excluding the outer $\sim$\,9\%\ of the coverage area.

When performing the co-addition of the GALEX images, each tile's contribution was appropriately weighted according to the exposure-time information provided with the tile.

The large numbers of pixels in shallow GALEX tiles that contain flux from few or zero photons also causes problems when attempting to match the background levels of tiles being co-added. The distribution of pixel values in shallow GALEX tiles is highly discrete and asymmetrical, especially at the distribution peak; this makes it difficult to ascertain the actual background level -- and hence difficult to match to the background levels of other tiles when co-adding. To address this, a copy of each tile was convolved with a 30\arcsec\ (20 pixel) diameter top hat kernel; this had the effect of smoothing out discrete pixel value levels. The pixel values of each smoothed map were iteratively sigma-clipped to a three-sigma threshold in order to remove particularly bright pixels; a flat plane was fit in a least-squares manner to the remaining pixels. The level of this plane was taken to represent the background level of the tile in question, and subtracted prior to co-addition.

The native GALEX tile pixel units of counts\,s$^{-1}$\,pix$^{-1}$ were translated to Jy\,pix$^{-1}$ using conversion factors of $1.076\times\,10^{-4}\,{\rm Jy\,counts^{-1}\,s}$ in the  FUV and $3.373\times\,10^{-5}\,{\rm Jy\,counts^{-1}\,s}$ in the NUV; these correspond to the standard GALEX AB magnitude zero points of 18.82\,mag and 20.08\,mag in FUV and NUV respectively \citep{Morrissey2007B}.

\subsubsection{SDSS} \label{Subsubsection:SDSS_Imagery}

The Sloan Digital Sky Survey (SDSS; \citealp{York2000B,Eisenstein2011B}) provides UV, optical, and NIR imaging of 35\%\ of the sky in the {\it ugriz} bands. SDSS data was acquired from SDSS DR12 \citep{Alam2015A}, hosted at the SDSS Science Archive Server\footnote{\url{http://dr12.sdss3.org/}}, yielding coverage for 656 (75\%) of the DustPedia galaxies.

The {\sc mArchiveList} tool, part of \montage, was used to identify all SDSS fields in the vicinity of the DustPedia galaxies. For each band, most places in the SDSS footprint have been observed more than once; however, for each point, a particular field -- the one deemed of the best quality -- is designated to be primary. Primary frames are those used by SDSS for photometry, mosaic generation, etc. Only fields designated as being primary by the SDSS were selected to be retrieved and co-added. By only using primary fields we ensure that only science-quality SDSS data is used.

Once the primary SDSS fields had been retrieved, they were co-added in line with the process described in Section~\ref{Subsection:Ancillary_Imagery}, with the exception that the final cutouts have pixel sizes of 0.45\arcsec. This was done to minimise the appearance of the conspicuous moir\'e patterns that otherwise occured when reprojecting to a North-East orientation. Given the median seeing full-width half-maximum (FWHM) of 1.43\arcsec\ for SDSS imaging fields\footnote{\url{http://www.sdss.org/dr12/imaging/other_info/}}, a pixel size of 0.45\arcsec\ is within the limit required for Nyquist sampling. Indeed, given that Nyquist sampling is achieved in diffraction-limited observations provided that pixel sizes are smaller than ${\it FWHM}/2.44$, the 0.45\arcsec\ pixels are suitable for frames with seeing-limited resolution \textgreater\,1.09\arcsec, which is the case for \textgreater\,97\%\ of SDSS frames\footnote{\url{http://www.sdss.org/wp-content/uploads/2014/10/psfhist.pdf}}.

Maps retrieved from the SDSS DR12 Science Archive Server have native units of `nanomaggies' (an SDSS convenience unit of linear flux\footnote{\url{http://www.sdss.org/dr12/algorithms/magnitudes/}}), scaled such that a flux in nanomaggies can be converted to an SDSS magnitude using a zero point magnitude of 22.5\,mag. SDSS magnitudes were designed to be AB magnitudes, but in practice have been found to exhibit offsets\footnote{\url{http://www.sdss.org/dr12/algorithms/fluxcal/\#SDSStoAB}} of $-0.04$\,mag offset in {\it u}-band (such that $u_{\rm AB} =  u_{\rm SDSS} - 0.04$), and $0.02$\,mag in {\it z}-band (such that $z_{\rm AB} =  z_{\rm SDSS} + 0.02$). We corrected the pixel units for these offsets, scaling them to the AB magnitude scale, and thereby allowing conversion to Jy\,pix$^{-1}$.

\subsubsection{DSS} \label{Subsubsection:DSS_Imagery}

The Digitized Sky Survey (DSS) provides UV, optical, and NIR coverage of the entire sky, in the form of scans of photographic survey plates from the Samuel Orschin telescope at Palomar Observatory, and the UK Schmidt Telescope at the Anglo-Australian Observatory.

As the DSS is produced from photographic plates, the pixel values are not in units of linear flux; rather, the units are `photographic densities'. The photographic density scale is not constant across the DSS, and varies from plate to plate. Whilst there are some general prescriptions for photometric calibration of the DSS, these only apply to individual contributing sub-surveys, and typically have calibrations no better than 0.5\,mag \citep{Doggett1996A}. Whilst a photometric calibration can be determined for each individual plate by comparison to standard reference stars, great care needs to be taken when doing so, particularly with regards to saturation of reference stars, and whether or not any portion of each target galaxy is also saturated. As such, we opt not to attempt a calibration of the DSS imagery, instead leaving it in its native units of photographic densities. However, even without photometric calibration, the DSS data is valuable, as it provides consistent optical coverage of our entire sample - including the 25\% of DustPedia not covered by the SDSS. For example, the DSS data can still be used to work out the optical axial ratio, position angle, and extent (not in an absolute isophotal sense, but using signal-to-noise analysis) of a galaxy; all useful information when determining apertures for our multiwavelength photometry (see Section~\ref{Subsection:Aperture_Fitting}).

DSS {\it B}-band and {\it R}-band imagery for the DustPedia \hersc\ galaxies was retrieved from the NASA SkyView service\footnote{\url{http://skyview.gsfc.nasa.gov/current/cgi/query.pl}}, and used to produce standard cutouts as described in Section~\ref{Subsection:Ancillary_Imagery}. However, given the complications detailed above, we made no attempt to convert the pixel values to units of Jy\,pix$^{-1}$; instead, they were left as photographic densities. Whilst the DSS provides UV and NIR imagery for some of the sky, we did not make use of these data, as GALEX, SDSS and 2MASS provide vastly superior coverage. The DSS is divided into two phases, DSS1 and DSS2, each made up of different sets of contributing sub-surveys. As the different contributing sub-surveys operate at a range of resolutions, no-attempt was made to co-add the DSS1 and DSS2 imagery; rather, cutouts for each were produced separately. As such, many sources have both DSS1 and DSS2 cutouts available. The DSS cutouts have pixel sizes of either 1\arcsec\ or 1.7\arcsec, depending upon which contributing sub-survey the relevant photographic plate scans came from.

\subsubsection{2MASS} \label{Subsubsection:2MASS_Imagery}

The 2 Micron All-Sky Survey (2MASS; \citealp{Skrutskie2006A}) provides NIR imaging of the entire sky in {\it J}-band, {\it H}-band, and {\it K$_{\it S}$}-band. 2MASS tiles in the vicinity of the DustPedia \hersc\ galaxies were identified and acquired from the NASA/IPAC Infrared Science Archive (IRSA\footnote{\url{http://irsa.ipac.caltech.edu/frontpage/}}). The retrieved 2MASS tiles were then co-added in line with the process described in Section~\ref{Subsection:Ancillary_Imagery}. The final cutouts retain the standard 2MASS All-Sky Data Release pixel size of 1\arcsec; the pixel units were rendered in Jy\,pix$^{-1}$ according to the zero points provided in the headers of the input archival maps. 

Note that severe sky brightness can give rise to significant aperture noise in the 2MASS bands -- especially in {\it H}-band, primarily due to OH emission in the atmosphere \citep{Jarrett2003A}. The sky brightness is typically seen to vary across angular scales of a few arcminutes, making it particularly troublesome for photometry of nearby galaxies. Because 2MASS observed in {\it JHK$_{S}$} concurrently, an almost identical sky brightness structure will be seen in all three bands, varying only in magnitude. This means that the resulting aperture noise will be strongly correlated between bands. If a given {\it H}-band flux is erroneously high, the {\it J}- and {\it K$_{S}$}-band fluxes are likely to be too high also -- although usually not to the same degree as in {\it H}-band, given the much worse sky brightness {\it H}-band tends to suffer.

\subsubsection{WISE} \label{Subsubsection:WISE_Imagery}

The Wide-field Infrared Survey Explorer (WISE; \citealp{Wright2010F}) provides NIR and MIR coverage of the entire sky, observing at 3.4\,\micron, 4.6\,\micron, 12\,\micron, and 22\,\micron. WISE imagery was obtained from the AllWISE data release Image Atlas, which combines WISE cryogenic and NEOWISE (Near Earth Object WISE; \citealp{Mainzer2011K}) survey phases. AllWISE Atlas images in the vicinity of the DustPedia \hersc\ galaxies were identified and retrieved from IRSA, and co-added in line with the process described in Section~\ref{Subsection:Ancillary_Imagery} (however, given the large area covered by each AllWISE Atlas image, co-addition was very rarely necessary). 

The final WISE cutouts retain the standard AllWISE Image Atlas pixel size of 1.375\arcsec. The map pixel units were rendered in Jy\,pix$^{-1}$ by using the zero-point magnitudes provided in the headers of the AllWISE Atlas images to convert pixel units to Vega magnitudes, which were then converted to AB magnitudes, and thence to Jy, according to the conversions given in Table~1, Section~IV.3.a of the Explanatory Supplement to the AllWISE Data Release Products \footnote{\url{http://wise2.ipac.caltech.edu/docs/release/allwise/expsup/}}. Extended-source corrections need to be applied when performing aperture photometry upon WISE maps, as per Table~5, Section~IV.4.c.vii of the WISE All-Sky Data Release Explanatory Supplement\footnote{\url{http://wise2.ipac.caltech.edu/docs/release/allsky/expsup/}}, to account for the fact that the calibration of WISE data is based upon profile fitting of point sources.


Noticeable image artefacts are occasionally encountered in the final WISE cutouts. However, as these are present in the input AllWISE Atlas images, and as there is typically only a single AllWISE Atlas image covering a given location on the sky, there is little that can currently be done to address them. The artefacts commonly encountered in WISE data are discussed in Sections IV.4.g and VI.2.c of the WISE All-Sky Data Release Explanatory Supplement.

It should also be noted that \citet{Wright2010F} and \citet{Brown2014C} have found evidence that the in-orbit spectral response functions of the WISE W3 (12\,\micron) and and W4 (22\,\micron) filters deviate from the laboratory-measured pre-launch functions. For monochromatic fluxes, \citet{Brown2014C} report that W4 filter's effective wavelength should be revised from 22.1\,\micron\ to 22.8\,\micron. However, the change for any given source depends on its spectral index in wavelength range of the W3 and W4 filters. Ultimately, this means W3 and W4 bands require more severe colour-correction than would otherwise have been necessary. As for other instruments, we perform no WISE colour-correction ourselves, as the full SED modelling necessary to determine the appropriate corrections is beyond the scope of this work. However, users of our WISE data products are reminded to be aware of the particular considerations necessary for the WISE 12 and 22\,\micron\ bands, and to ensure that they are correctly accounting for the filters' spectral response functions. WISE maps are calibrated assuming a $\nu^{-2}$ reference spectrum \citep{Wright2010F}.

\subsubsection{\textit{Spitzer}} \label{Subsubsection:Spitzer_Imagery}

The \spitz\ Space Telescope \citep{Werner2004B} provides imaging across 7 photometric bands in the NIR to FIR, observed using two cameras. The InfraRed Array Camera (IRAC; \citealp{Fazio2004G}) observes at 3.6\,\micron, 4.5\,\micron, 5.8\,\micron, and 8.0\,\micron; whilst the Multiband Imager for Spitzer (MIPS; \citealp{Rieke2004K}) observed at 24\,\micron, 70\,\micron, and 160\,\micron. Since depletion of the helium cryogen in 2009, only the IRAC 3.6\,\micron\ and 4.5\,\micron\ bands are able to operate. For each DustPedia \hersc\ galaxy, we acquired \spitz\ data from the best archive source available. For both IRAC and MIPS, an order of preference was used to decide which data to use, described below. Once data had been retrieved from the preferred source, it was otherwise handled as per Section~\ref{Subsection:Ancillary_Imagery}. The pixel units of all maps were converted from MJy\,sr$^{-1}$ to Jy\,pix$^{-1}$.

In total, 826 of the DustPedia galaxies (94\%\ of the total) have \spitz\ coverage in at least one photometric band (with data in 4 \spitz\ bands being the median for those sources that have coverage); 808 (92\%) have IRAC coverage, whilst 493 (56\%) have MIPS coverage.

\paragraph{\textit{Spitzer}-IRAC} \label{Subsubsubsection:Spitzer_IRAC_Imagery}

For IRAC, the preferred data source was the \spitz\ Survey of Stellar Structure in Galaxies (S$^{4}$G\footnote{\url{https://irsa.ipac.caltech.edu/data/SPITZER/S4G/}}; \citealp{Sheth2010B,Munoz-Mateos2013E,Querejeta2015E}), hosted by IRSA. S$^{4}$G was a \spitz\ post-cryogenic Exploration Science Legacy Program which provides high-quality dedicated imaging of nearby galaxies at 3.6\,\micron\ and 4.5\,\micron.

When S$^{4}$G data was not available, the next preferred IRAC data source was the \spitz\ Enhanced Imaging Products (SEIP), produced and hosted by IRSA. SEIP images combine \spitz\ observations made of a given part of the sky into `Super Mosaics', reduced using the latest version of the \spitz\ pipeline. Whilst SEIP data provides consistent data products for a large fraction of all \spitz\ observations, conspicuous imaging artefacts are sometimes encountered; these are most often due to the co-addition and mosaicing process being confounded by edge-of-field effects and/or by artefacts around bright sources. 

When SEIP data was not available, IRAC data was retrieved from the \spitz\ Heritage Archive (SHA\footnote{\url{http://sha.ipac.caltech.edu/applications/Spitzer/SHA/}}). When using SHA data, the procedure laid out in Section~\ref{Subsection:Ancillary_Imagery} was followed to produce the final cutouts; this process quite closely mimics the way in which the SEIP `Super Mosaics' were produced\footnote{\url{https://irsa.ipac.caltech.edu/data/SPITZER/Enhanced/SEIP/docs/seip_explanatory_supplement_v3.pdf}}. However, SEIP was preferred over SHA, as when IRSA produced the SEIP maps, they were able to  tailor their mosaicing process to partially mitigate the effects of \spitz-specific pathologies (eg, limiting the influence of artefacts during co-addition).

When performing aperture photometry on IRAC maps, it is necessary to perform calibration corrections, as given in Table~4.8, Section~4.11.2 of the IRAC Instrument Handbook\footnote{\url{https://irsa.ipac.caltech.edu/data/SPITZER/docs/irac/iracinstrumenthandbook/}}, to account for the fact that IRAC maps are calibrated using aperture photometry of point sources, using apertures of 12\arcsec\ radius. The calibration corrections ensure that the correct fluxes are measures when performing photometry of non-point sources, and/or when using apertures with radii other than 12\arcsec. Additionally, users who wish to perform surface-brightness analyses with the maps will have to consider the surface-brightness corrections described in Section~4.11.3 of the IRAC Instrument Handbook. IRAC maps are calibrated assuming a flat $\nu S_{\nu}$ reference spectrum, as described in Section~4.4 of the IRAC Instrument Handbook. 

\paragraph{\textit{Spitzer}-MIPS} \label{Subsubsubsection:Spitzer_MIPS_Imagery}

For MIPS, the preferred data source was the MIPS Local Galaxies Program \citep{Bendo2012C}, hosted by IRSA, which compiled and re-reduced all \spitz\ archive observations of local galaxies that were selected to be observed by 4 \hersc\ key programs -- \hersc\ Virgo Cluster Survey (HeViCS; \citealp{Davies2010A,Davies2012A}), the Dwarf Galaxy Survey (DGS; \citealp{Madden2013B}), the Very Nearby Galaxy Survey (VNGS), and the \hersc\ Reference Survey (HRS; \citealp{Boselli2010}) -- in order to provide standardised ancillary data for those surveys. This makes the MIPS Local Galaxies Program the ideal source of MIPS data for the sources it covers (25\%\ of the DustPedia \hersc\ galaxies).

When MIPS Local Galaxies Program data was not available, the next preferred source was data provided by the \spitz\ Legacy/Exploration Science Programs, hosted by IRSA. These are the final data products used and supplied by each program's team, and are consistently free of artefacts. We made use of MIPS data provided by the C2D (from molecular Cores To planet-forming Disks; \citealp{Evans2003B,Evans2009B}), SEP (\spitz\ {\sc mips} 24 and 70\,\micron\ imaging near the south Ecliptic Pole; \citealp{Scott2010B}), and SWIRE ({\sc Sirtf} Wide-area Infrared Extragalactic Survey; \citealp{Lonsdale2003B}\footnote{SWIRE DR2 documentation: \url{http://swire.ipac.caltech.edu/swire/astronomers/publications/SWIRE2_doc_083105.pdf}}) programs.

When neither MIPS Local Galaxies Program nor \spitz\ Legacy/Exploration Science Program MIPS data was available, we made use of SEIP MIPS data. Note that SEIP only includes 24\,\micron\ MIPS data.

If no other MIPS data source is available, we make use of SHA data in the same manner as for IRAC.

The reference spectrum for MIPS maps has the shape of a 10$^{4}$\,K blackbody, as described in Section~4.3.2 of the MIPS Instrument Handbook\footnote{\url{http://irsa.ipac.caltech.edu/data/SPITZER/docs/mips/mipsinstrumenthandbook}}.

\subsubsection{\textit{Planck}} \label{Subsection:Planck_Imagery}

The \planck\ satellite \citep{Planck2011I} mapped the entire sky in 9 bands, in the 350\,\micron\ to 1\,cm wavelength range, making \planck\ well-suited to detecting dust, free-free, and synchrotron emission from nearby galaxies. However, \planck\ has extremely low resolution compared to the other datasets we employ, with a beam FWHM ranging from 5--32\arcmin\ in size; a factor of 8--54 worse than the 36\arcsec\ resolution of the SPIRE 500\,\micron\ band (which is the lowest-resolution imagery we present aside from \planck). This stark difference in resolution makes it impractical to perform aperture-matched \caapr\ photometry with \planck\ maps in the same manner as for the other bands; instead, we will use catalogue photometry provided by the \planck\ collaboration (see Section~\ref{Section:Supplementary_Photometry}).

However, for completeness, we nonetheless produced \planck\ imaging cutouts for the DustPedia galaxies. \planck\ data was accessed from the NASA SkyView service, which uses the all-sky maps\footnote{Although the \planck\ all-sky maps are in the HEALPIX projection \citep{Gorski2005A}, the maps provided by SkyView are in the standard gnomonic TAN projection.} from the \planck\ Public Data Release (\citealp{Planck2013I}).

Due to the exceptionally large size of the \planck\ beam, our \planck\ imaging cutouts have diameters 4 times larger than our cutouts for other instruments; hence galaxies with optical angular sizes \textless\,6\arcmin\ are in 2\textdegree\,$\times$\,2\textdegree\ cutouts, whilst those with optical angular sizes \textgreater\,6\arcmin\ are in 4\textdegree\,$\times$\,4\textdegree\ cutouts.

All \planck\ maps are in our standard units of Jy\,pix$^{-1}$. The 350\,\micron\ (857\,GHz) and 550\,\micron\ (545\,GHz) cutouts were converted from their native units of MJy\,sr$^{-1}$, whilst the other 7 bands were converted from their native units of ${\rm K_{CMB}}$ according to the conversions provided by the Explanatory Supplement to the \planck\ Catalogue of Compact Sources\footnote{\url{https://irsa.ipac.caltech.edu/data/Planck/ercsc_v1.3/explanatory_supplement_v1.3.pdf}}. For the 350\,\micron--3.00\,mm \planck\ bands, the data is calibrated assuming a constant $\nu S_{\nu}$ reference spectrum \citep{Planck2015VII}; for the 4.26--10.60\,mm \planck\ bands, the reference spectrum is the Cosmic Microwave Background (CMB), a 2.73\,K blackbody \citep{Planck2015II}.

\section{The CAAPR Photometry Pipeline} \label{Section:Pipeline}

Here we describe the Comprehensive \& Adaptable Aperture Photometry Routine (\caapr), which we use to produce consistent multiwavelength aperture-matched photometry of the DustPedia galaxies from our GALEX, SDSS, 2MASS, WISE, \spitz, PACS, and SPIRE imagery.

\caapr\ is a development of the photometry pipeline employed by \citet{CJRClark2015A}, \citet{DeVis2017A}, \citet{DeVis2017B}, and Keenan et al. (in prep.). Its development and functionality are driven by three main motivations. 

Firstly, \caapr\ was created to be able to generate aperture-matched photometry that is cross-comparable, even in the face of the broad gamut of data that are required for multiwavelength astronomy -- producing fluxes {\it and uncertainties} in a consistent manner, despite the great variation in the characteristics of observations ranging from the UV to the IR to the submm.

Secondly, \caapr\ allows automated standardised photometry for local Universe surveys, as has long been the norm for intermediate-to-high redshift astronomy. Photometry of nearby galaxies generally entails a large degree of `by hand' tweaking, such as altering aperture dimensions to maximise flux and signal-to-noise ratio (S/R), masking of contaminating foreground and background sources, and manually selecting an appropriate background region. This manual approach was taken by, for instance, the \hersc\ Reference Survey (HRS, see Section~\ref{Subsubsection:HRS_Validation}), the Third Reference Catalogue \citep{deVaucouleurs1991A}, the SCUBA Local Universe Galaxy Survey (SLUGS; \citealp{Dunne2000A}), the 2MASS Large Galaxy Atlas \citep{Jarrett2003A}, the \spitz\ Infrared Nearby Galaxy Survey (SINGS; \citealp{Dale2005D}), the GALEX Ultraviolet Atlas of Nearby Galaxies \citep{GilDePaz2007A}, the Key Insights on Nearby Galaxies Far-Infrared Survey with \hersc\ (KINGFISH; \citealp{Dale2012A}), the WISE High-Resolution Galaxy ATLAS \citep{Jarrett2012A}, {\it ad nauseum}. However, this manual approach is a luxury we cannot afford with DustPedia, as we deal with  a larger number of targets, and a much more extensive range of bands, than is typically the case for nearby galaxy surveys. As such, \caapr\ requires minimal human interaction once started (except for uncommon occasions where visual inspection leads us to exclude a certain band from contributing to a source's aperture fitting, or disable star removal for a particular map, etc). This automated approach allows us to handle the scale of the DustPedia dataset, and be confident that all of our measurements have been conducted in a consistent and cross-comparable way. 

Thirdly, \caapr\ is designed to be able to handle the wide range of shapes, sizes, brightness distributions, and other morphological traits found in nearby galaxies (and to do so across the very different resolutions and noise environments found in multiwavelength datasets). Naturally, aperture photometry requires that the shape and size of a target source is well constrained, in order to construct an appropriate aperture -- however this is notoriously difficult for nearby galaxies, which are often `shredded' (ie, erroneously identified as multiple separate sources). The widely-used {\sc SExtractor} software \citep{Bertin1996A} is prone to shredding very extended sources \citep{Hammer2010B,Hammer2010D,Wright2016B}, and the shredding of nearby galaxies in the SDSS is well-documented \citep{Blanton2001A,Hall2012A,Budavari2009A}. Indeed, when describing the {\sc lambdar} photometry pipeline (which is designed for higher-redshift galaxies), \citet{Wright2016B} give nearby galaxy NGC\,5690 as an example of where their multi-pass {\sc SExtractor}-derived aperture generation fails -- even though NGC\,5690 is relatively `well-behaved' from the standpoint of nearby galaxy aperture fitting, being bright, edge-on, and only somewhat flocculent (by coincidence, NGC\,5690 is one of the galaxies in the sample of \citealp{CJRClark2015A}, for which \caapr\ was originally developed, and where \caapr's aperture generation performs successfully).

\caapr\ takes the form of a Python 2.7 package, and is available on GitHub\footnote{\url{https://github.com/Stargrazer82301/CAAPR}}. \caapr\ is operated via a single high-level function, with two primary inputs -- a table describing the properties of each source, and a table describing the properties of each band that is to be processed -- along with a number of optional arguments (eg, stating whether \caapr\ is to operate using parallel processing, or specifying that only the aperture-fitting phase of the pipeline should be run). The input tables allow the user to alter how the pipeline runs for individual sources (eg, the user might want to exclude the $r$-band map of a given source from contributing to the shape of the final photometric aperture, due to contamination from a satellite trail), and set up the pipeline to run appropriately in each band (eg, the user is likely to want to enable foreground star removal at 3.4\,\micron, but disable it at 500\,\micron). 

\subsection{Preprocessing} \label{Subsection:Preprocessing}

\caapr\ can operate with separate input maps provided for each target in a given band, or with a single input map for all targets in a given band. In either case, the user has the option of extracting a cutout map centred, on the target source, which will then be used throughout the rest of the pipeline (this is often necessary if the input map is 
particularly large, and hence memory-intensive); this cutout will be reprojected to a gnomonic TAN projection. Note that for purposes of consistency, the maps being processed by \caapr\ will be referred to as cutouts for the rest of this section.

\subsection{Foreground Star Removal} \label{Subsection:AstroMagic}

\begin{figure}
\begin{center}
\includegraphics[width=0.475\textwidth]{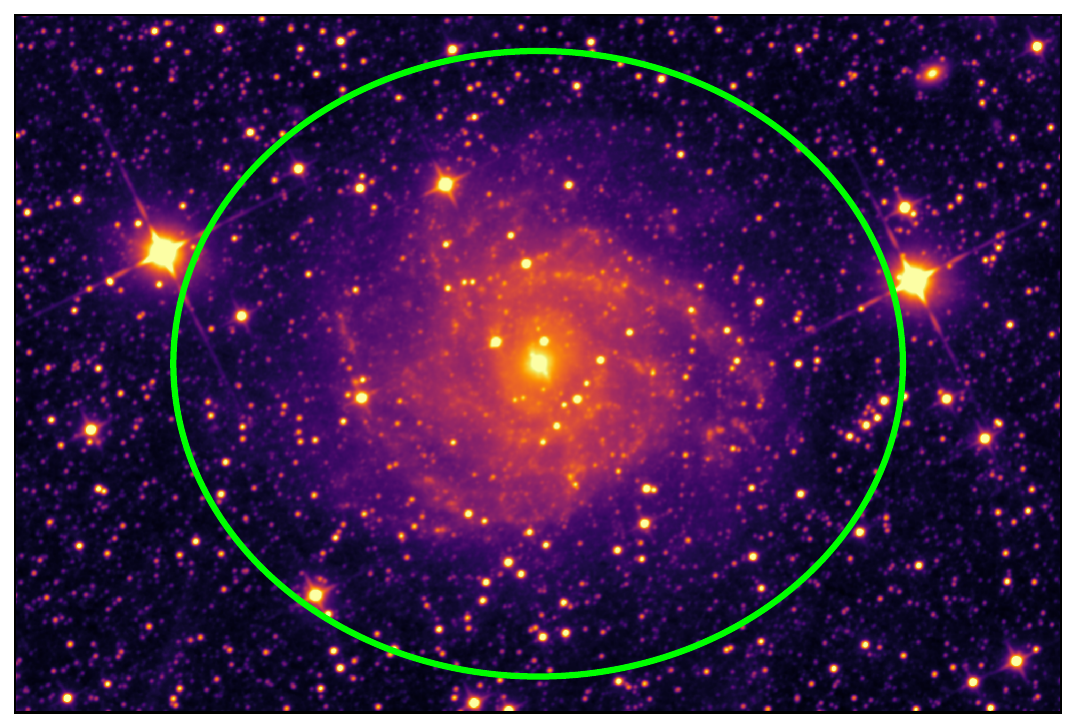}
\includegraphics[width=0.475\textwidth]{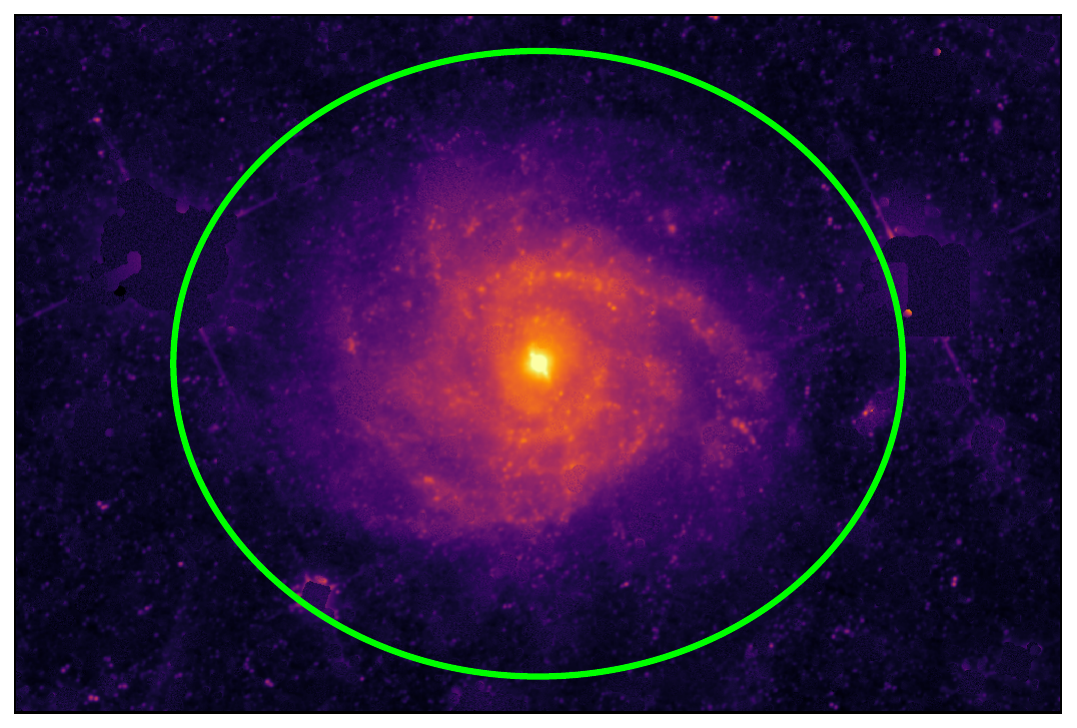}
\caption{Demonstration of our foreground star removal process, using the example of WISE 3.4\,\micron\ observations of IC\,0342, over a 45\arcmin\,$\times$\,30\arcmin\ area of sky, shown before ({\it upper}) and after ({\it lower}) undergoing star removal. Both images are displayed using the same logarithmic brightness scale, to allow both bright and faint features to be seen. Photometric master aperture ellipse, as determined by \caapr, is marked in green. IC\,0342 lies in the plane of the Milky Way, well within the zone of avoidance ($b = 10$\textdegree), leading to a high degree of foreground stellar contamination. Nonetheless, our foreground star removal process successfully subtracts the vast majority of all emission from foreground stars in the vicinity of IC\,0342.}
\label{Fig:StarSub_Example}
\end{center}
\end{figure}

Contamination from foreground stars in UV--MIR bands is minimised using the foreground star removal functionality included in the Python Toolkit for SKIRT (PTS\footnote{\url{http://www.skirt.ugent.be/pts/}}; \citealp{Camps2015C}, Verstocken et al.\,in prep.).

The foreground star removal procedure works on a band-by-band basis and runs semi-automatically, requiring a small amount of adjustment of the configuration parameters. For a particular image, the appropriate entries from the 2MASS All-Sky Catalog of Point Sources \citep{Cutri2003D} are retrieved as a starting point. The catalogue is queried through the VizieR interface of Astroquery\footnote{\url{https://github.com/astropy/astroquery}} Python library. Around each catalogued position, a small patch of the image is subtracted by the estimated background in that area and subsequently a local peak is searched, using tools from the Photutils\footnote{\url{https://github.com/astropy/photutils}} Python package. If a peak with a reasonable S/N is not found within a radius of a few pixels, the position is ignored. If a matching peak is found, it has to show no deviations from being a true point source (other than the peak exhibiting saturation in the case of the brightest stars), otherwise it will also be ignored. This is an effective way of preventing most compact sources in the target galaxies, primarily H{\sc ii} regions, from being removed -- as such objects tend to lie within local peaks in a galaxy's brightness distribution. For example, even in the uncommon cases where H{\sc ii} regions display no extended nebulosity at all, they generally lie along the brightness `ridge' of a galaxy's spiral arms; this makes their brightness profile deviate from being a true point source.

The FWHM of a positively identified point source determines the size of the area which will be masked and replaced with an interpolation. For the brightest point sources, image segmentation is used to detect saturation bleed, ghosts, and diffraction artefacts. The segmentation is also replaced with an interpolation. This interpolation is performed by approximating the source neighbourhood by a 2-dimensional polynomial function, whilst including Gaussian noise based on the average deviation of the pixel values from this polynomial. Figure~\ref{Fig:StarSub_Example} shows a WISE 3.4\,\micron\ map of IC\,0342 before and after undergoing the foreground star removal process.

The effectiveness of the foreground star removal is explored in-depth in Section~\ref{Section:Astromagic_Validation}. In uncommon instances where visual inspection revealed that the foreground star removal caused problems (such as confusing a bright H{\sc ii} region in the target galaxy for a star, and hence removing it), then \caapr\ was run with the star removal disabled; this was required for approximately 1--2\%\ of cases.

\subsection{Polynomial Sky Flattening} \label{Subsection:Polynomial_Sky_Flattening}

\begin{figure*}
\begin{center}
\includegraphics[width=0.32\textwidth]{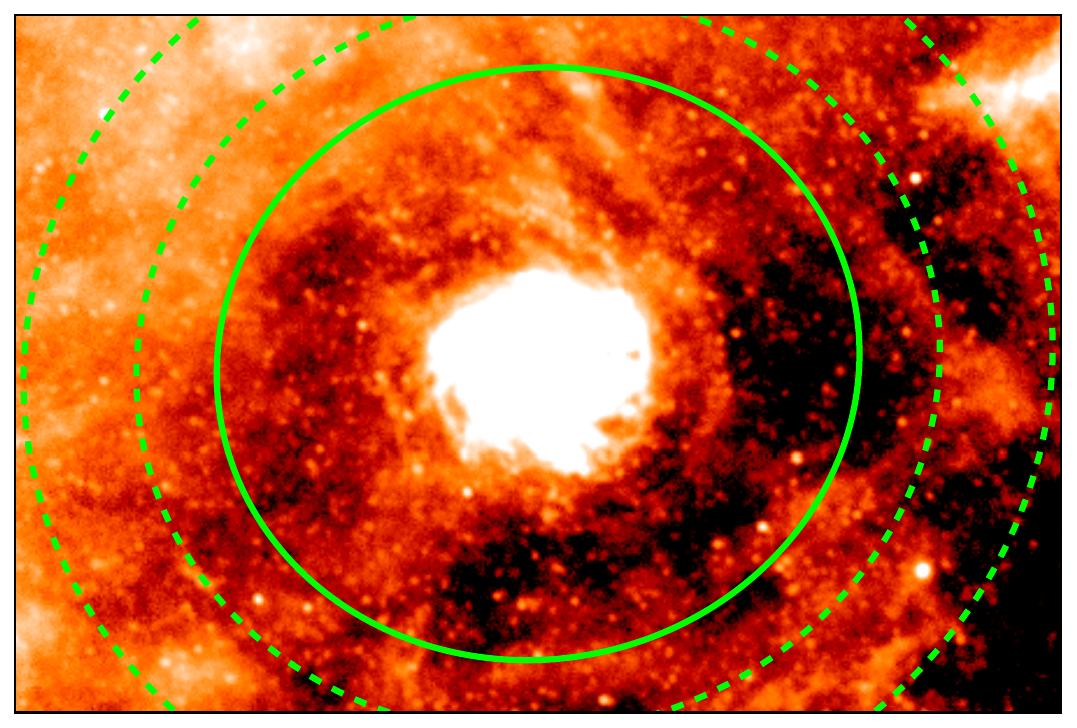}
\includegraphics[width=0.32\textwidth]{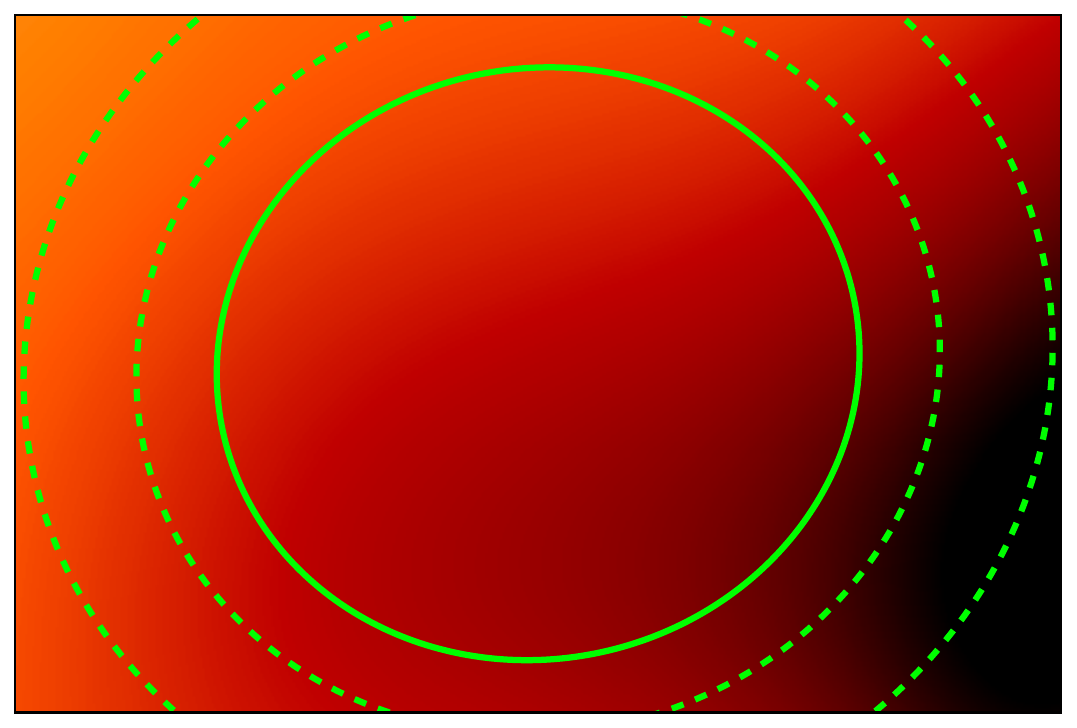}
\includegraphics[width=0.32\textwidth]{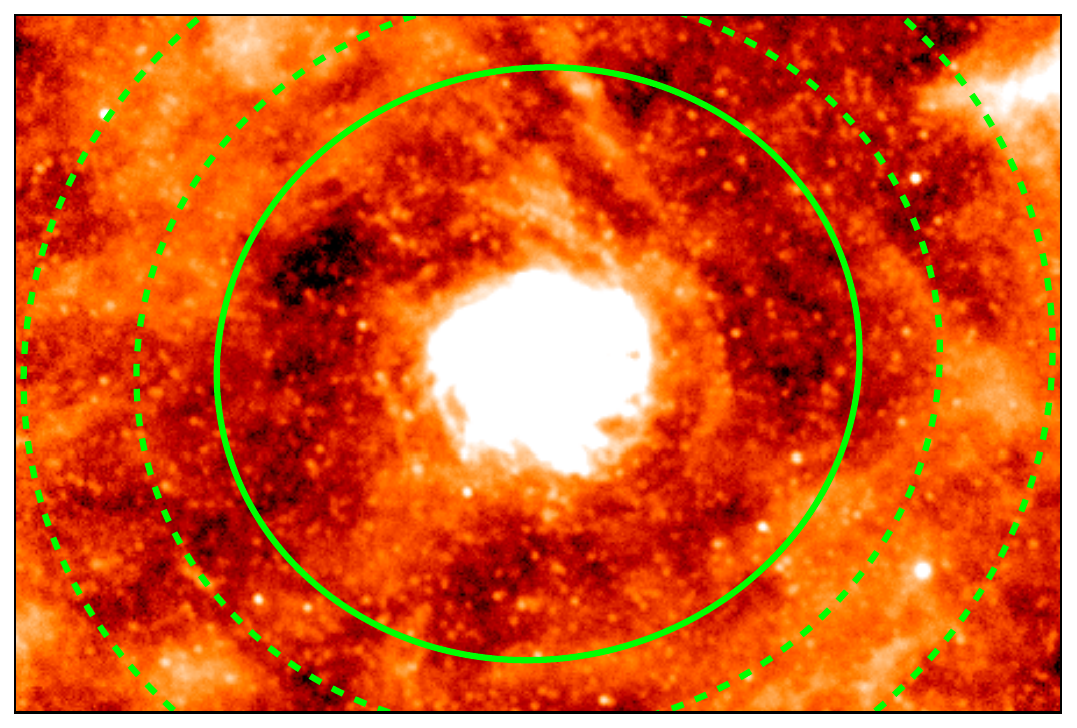}
\caption{Illustration of our polynomial sky subtraction, with our SPIRE 250\,\micron\ map of NGC\,6946, over a 45\arcmin\,$\times$\,30\arcmin\ portion of the map. As can be seen, NGC\,6946 suffers from a high degree of contamination from foreground Galactic cirrus. Photometric master aperture ellipse, as determined by \caapr, indicated by solid green line; background annulus is demarked by dashed green lines. {\it Left:} Original, un-subtracted map. {\it Central:} Best-fit 2-dimensional 5\textsuperscript{th}-order polynomial sky model used to remove large-scale emission, {\it Right:} Map after undergoing polynomial sky subtraction (displayed using same colour scale as in the left panel); note that background has been appreciably flattened.} 
\label{Fig:PolySub_Example}
\end{center}
\end{figure*}

After star-subtraction, \caapr\ removes large-scale sky structure in each cutout, if necessary, by means of fitting and subtracting a 2-dimensional 5\textsuperscript{th}-order polynomial to the sky in the cutout (following \citealp{Auld2013A}, a 5\textsuperscript{th}-order polynomial provides the flexibility to model relatively a complex sky, whilst mitigating the risk of over-fitting). This is useful for removing foreground emission (eg, Galactic cirrus in MIR--submm, sky brightness in NIR), or an instrumental gradient (which is particularly common in GALEX and \spitz\ maps). The target source, and any other bright sources, are masked in order to prevent them from influencing the sky polynomial fitting.

To mask the target source, a preliminary run of the aperture-fitting process (which is described fully in Section~\ref{Subsection:Aperture_Fitting}) is conducted; the semi-major and -minor axes of the fitted ellipse are doubled, and this region is then masked. To mask any other bright sources, the pixel values in the cutout are run through an iterative sigma-clip with a 3\,$\sigma$ threshold; after convergence, all pixel values that lie above the final threshold are masked. Once masking is completed, \caapr\ fits the sky polynomial to all remaining un-masked pixels in the cutout via least-squares minimisation.

However, in cutouts where the sky is already flat and well-behaved, subtracting the polynomial sky fit is undesirable; at best it is unnecessary, and at worst it may actually degrade the sky quality. Therefore, \caapr\ evaluates the cutout's pixel distribution before and after the sky polynomial is subtracted. For each distribution, only pixel values below the peak are considered in this evaluation, as these values should represent pixels corresponding to `empty' sky. In cases where the sky polynomial subtraction improves the sky quality -- flattening the sky -- the range of pixels values below the distribution peak should be narrowed, as there should be a narrower range of pixel values corresponding to empty sky. To quantify this width, \caapr\ finds the root-mean-squared (RMS) deviation between the peak pixel value and all of the values beneath it, for both the polynomial-subtracted and polynomial-unsubtracted distributions. If the RMS width for the subtracted distribution is at least 10\%\ smaller than that of the unsubtracted distribution, it is assumed that the polynomial subtraction has lead to the sky being appreciably flattened, and the polynomial-subtracted cutout is utilised for all subsequent stages of the pipeline; otherwise, the original cutout is used, unaltered. The polynomial sky subtraction process is demonstrated in Figure~\ref{Fig:PolySub_Example}, using the example of NGC\,6946. Note that the polynomial sky flattening is only conducted as part of our \caapr\ photometry; the FITS images provided in our data products (see Section~\ref{Subsection:FITS Images}) have not undergone this process.

\begin{figure*}
\begin{center}
\includegraphics[width=0.195\textwidth]{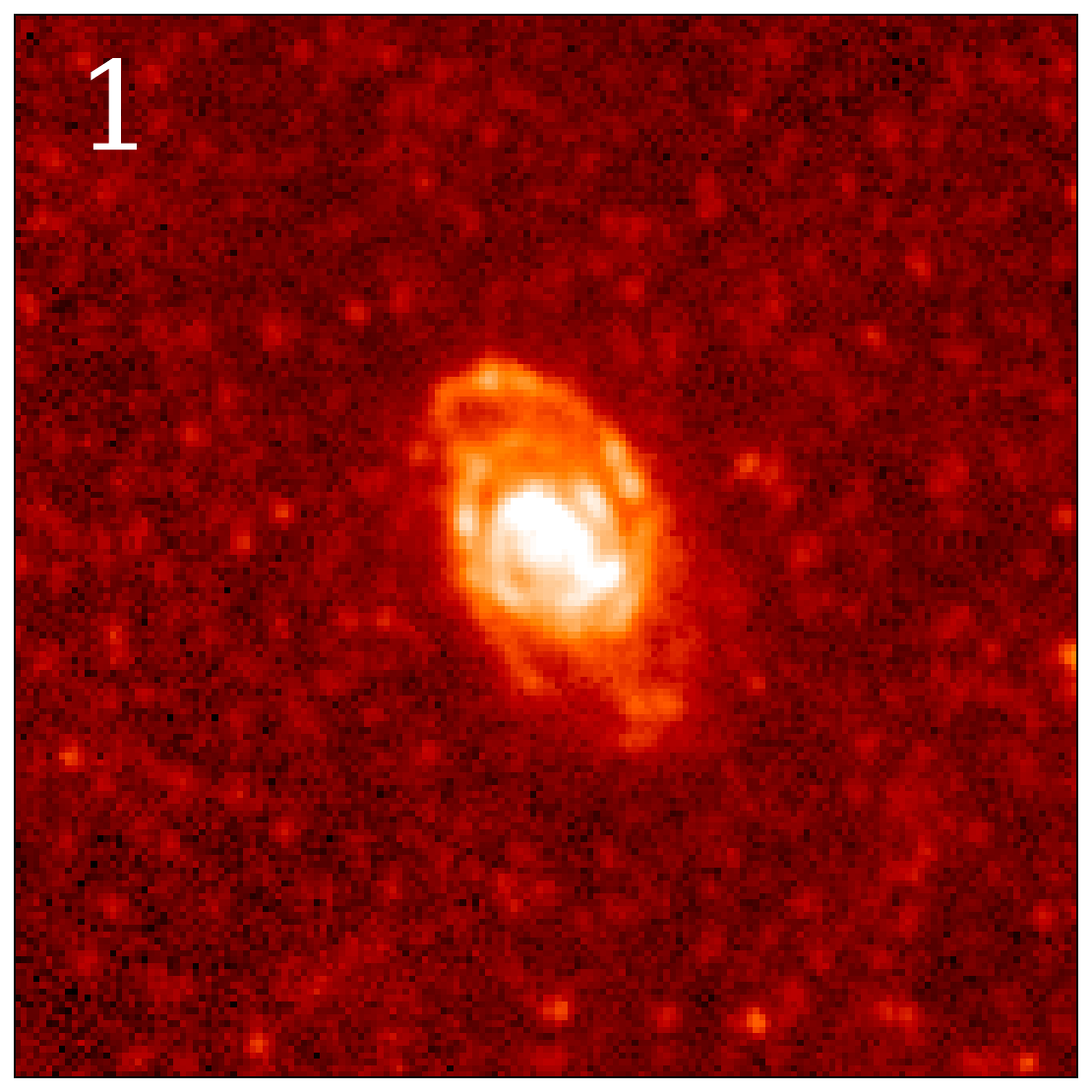}
\includegraphics[width=0.195\textwidth]{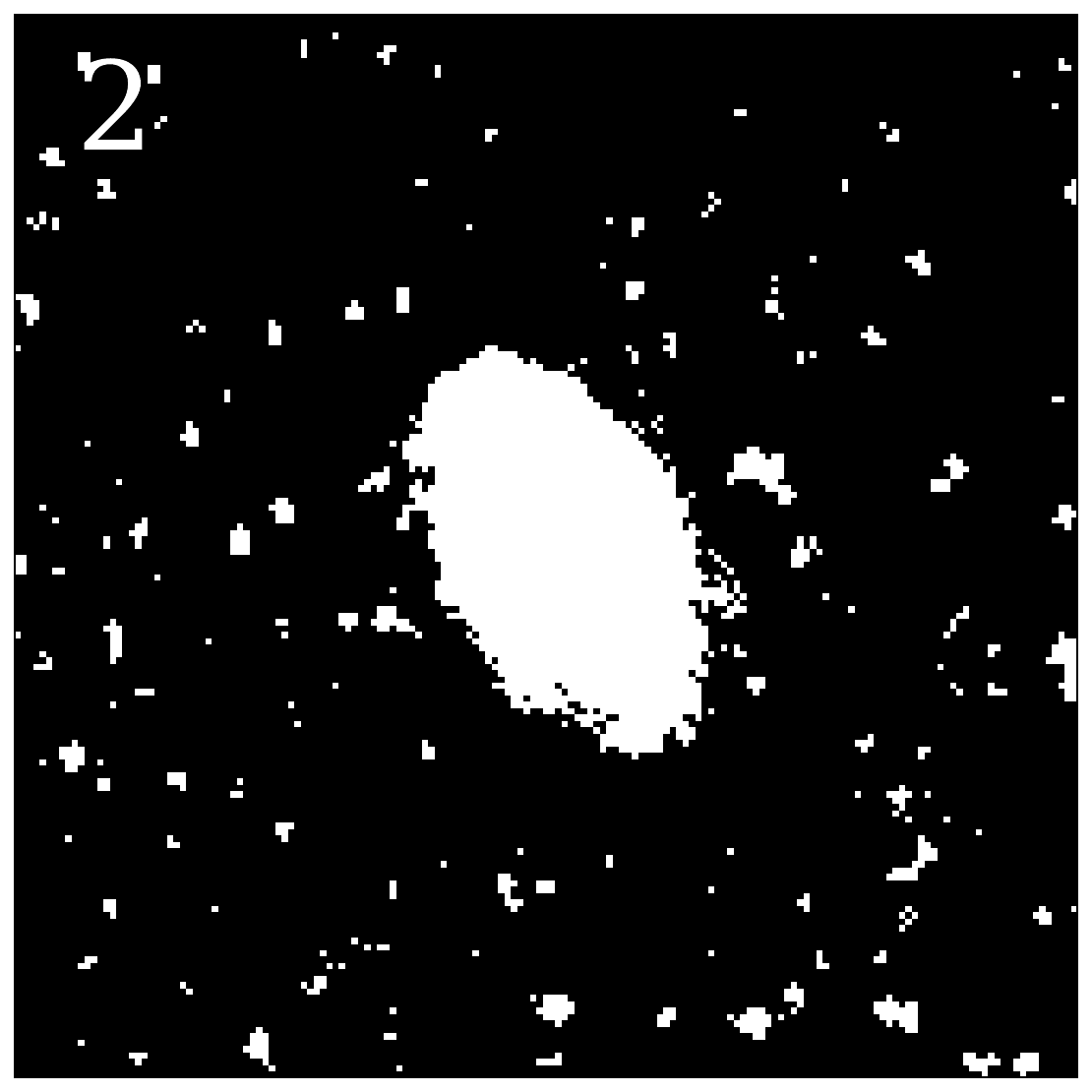}
\includegraphics[width=0.195\textwidth]{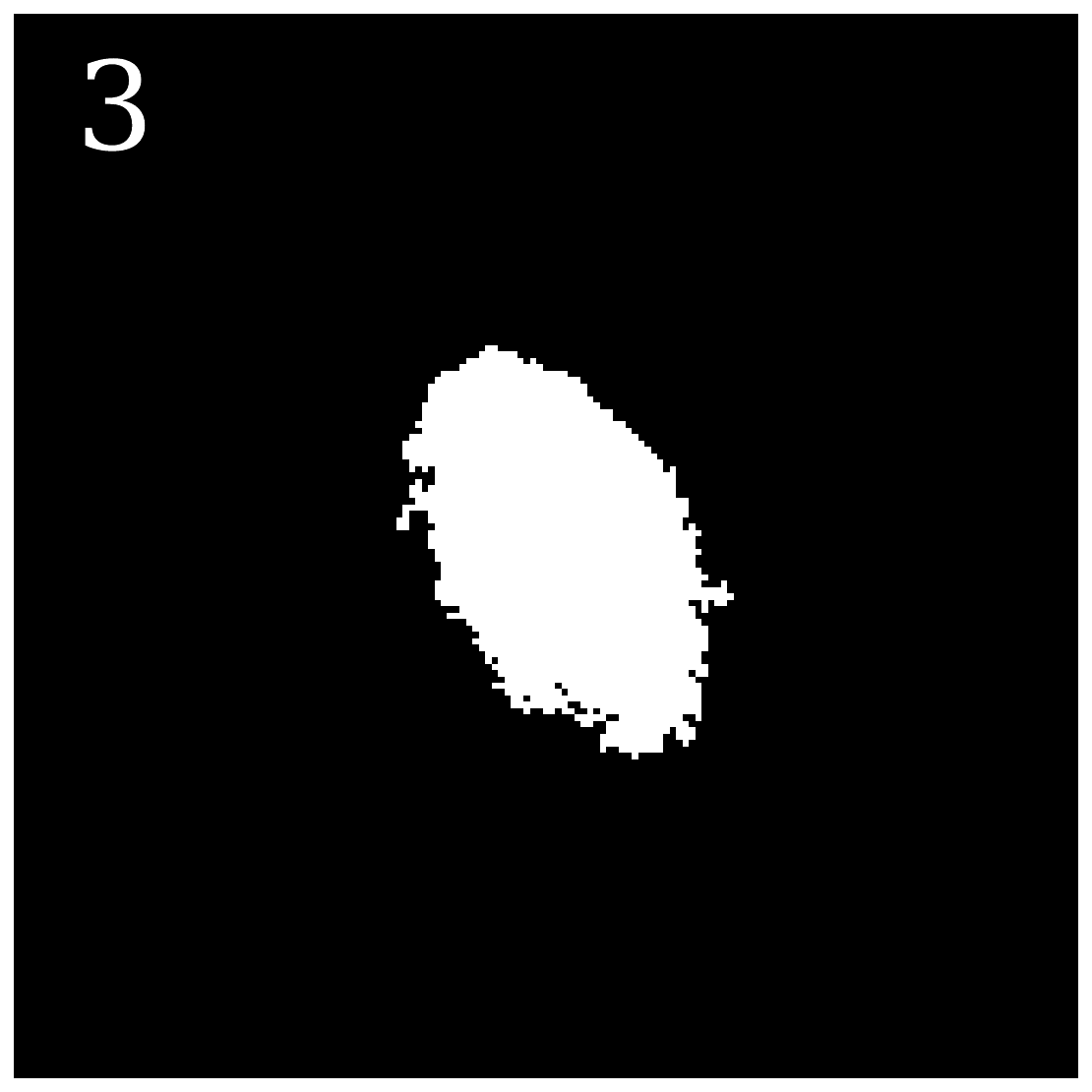}
\includegraphics[width=0.195\textwidth]{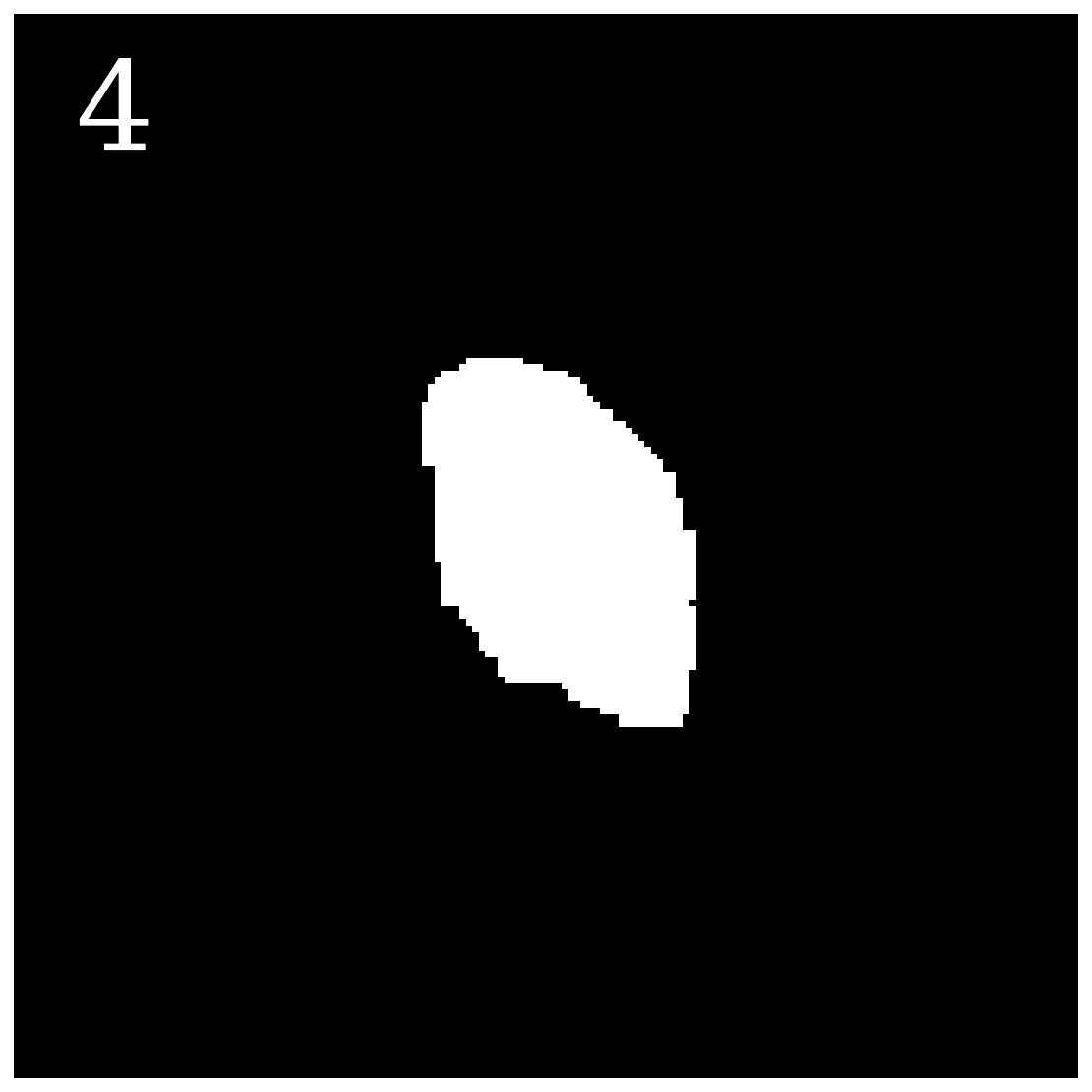}
\includegraphics[width=0.195\textwidth]{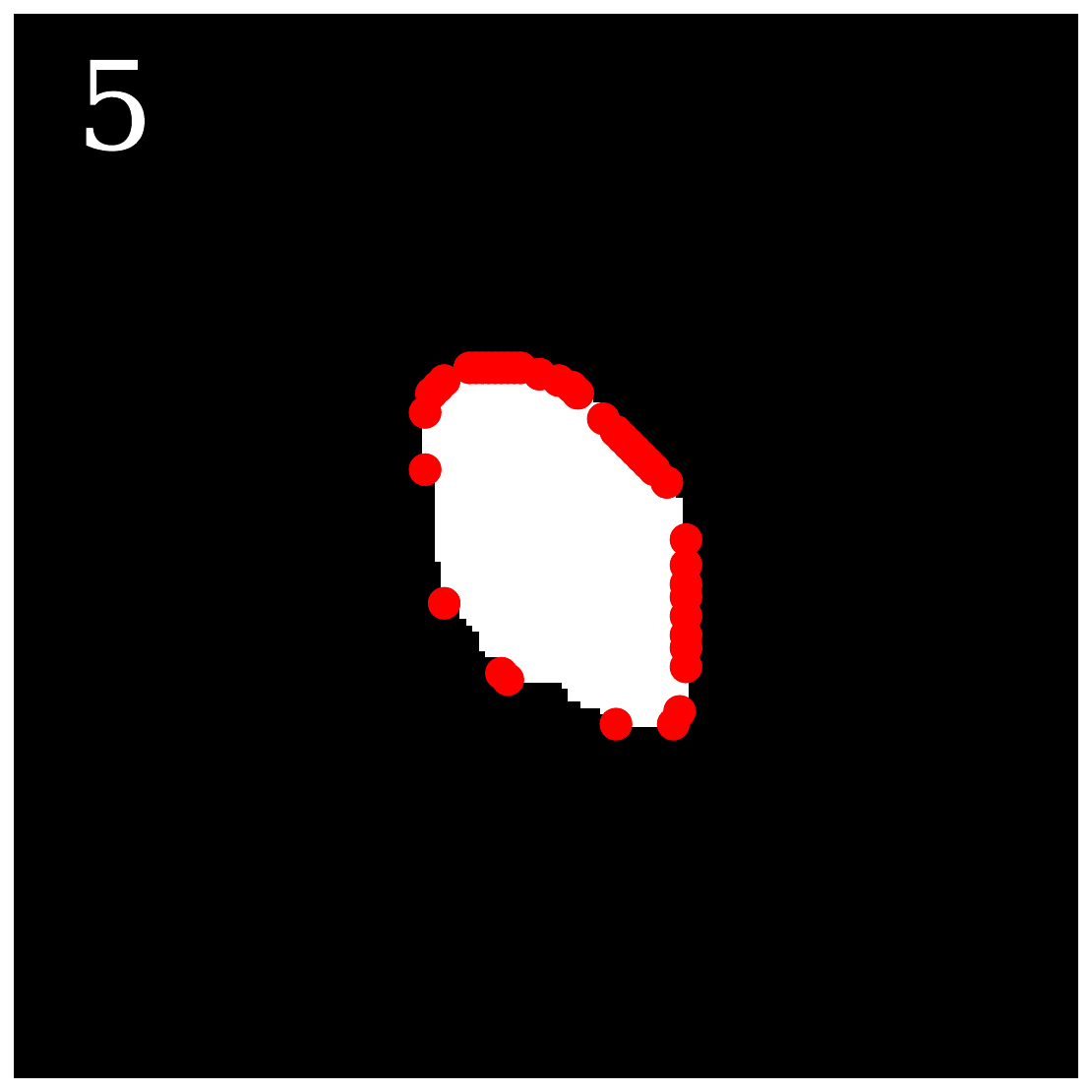}
\includegraphics[width=0.195\textwidth]{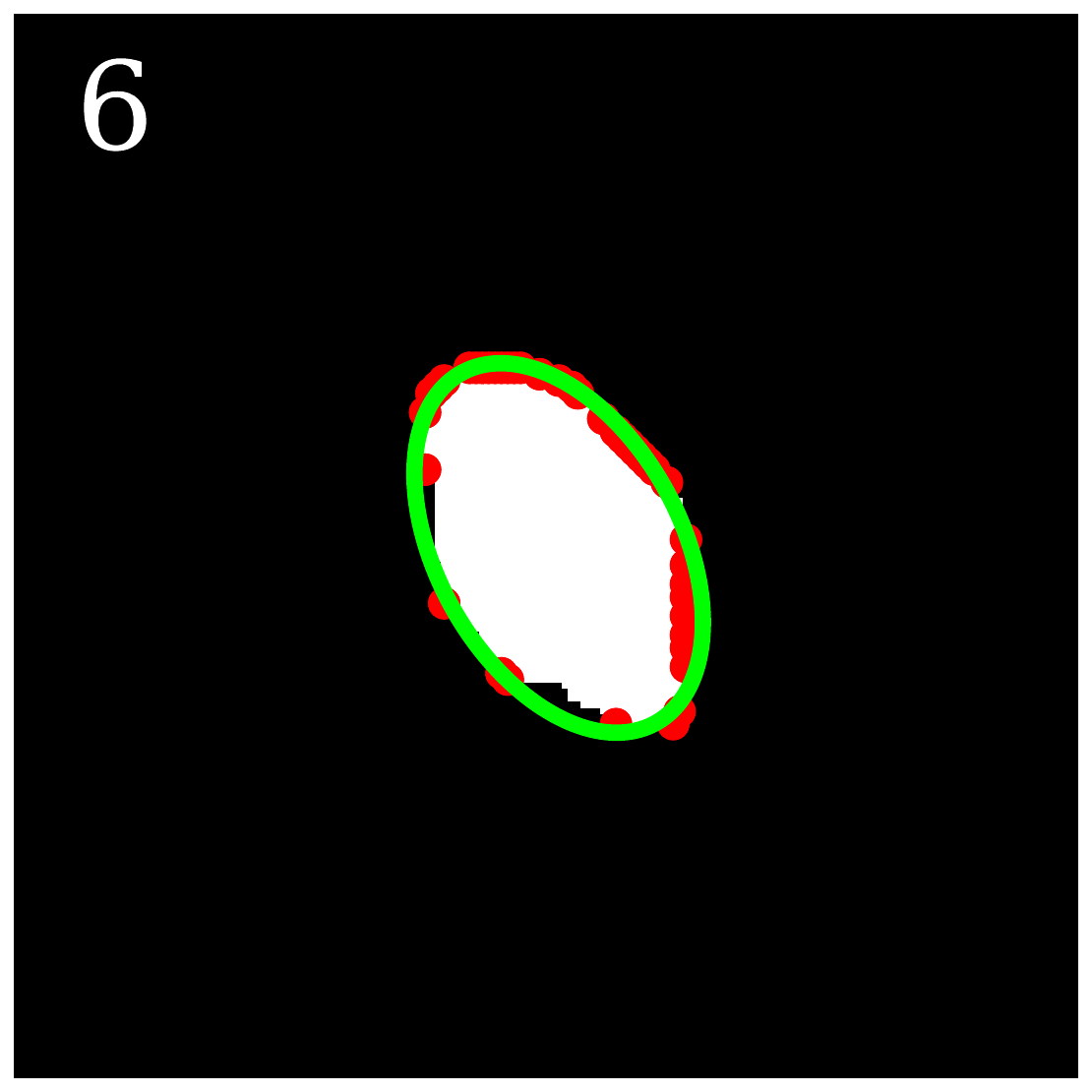}
\includegraphics[width=0.195\textwidth]{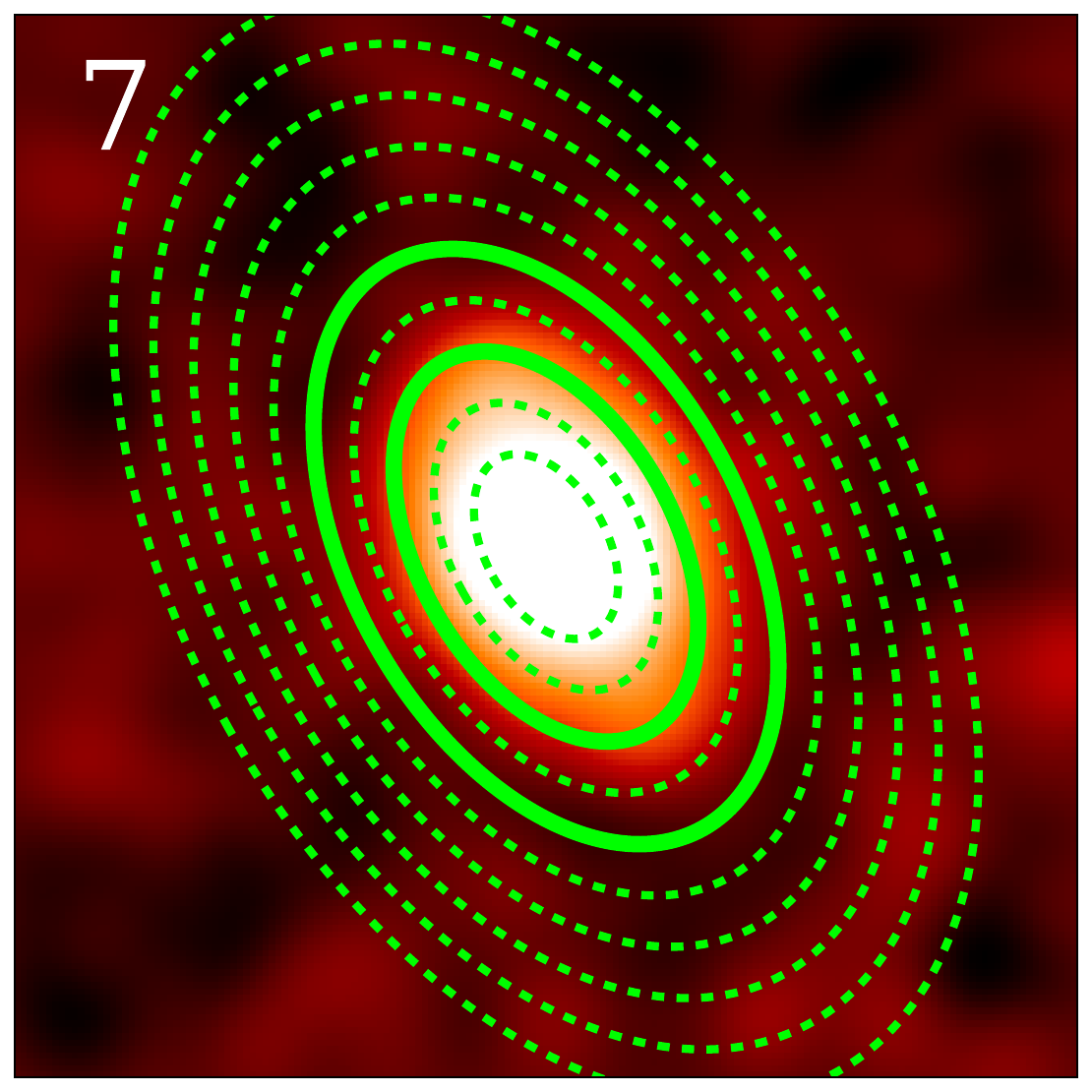}
\includegraphics[width=0.195\textwidth]{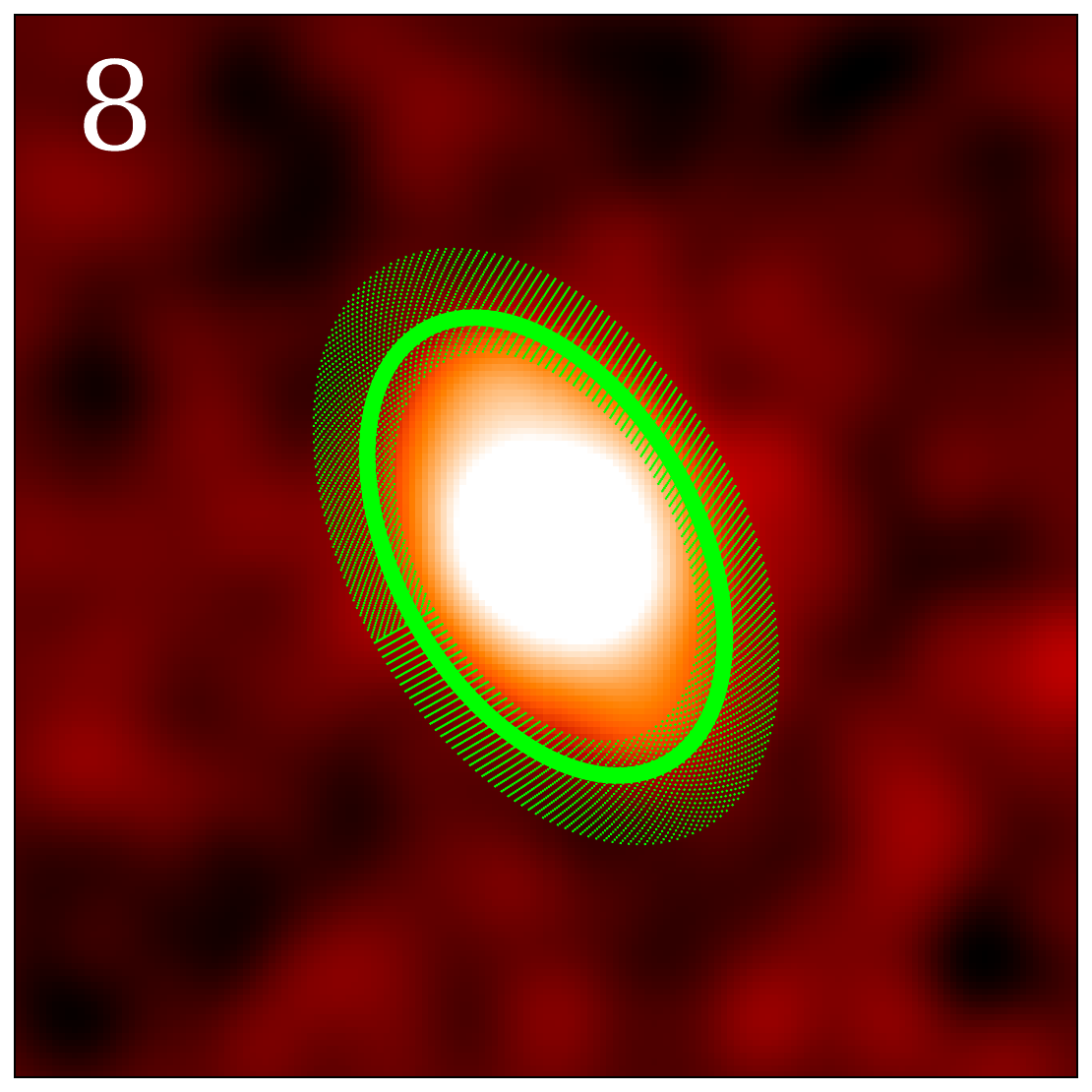}
\includegraphics[width=0.195\textwidth]{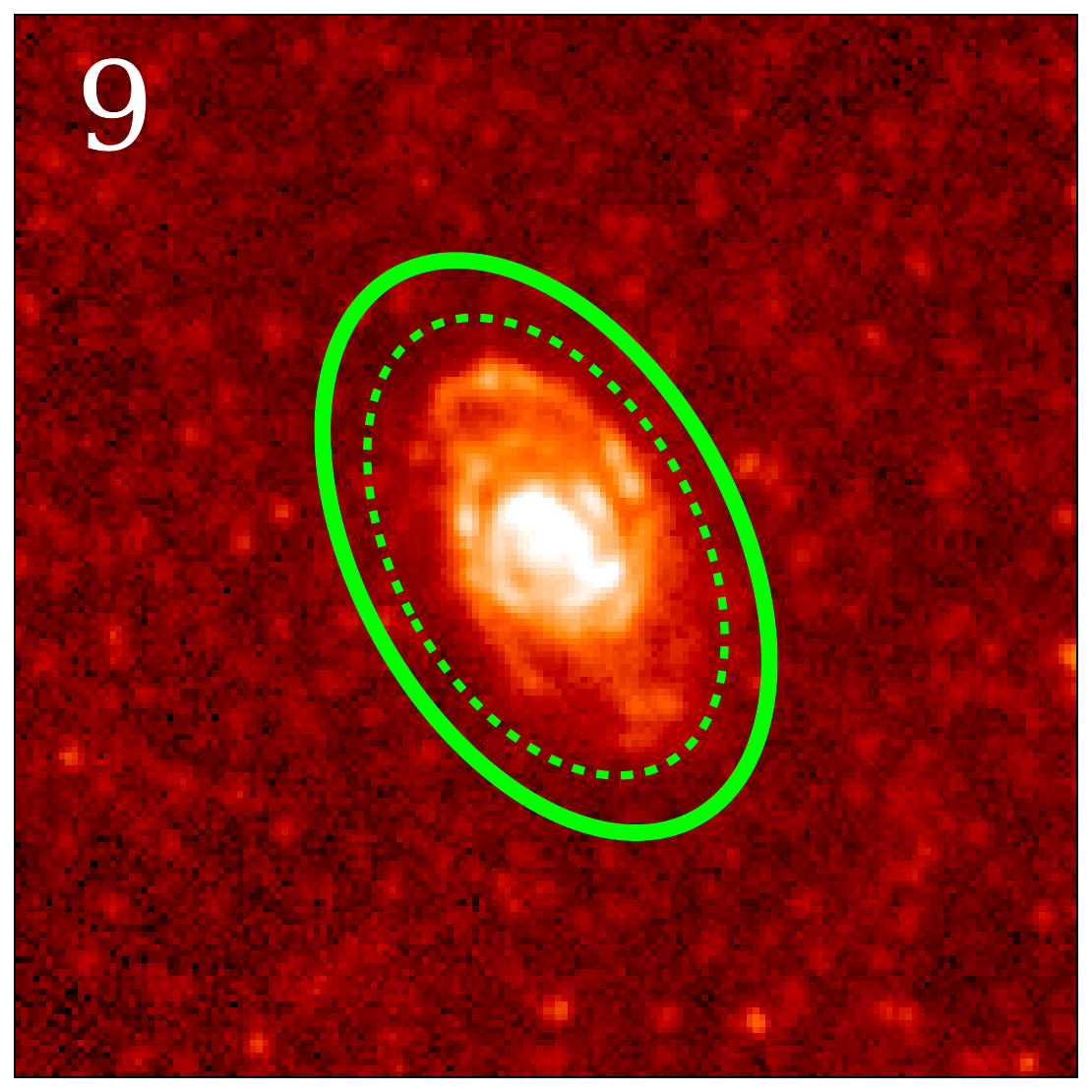}
\includegraphics[width=0.195\textwidth]{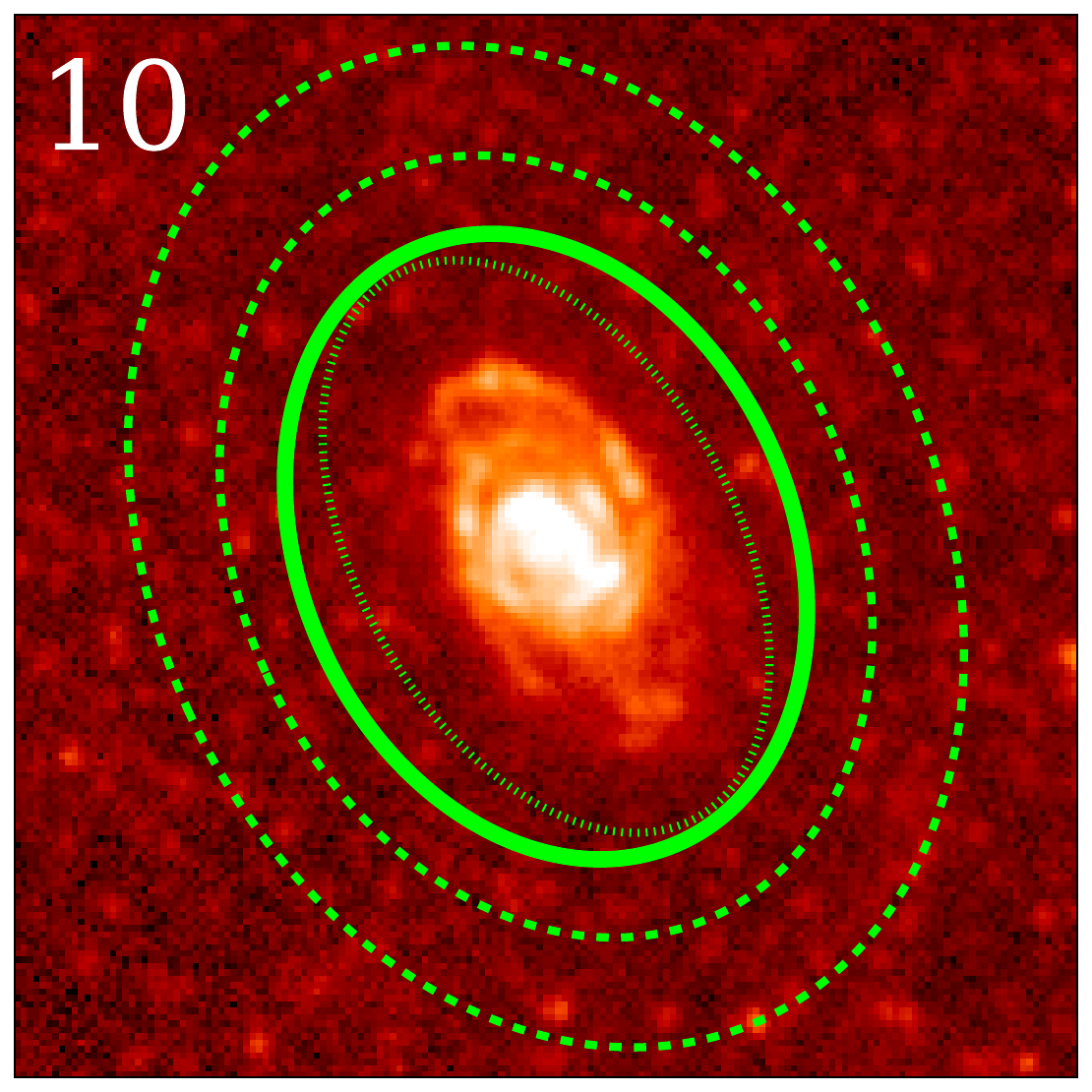}
\caption{Stages of the \caapr\ aperture-fitting process, using SPIRE 250\,\micron\ data for DustPedia galaxy NGC\,5364 as an example. {\it Panel~1:} Central $1000\arcsec\times 1000\arcsec$ portion of map. {\it Panel~2:} All of the pixels with S/R\,\textgreater\,3.{\it Panel~3:} Contiguous set of significant pixels isolated as being associated with target source. {\it Panel~4:} Isolated pixels after being subjected to binary opening. {\it Panel~5:} Vertices of convex hull of isolated pixels (red dots). {\it Panel~6:} Best-fit ellipse to convex hull of isolated pixels (solid line). {\it Panel~7:} Smoothed copy of map, with coarse radial profiling (dashed lines) to find interval within which S/R\,=\,2 (solid lines). {\it Panel~8:} Region that undergoes detailed radial profiling (dotted shading) to find exact semi-major axis at which S/R\,=\,2 (solid line). {\it Panel~9:} Resulting elliptical aperture, before (dashed line) and after (solid line) being multiplied by 1.25 expansion factor. {\it Panel~10:} Final photometric aperture (solid line), found by combining aperture fitted in this band (dotted line) with those fitted in all other bands; final background annulus is also shown (dashed lines).}
\label{Fig:CAAPR_Aperture_Fitting_Example}
\end{center}
\end{figure*}

\subsection{Aperture Fitting} \label{Subsection:Aperture_Fitting}

\caapr's aperture fitting process generates an elliptical aperture in every band for a given target; these apertures are then combined (as described in Section~\ref{Subsection:Aperture_Combining}) to provide a final aperture for the target. The aperture fitting process consists of two distinct stages; first determining the position angle and axial ratio (ie, the {\it shape}) of the aperture ellipse, then second determining the semi-major axis (ie, the {\it size}) of the aperture ellipse. The stages of the aperture-fitting process are illustrated in Figure~\ref{Fig:CAAPR_Aperture_Fitting_Example}.

To determine the position angle and axial ratio of the aperture ellipse, \caapr\ starts by estimating the noise in the cutout, by finding the iteratively sigma-clipped (with a 3\,$\sigma$ threshold) standard deviation of the cutout's pixel values. \caapr\ then identifies all contiguous groups of five or more pixels with a Signal to Noise Ratio (S/R)\,\textgreater\,3 (Panel~2 of Figure~\ref{Fig:CAAPR_Aperture_Fitting_Example}); if there is a group of contiguous S/R pixels associated with the target source, it is isolated (Panel~3 of Figure~\ref{Fig:CAAPR_Aperture_Fitting_Example}). If no group of S/R\,\textgreater\,3 pixels is found at the coordinates of the target source, it is assumed that the source is non-detected and \caapr\ defaults to a circular aperture (ie, an axial ratio of 1, and a position angle of 0\,deg). 

If \caapr\ succeeds in finding and isolating a group of contiguous S/R\,\textgreater\,3 pixels associated with the target source, then \caapr\ subjects that group of pixels to binary opening (ie, a succession of binary erosion then binary dilation), with an erosion element of size equal to the FWHM of the band's beam (Panel~4 of Figure~\ref{Fig:CAAPR_Aperture_Fitting_Example}). This removes small protrusions from the periphery of the group of pixels, as such features are often due to background sources, foreground stars, map artefacts, etc. For example, in Panel~3 of Figure~\ref{Fig:CAAPR_Aperture_Fitting_Example} the removed feature at the north-easterly edge of the isolated group of pixels is associated with a background galaxy.

Next, \caapr\ finds the convex hull (the tightest polygon that will contain a set of points) of the remaining pixels (Panel~5 of Figure~\ref{Fig:CAAPR_Aperture_Fitting_Example}), then fits an ellipse to the vertices of the convex hull (Panel~6 of Figure~\ref{Fig:CAAPR_Aperture_Fitting_Example}); the position angle and axial ratio of this ellipse will be used for the final aperture ellipse for the band in question. The semi-major axis of this ellipse will be employed by \caapr\ as an initial reference value for determining the size of the aperture ellipse, as described below.

Before determining the semi-major axis of the aperture ellipse, \caapr\ convolves a copy of the cutout with a Gaussian filter, of FWHM twice the beam width of the band in question. This serves to enhance any lower-surface-brightness features associated with the target source. This is particularly important for sources that are very well resolved, where individual regions of emission within the source can be separated by apparently `empty' sky (eg, individual star-forming regions in FUV for well-resolved face-on spirals); the Gaussian smoothing helps to mitigate this effect.

The semi-major axis of the aperture ellipse is determined by placing concentric elliptical annuli on the smoothed cutout, centred at the coordinates of the target source, with the position angle and axial ratio determined previously. Initially, 10 such elliptical annuli are placed, with semi-major axes ranging from 0.5--3.0 times the initial reference value determined above, and with widths such that the boundaries of each subsequent annulus are in contact. The average per-pixel S/R in each elliptical annulus is calculated, to identify the pair of annuli that bracket the semi-major axis at which S/R\,=\,2 (Panel~7 of Figure~\ref{Fig:CAAPR_Aperture_Fitting_Example}). Within this range of semi-major axes, a differential evolution optimisation \citep{Storn1997}, using elliptical annuli of one beam-width, is then used to find the precise semi-major axis at which S/R\,=\,2 (Panel~8 of Figure~\ref{Fig:CAAPR_Aperture_Fitting_Example}). This two-step approach is much faster than a full high-resolution radial profiling. The semi-major axis of the S/R\,=\,2 isophote is then multiplied by an expansion factor of 1.25\footnote{This is the default expansion factor, and the one used for the DustPedia photometry; however it can be altered by the user if they wish.} to give the final semi-major axis of the aperture ellipse for the band in question (Panel~9 of Figure~\ref{Fig:CAAPR_Aperture_Fitting_Example}); applying this expansion factor accounts for the fact that some fraction of a source's flux will always fall outside any practical S/R cutoff \citep{Overcast2010,Blanton2001A}. If no annulus had S/R\,\textgreater\,2, then we revert to a default semi-major axis of twice the beam FWHM in the current band (note that, by design, all galaxies in the DustPedia sample are detected in at least one band).

\subsection{Aperture Combining} \label{Subsection:Aperture_Combining}

\begin{figure*}
\begin{center}
\includegraphics[width=0.975\textwidth]{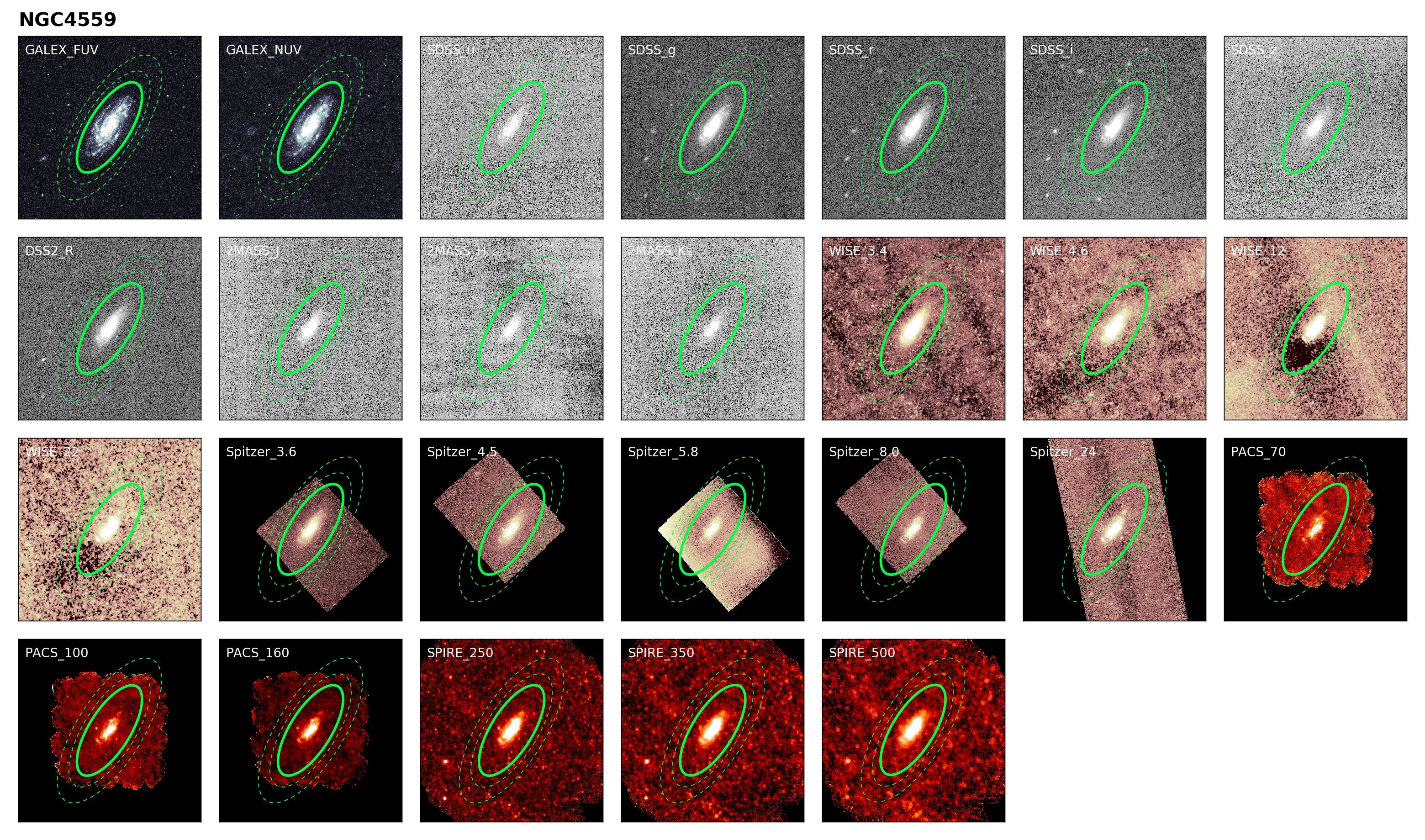}
\caption{Photometry thumbnail image grid for NGC\,4559. \caapr\ produces an image like this for every source. Solid lines show master aperture, whilst dashed lines demark background annulus. Several of the fluxes for NGC\,4559 have flags (flags are defined in Section~\ref{Section:Photometry_Flagging}). WISE 12\,\micron\ flux has `A' flag due to negative flux artefact in master aperture, whilst WISE 22\,\micron\ flux has `a' flag, due to a less pronounced version of the same feature. The \spitz\ 3.6, 4.5, 5.8, and 8.0\,\micron\ fluxes have `N' flags (although these are borderline cases), as \textless\,40\%\ of pixels in their sky annuli have coverage, so it is assumed that any background estimate will be unreliable (similarly, the lack of coverage in their master apertures would, independently, warrant `n' flags). \spitz\ 24\,\micron\ and PACS 70, 100, and 160\,\micron\ flues have `n' flags, as \textless\,80\%\ of pixels in their sky annuli have coverage, potentially reducing reliability of background estimate.}
\label{Fig:Example_Thumbnail_Grid}
\end{center}
\end{figure*}

The aperture-fitting process described in Section~\ref{Subsection:Aperture_Fitting} is repeated for every band for which data is available for a given target. We include DSS data in our aperture-fitting, as it provides consistent optical coverage for all of our galaxies (particularly useful for determining the stellar dimensions of galaxies without GALEX and/or SDSS data); although we exclude DSS data from the actual photometry as the maps are in photometrically uncalibrated units of photographic densities. To allow a valid comparison despite a potentially wide range of resolutions, the beam-width in each band is subtracted in quadrature from the major and minor axes of that band's aperture ellipse (the axial ratio is recomputed accordingly), effectively deconvolving them. These corrected ellipses are compared, to find a `super-ellipse' -- the smallest ellipse which will contain all of the deconvolved ellipses. This is the master aperture for this source (Panel~10 of Figure~\ref{Fig:CAAPR_Aperture_Fitting_Example}), to be used for the aperture photometry and uncertainty estimation described in Sections~\ref{Subsection:Aperture_Photometry} and \ref{Subsection:Uncertainty_Estimation}.

The user can intervene to exclude specific bands from the aperture fitting and combining process, either for an individual target or all targets, by adjusting the corresponding entries in the input tables passed to \caapr. 

An example of the final master aperture for a given source, applied to all bands, is shown in Figure~\ref{Fig:Example_Thumbnail_Grid} for NGC\,4559.

In general, for earlier-type galaxies (types E--Sa), the final master aperture size was often driven by the WISE 3.4 and 4.6\,\micron\ bands, thanks to their sensitivity to low-surface-brightness NIR emission (earlier-type galaxies being dominated by evolved stellar populations which are luminous in the NIR). Conversely, the master apertures of later-type galaxies (types Sb--Sm) were typically dictated by the apertures fit in the GALEX and UV--optical SDSS bands.

\subsection{Aperture Photometry} \label{Subsection:Aperture_Photometry}

The photometric measurements made by \caapr\ take the form of standard aperture photometry, with local background subtraction conducted using an elliptical background annulus around the elliptical master aperture, both centred on the coordinates of the target source. The user can opt to use apertures generated by \caapr\ as per Sections~\ref{Subsection:Aperture_Fitting} and \ref{Subsection:Aperture_Combining}; or the user can provide their own apertures.

The defined master aperture for a given target is used in all bands; however, in each band, the semi-major and -minor axes of the master aperture is added in quadrature to the beam FWHM in that band, effectively convolving the aperture with that band's beam. As such the master aperture, as employed in in each band, is sampling the same effective region of sky once resolution is accounted for, rendering the final photometry aperture-matched. This approach avoids the computational cost of convolving thousands of large-area high-resolution optical maps with the low-resolution 500\,\micron\ beam, and averts the risk of flux from background objects being spread into the aperture. We tested that photometry obtained in this way agrees with photometry obtained by convolving all maps to the same resolution. We did this by repeating some of the photometry with the maps convolved to the coarsest 36\arcsec\ resolution of the SPIRE 500\,\micron\ data (in particular, we repeated the photometry for all sources in the GALEX FUV and WISE 12\,\micron\ bands; these data have sufficiently high resolution for convolution to 36\arcsec\ to represent a significant change, without being computationally excessive). The photometry obtained in these repeats agreed to within 1\%, with the typical difference in flux being 0.4\,$\sigma$.

The flux in a source aperture is simply the sum of the values of all the pixels it contains. Bands with larger pixel sizes require consideration of partial pixels; \caapr\ determines what fraction of each pixel's flux falls within the source aperture by dividing each pixel into a number of sub-pixels (a standard technique; see \citealp{Bourne2013B}, \citealp{Wright2016B}, and references therein). For the DustPedia photometry we divide each input pixel into $10 \times 10$ grids of 100 sub-pixels for the \spitz\ 70 and 160\,\micron\ bands, all three PACS bands, and all three SPIRE bands.

The background annulus for a given target and band has the same axial ratio and position angle as the master aperture it surrounds. The inner and outer semi-major axes of the annulus are at 1.250 and 1.601 times the beam-adjusted semi-major axis of the source aperture (Panel~10 of Figure~\ref{Fig:CAAPR_Aperture_Fitting_Example}); these proportions mean that the background annulus has the same area as the master aperture (although they can be adjusted by the user). The average background level in the annulus is determined by taking the iteratively sigma-clipped (with a 3\,$\sigma$ threshold) mean of the pixel values contained within the annulus, with partial pixels considered in the same manner as for the master aperture. This average background is then subtracted from the flux measured in the master aperture, appropriately weighted for the aperture's area. 

For maps so small that the background annulus lies wholly outside their coverage (for example, in the instance of some \spitz\ and \hersc\ observations which observed only the nucleus of the target galaxies), then no background can be measured, and so a null flux is recorded for the source.

It is possible for a `negative' flux to be recorded, if the locally-determined background is brighter than the target source. This is to be expected; consider sources where there is essentially no flux present, such as in SPIRE observations of dust-free early-type galaxies. In these cases, the flux level measured in the background annuli should, on average, be the same as the flux level measured in the elliptical master apertures. However, they will never be {\it perfectly} identical for any individual instance, due to aperture noise - so 50\%\ of the time there will be a brighter flux level in the background annulus than in the elliptical master aperture, which will hence result in a negative flux measurement. Negative fluxes can also occur when measuring  faint sources in fields with complex backgrounds (eg, Galactic cirrus in MIR--submm bands). The vast majority of negative flux measurements are compatible with a flux of 0 when the uncertainty is considered.

For bands with beam FWHM\,\textgreater\,10\,\arcsec, an aperture correction is applied to account for the fraction of the source flux spread outside the master aperture by the PSF. Most instrument handbooks only provide such corrections for point sources, as corrections for extended sources (such as the DustPedia galaxies) require a model for the underlying unconvolved flux distribution. \caapr\ assumes that each target galaxy, as observed in a given band, can be approximated as a 2-dimensional S\'ersic distribution \citep{Sersic1963C} convolved with the band's PSF. Therefore \caapr\ fits a 2-dimensional convolved-S\'ersic model to the map, and uses the (unconvolved) S\'ersic distribution of the best-fit model to estimate the factor by which the measured flux is altered by the PSF. This factor is used to correct the measured flux accordingly. For this, we used the circularised PSF kernels of \citet{Aniano2011A}\footnote{\url{http://www.astro.princeton.edu/~ganiano/Kernels.html}}, to allow consistency between instruments. No attempt to apply aperture corrections is made for sources with S/R\,\textless\,3, as the results of the fit were too likely to be spurious, and the spreading of flux into the aperture from background sources would confuse the result. Note that our aperture corrections evaluate and account for the amount of source emission that is spread into the background annulus (which hence artificially inflates the estimated background level).

Fluxes at wavelengths \textless\,10\,\micron\ are corrected for Galactic extinction according to the prescription of \cite{Schlafly2011C}, using the IRSA Galactic Dust Reddening and Extinction Service\footnote{\url{https://irsa.ipac.caltech.edu/applications/DUST/}}.

\subsection{Uncertainty Estimation} \label{Subsection:Uncertainty_Estimation}

\begin{figure}
\begin{center}
\includegraphics[width=0.475\textwidth]{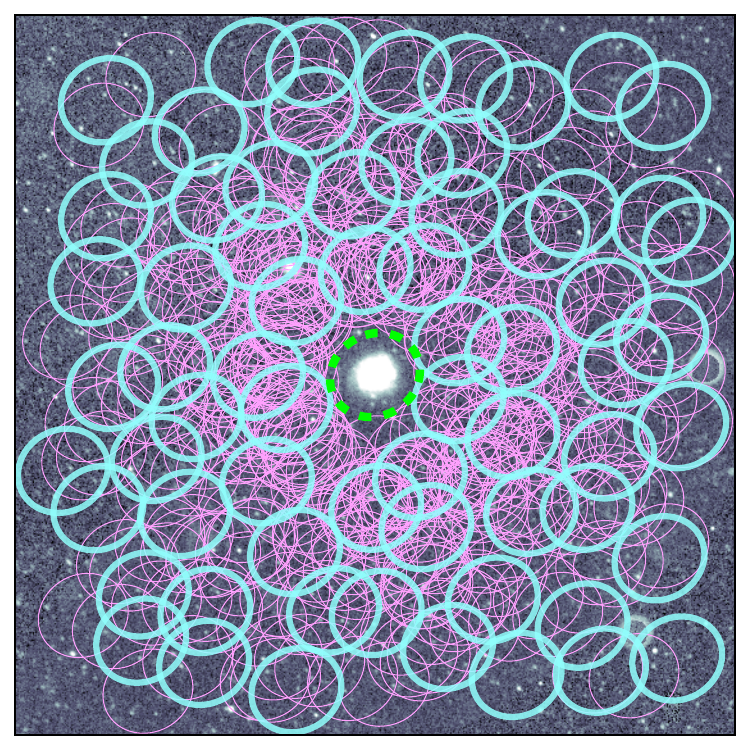}
\caption{Illustration of how \caapr\ determines the aperture noise in map, by placement of randomly positioned sky apertures, for our GALEX NUV map of NGC\,5147 (full 0.5\textdegree\,$\times$\,0.5\textdegree\ cutout shown). Thick dashed green ellipse around NGC\,5147 is master aperture, within which random sky apertures may not be placed. Thin pink ellipses show all of the candidate randomly positioned sky apertures generated by \caapr; as can be seen, they are biased towards being located closer to the target source, with distribution thinning out at greater distances as per a Gaussian distribution. Thick blue ellipses are accepted random sky apertures used to calculate aperture noise.}
\label{Fig:Random_Apertures_Example}
\end{center}
\end{figure}

Multiwavelength photometry requires determining photometric uncertainties in a wide range of noise regimes; from photon-arrival Poisson noise in the UV, to sky brightness in the NIR, to instrumental noise in the MIR, to confusion noise in the submm. \caapr\ needs to not only be able to measure photometric uncertainties in all of these situations, but to do so in a legitimately cross-comparable manner. As such, \caapr\ takes a robust `brute force' approach to estimating uncertainties, by measuring the variation in the fluxes recorded in sky apertures placed in the vicinity of the target master aperture. The aperture noise measured in this way will encompass the uncertainty contributions arising from instrumental variation in pixel values, and from confusion, and from the risk of unrelated sources contaminating the aperture, and from large-scale noise such as sky brightness and Galactic cirrus.

The random sky apertures all have the same area as the master aperture, and are background subtracted using an encircling background annulus in the same manner as for master aperture. \caapr\ places each random sky aperture quasi-randomly on the cutout, with its coordinates being drawn from a Gaussian distribution centred on the target source, with a standard deviation five times the semi-major axis of the source aperture. This distribution means that the random sky apertures are most likely to be positioned in the immediate vicinity of the master aperture, where the measured aperture noise will be most pertinent.

However, each randomly generated sky aperture has to pass three tests before it is accepted for use. Firstly, no random sky aperture is permitted to overlap with the master aperture; otherwise the inclusion of the target source in the aperture noise estimation would artificially drive up the reported uncertainty. Secondly, no random sky aperture is permitted to have more than 50\%\ of its pixels be sampled by any other random sky aperture; if a large number of random sky apertures happened to be positioned in the same location, this would artificially reduce the reported aperture noise. Thirdly, no random sky aperture may extend beyond the boundaries of the cutout. Figure~\ref{Fig:Random_Apertures_Example} shows the distribution of accepted and rejected random sky apertures for the example case of NGC\,5147.

\caapr\ keeps generating random sky apertures, and accepting those that pass the three tests listed above, until certain criteria are met. If 100 random sky apertures are successfully generated, then the generation process stops and those apertures will be used to determine the aperture noise. However, if the situation arises where 250 random apertures have been generated in a row without a single one of them being accepted, it is assumed that there are no more valid positions left in the cutout for any more apertures to be successfully placed; if this occurs, one of two things can happen. If at least 50 random sky apertures have already been accepted, these apertures will be used to determine the aperture noise. But if fewer than 50 random sky apertures have been accepted, this is deemed an insufficient number, and aperture noise will instead be determined using the mini-aperture extrapolation method described in Section~\ref{Subsection:Aperture_Noise_Extrapolation}.

This method of determining aperture noise has the additional benefit of being highly stable. Running repeated aperture noise estimates on the same cutout yields aperture noise estimates that typically vary by \textless\,10\%. Compare this to the high degree of variation -- as much as a factor of several -- that can be suffered when placing sky apertures across a square cutout totally at random \citep{CJRClark2014A}.

Once a sufficient number of random sky apertures has been successfully generated, \caapr\ calculates the iteratively sigma-clipped (with a 3\,$\sigma$ threshold) standard deviation of the distribution of flux values measured from those apertures. This value is taken to be the aperture noise for the aperture size and cutout in question.

The final uncertainty for a given flux is calculated by adding in quadrature the aperture noise and photometric calibration uncertainty of the band in question. The calibration uncertainties for all of our bands are listed in Table~\ref{Table:Band_Parameters}.

\subsection{Aperture Noise Extrapolation} \label{Subsection:Aperture_Noise_Extrapolation}

\begin{figure}
\begin{center}
\includegraphics[width=0.475\textwidth]{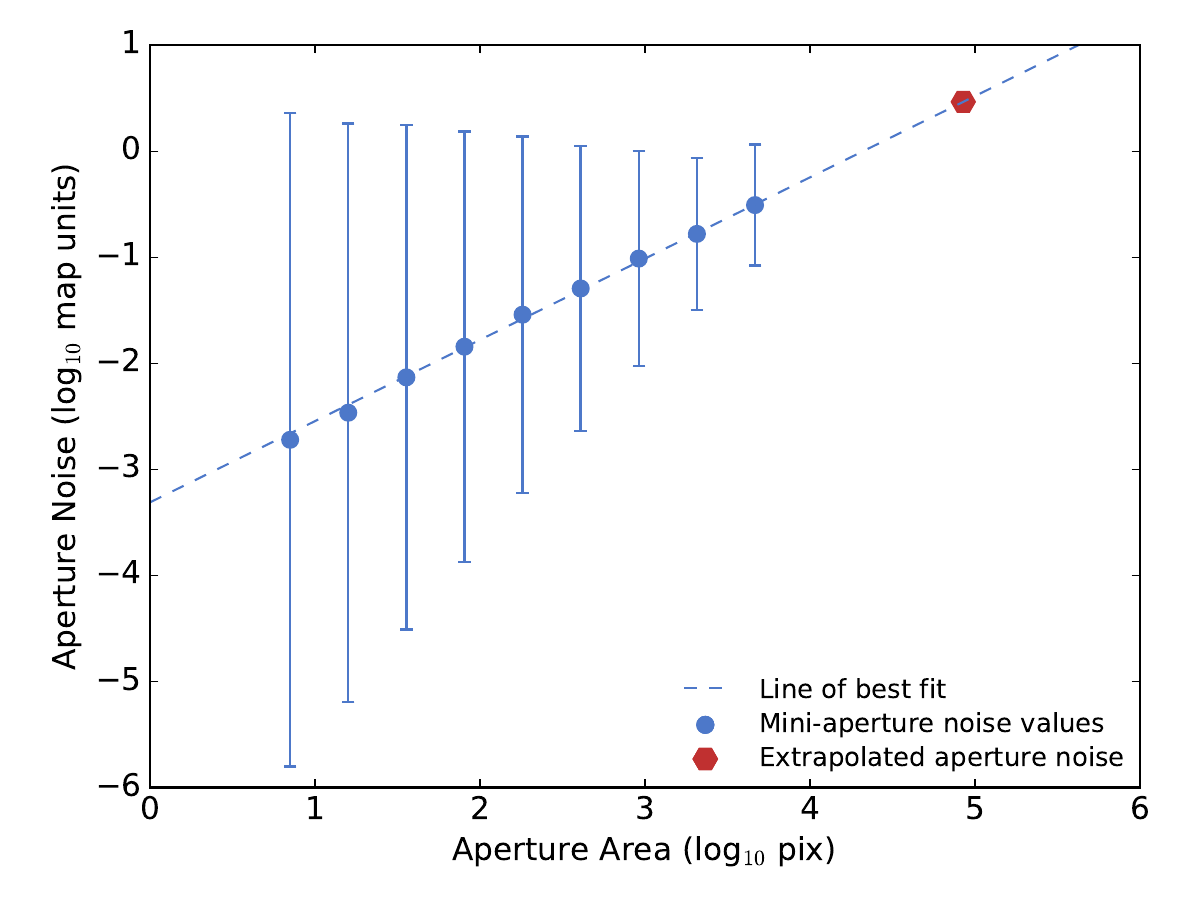}
\caption{Example of how \caapr\ extrapolates aperture noise, for SPIRE 250\,\micron\ photometry of M\,101 (NGC\,5457). Blue circles indicate aperture noise determined when using  mini-apertures of different sizes. Note that the error bars do not represent uncertainties, but rather indicate weightings assigned to each point, with greater weight given to mini-apertures closer in size to the master aperture. Blue dashed line indicates best-fit power law to trend exhibited by these points. Red hexagon indicates extrapolated aperture noise for M\,101, given area of its master aperture.}
\label{Fig:Example_Aperture_Noise_Extrapolation}
\end{center}
\end{figure}

The approach of determining aperture noise via the placement of randomly-positioned sky apertures has the virtue of being effective for essentially any noise regime (eg, instrumental noise, confusion noise, contaminating foreground/background sources, Galactic cirrus, etc). However, it has the distinct disadvantage that it requires maps with a sufficient amount of coverage area around the target source to permit the placement of enough random sky apertures.

Moreover, consider the case of an extremely extended nearby galaxy, such as M\,101 (NGC\,5457), for which we employ a photometric aperture with a major axis of 35\arcmin. For a map of M\,101 to be large enough for the placement of 50 random sky apertures (our threshold minimum acceptable number) around the target, with perfect tessellation, the map would have to cover an area of over 13\,sq\,deg. Whilst large-area surveys like SDSS might observe an area of that size, most observations by telescopes like \spitz\ and \hersc\ would not map even a fraction of that amount of sky when observing a galaxy like M\,101 (for instance, Figure~\ref{Fig:Example_Thumbnail_Grid} shows how small the \spitz\ and \hersc\ maps of NGC\,4559 are in relation to the size of the actual galaxy). And even if maps {\it were} consistently that large, it would not be appropriate to use such a vast area of sky for determining aperture noise; an aperture noise estimate determined using random sky apertures placed several degrees away from the target will not be representative of the aperture noise at the location of the target itself. To remedy this, \caapr\ features a novel method of {\it extrapolating} the aperture noise, for maps of any size. 

In instances where \caapr\ is unable to place a sufficiently large number of full-size random sky apertures on a cutout (as described in Section~\ref{Subsection:Uncertainty_Estimation}), \caapr\ will instead switch to placing random `mini-apertures', of a range of sizes, across the map. \caapr\ first finds the largest size of random mini-aperture that {\it can} successfully be placed in sufficiently large numbers, and then measures the aperture noise for that size mini-aperture, doing so in the exact same way as laid out in Section~\ref{Subsection:Uncertainty_Estimation} -- with the only difference being that a smaller radius is being used. \caapr\ then places a set of even smaller mini-apertures across the map, and likewise determines the aperture noise for mini-apertures of that size. \caapr\ repeats this process, measuring the aperture noise for successively smaller sets of mini-apertures. The radii of each successive set of mini-apertures are evenly-spaced in log space, following a power law of index 1.2 (an empirically-selected value, which provides sampling dense enough to be effective, whilst coarse enough to still be reasonably fast).

\caapr\ keeps on calculating the aperture noise associated with smaller and smaller sizes of mini-aperture, stopping when one of two conditions is reached; until 10 increments of mini-aperture size have been successfully processed, or until the mini-aperture radius has fallen to 1 pixel (making smaller apertures impossible).

\caapr\ then produces a log-log plot of the aperture noise determined for each set of random mini-apertures, against the mini-aperture area used for each set. An example of such a plot, for the SPIRE 250\,\micron\ photometry of M\,101 (NGC\,5457) is shown in Figure~\ref{Fig:Example_Aperture_Noise_Extrapolation}. In these area--noise plots, there is consistently a strong correlation between the area of the random mini-apertures, and their associated aperture noise. \caapr\ uses these area-noise plots to extrapolate what the aperture noise associated with the area of the full-size random sky apertures would have been, had there been sufficient map area available; this extrapolation is performed by fitting a power law to the area--noise points, in a chi-squared minimising manner. For fitting, the error bar on each point is derived from the Poisson uncertainty given the number of random mini-apertures employed, which is then multiplied by a weighting that takes the form $1 + \log_{10}( \Delta_{A} )$, where $\Delta_{A}$ is the factor difference between the areas of the random mini-apertures and the master aperture. This empirical weighting accounts for the fact that random mini-apertures with areas more similar to that of the master aperture should have greater influence over the extrapolated aperture noise than random mini-apertures with areas a lot smaller than the master aperture.

We demonstrate the validity of this approach, and quantify its effectiveness, in Section~\ref{Section:Extrapolation_Validation}. This method allows \caapr\ to produce reliable estimates of aperture noise for almost any photometric measurement; if there is sufficient map area to determine a flux, there is sufficient map area to estimate the aperture noise, in all but the most extreme cases. In the very rare instances where there is not enough map area for even the  mini-aperture approach to work, we record a negative uncertainty in our photometry; this indicates that that uncertainty represents the calibration uncertainty alone.

\section{Photometry Flagging} \label{Section:Photometry_Flagging}

The \caapr\ photometry of the DustPedia galaxies provides an enormous number of fluxes, produced in a consistent manner between galaxies and across bands. The size of our dataset makes it impractical to manually adjust the photometry in the case of individual sources where issues may have been encountered (and regardless, such tweaking would risk introducing inconsistencies). Instead, we opted to manually inspect all of the photometry outputs produced by \caapr, and flag fluxes where there is a possible matter of concern.

For every DustPedia galaxy, \caapr\ produced a grid of thumbnail images, an example of which is shown in Figure~\ref{Fig:Example_Thumbnail_Grid} for NGC\,4559. At least two members of the DustPedia team inspected the thumbnail grid for each galaxy, along with the corresponding fluxes, and recorded flags for instances where issues were encountered.

Three different types of flag are used to indicate when there may be issues with a particular flux in our aperture-matched photometry (along with an additional flag used only for supplementary photometry, discussed below). Each flag is split into `minor' and `major' categories. Minor flags are indicated by lowercase letters, whilst major flags are indicated by uppercase letters. The types of flag are defined as follows:

\begin{description}
\item [{\bf a/A}] -- Artefact flag. This indicates that some sort of map artefact is present in the region of the target source, as determined by visual inspection. Examples of this include poor mosaicing in the archival data, satellite trails, or evidence of saturation in bright pixels.
\item [{\bf c/C}] -- Contamination flag. This indicates contamination of the photometry due to a nearby source, as determined by visual inspection. Examples of this include foreground stars that were not fully removed, or other galaxies close to the target source.
\item [{\bf n/N}] -- Null-coverage flag. This flag is automatically assigned where necessary. This indicates that the observation didn't provide full coverage of the area of the source, limiting \caapr's ability to suitably measure the source flux and/or background. If more than 5\%\ of the pixels in the master aperture and/or more than 20\%\ of the pixels in the background annulus are found to be NaN -- `Not a Number' -- then a minor flag is raised. If more than 20\%\ of the pixels in the master aperture and/or more than 60\%\ of the pixels in the background annulus are found to be NaN, a major flag is raised. Instruments for which small maps are common (such as \spitz\ and PACS) are particularly likely to have fluxes with this flag.
\item [{\bf e/E}] -- Excluded extended emission flag. This flag is only used with our supplementaty photometry (as described in Section~\ref{Section:Supplementary_Photometry}), and is not assigned to any of our aperture-matched \caapr\ fluxes. It indicates that the target is sufficiently extended that there is a risk some of its emission will have been excluded from the measurement in question, which would lead to the flux being an under-estimate.
\end{description}

A major flag indicates that the associated flux probably is not suitable for general use -- although it may still be useful for specific considerations. For example, the most common reason for a `C' flag is because the target galaxy is interacting with another galaxy inside the master aperture; although the recorded flux will not be of use as a measure of the emission from the target galaxy alone, it will nonetheless be useful to anyone interested in the flux of the overall system. 

A minor flag indicates that the associated flux is probably still useful for most purposes -- although it is advisable for users to inspect the corresponding data to confirm that the flux is still valid for their specific applications. 

Some sources have a `global' flag applied. In this case, an issue affects so many bands that {\it all} bands receive the relevant flag. The issue may not be apparent in every band -- for example, if the target galaxy is a spiral that overlaps with an elliptical companion, the elliptical might not be visible in SPIRE bands. Nonetheless, if an issue affects enough bands, all bands get flagged by default. Only `c/C' flags can be applied globally.

The flagging is not intended to cover the types of minor issues which are encompassed by the quoted uncertainties -- minor background objects within the master aperture, foreground Galactic cirrus, etc. Rather, the flagging is designed to highlight aberrations that uncertainties would not reflect.

Amongst the small number of our sources for which a negative flux is measured (see Section~\ref{Subsection:Aperture_Photometry}), the flux can occasionally be `significantly' negative, and hence unphysical. This can arise due to a particularly inopportune distribution of Galactic cirrus surrounding the target source, for example. In the rare instances where the flux has a S/R\,\textless\,$-2$, we automatically assigned an `A' flag. This was necessary for 0.4\%\ of our fluxes.

In total, 4,081 (22.4\%) of our aperture-matched \caapr\ fluxes have flags associated with them; 2,262 (12.4\%) have major flags, whilst 1,819 (10.0\%) only have minor flags. Whilst this might seem high, a large fraction of these are due to `N' or `n' flags being automatically assigned to \spitz\ fluxes, because of the very small map sizes of many \spitz\ observations; excluding \spitz\ photometry, only 2,234 (14.6\%) of our fluxes possess flags, of which 1,202 (7.8\%) have major, and 1,032 (6.8\%) have only minor.

Some examples of issues for which flags are raised are illustrated in Figure~\ref{Fig:Example_Thumbnail_Grid}, for NGC\,4559.

\section{Supplementary Photometry} \label{Section:Supplementary_Photometry}

\subsection{\textit{Planck} CCS2} \label{Subsection:Supplementary_Planck}

\subsubsection{\textit{Planck} CCS2 Photometry} \label{Subsubsection:Supplementary_Planck_Photometry}

\planck\ \citep{Planck2011I} photometry was obtained from the Second \planck\ Catalogue of Compact Sources (\planck\ CCS2; \citealp{Planck2015XXVI}). The poor resolution of \planck\ made it impractical to include in our aperture-matched \caapr\ photometry. Entries in the \planck\ CCS2 (from the full range of Galactic latitudes) were matched to DustPedia sources within a matching radius equal to the FWHM of each \planck\ band.

The \planck\ CCS2 provides photometry in all 9 \planck\ bands, with a focus on source reliability. The \planck\ CCS2 provides 4 different flux measurements for each source. Of these different fluxes, \citet{Planck2015XXVI} find that their aperture photometry (the APERFLUX field in the published \planck\ CCS2 tables) is the measure that compares best with \hersc\ aperture photometry of HRS galaxies. Therefore the \planck\ CCS2 aperture photometry values are the fluxes we provide here. The \planck\ CCS2 fluxes are not colour corrected, and assume the standard \planck\ reference spectra (see Section~\ref{Subsection:Planck_Imagery}).

The calibration uncertainties of each \planck\ band (as provided in \citealp{Planck2015V,Planck2015VIII}, including the systematic uncertainties quoted for the 350 and 550\,\micron\ bands) are listed in Table~\ref{Table:Band_Parameters}, and were added in quadrature to the photometric uncertainties given for each source, to produce the final flux uncertainties.

In total, the \planck\ CCS2 provides an additional 1,079 fluxes for the DustPedia \hersc\ sample. The number of detections ranges from 394 (45\%\ of the DustPedia galaxies) at 350\,\micron, to just 11 at 4.26\,mm. Perhaps the most useful \planck\ band for our purposes is 850\,\micron, as it is the most sensitive \planck\ band to cover wavelengths longer than those observed by \hersc; 197 (22\%) of the DustPedia galaxies are detected at 850\,\micron.

\subsubsection{\textit{Planck} CCS2 Flagging} \label{Subsubsection:Suplementary_Planck_Flagging}

The \planck\ CCS2 aperture photometry uses circular apertures with radii equal to 1 FWHM in each band. Emission more extended than 1 FWHM from the target source's centre will therefore be lost (not counting flux that falls outside the aperture due to the instrumental PSF, which the CCS2 corrects for). To identify which of our sources are vulnerable to this effect, we first measured the angular scale of the FIR/submm emission of the DustPedia galaxies, by running the aperture-fitting phase of \caapr\ (see Section~\ref{Subsection:Aperture_Fitting}) on the SPIRE\,250\,\micron\ map of each target (or the PACS\,160\,\micron\ map for targets without SPIRE coverage); no expansion factor was applied, so the semi-major axes of the resulting ellipses represent the actual extent of the detectable FIR/submm emission associated with each target. In each \planck\ band, sources where the semi-major axis of this emission ellipse was larger than the CCS2 aperture radius (being the effective FWHM listed in Table~2 of \citealp{Planck2015XXVI}) were identified, and the corresponding fluxes flagged with an `e' flag -- indicating a source's emission may be extended beyond the region being measured, leading to missed flux. \planck's resolution at 4.26, 6.81, and 10.60\,mm is poor enough that no sources were identified as being at risk of losing flux in this way. At 1.380 and 2.100\,mm, \textgreater\,45\%\ of fluxes required flagging. In all other bands, \textless\,27\%\ of fluxes are affected. Note that the presence of an `e' flag does not necessarily mean any flux has been missed, as the emission scales in the \planck\ bands will not be identical to the FIR/submm flux extent. A difference in emission scales is particularly likely at the longer \planck\ wavelengths, where emission processes are different.

If a `C' or 'c' flag was associated with the SPIRE 350 or 500\,\micron\ photometry for a given source, as per Section~\ref{Section:Photometry_Flagging}, then that flag was propagated to all \planck\ submm fluxes (ie, 350, 500, or 850\,\micron). Additionally, we visually inspected the 350\,\micron\ SPIRE maps for each source with \planck\ submm photometry, to look for any potential contaminating source within the CCS2 aperture, and applied `c' and `C' flags as appropriate. We do not apply contamination flags derived from submm wavelengths to the longer-wavelength \planck\ bands, due to the very different emission processes that dominate there. Moreover, the much lower source density at the longer \planck\ wavelengths -- even when accounting for the larger \planck\ beam and CCS2 aperture sizes -- means that there is a much smaller contamination risk.

\subsection{IRAS SCANPI} \label{Subsection:Supplementary_IRAS}

\subsubsection{IRAS SCANPI Photometry} \label{Subsubsection:Supplementary_IRAS_Photometry}

Photometry from the InfraRed Astronomical Satellite (IRAS; \citealp{Neugebauer1984}) was acquired using the Scan Processing and Integration Tool (SCANPI\footnote{\url{http://irsa.ipac.caltech.edu/applications/Scanpi/}}) IRAS has extremely poor resolution, and the highly asymmetrical PSF, made it impractical to perform aperture-matched photometry using IRAS maps. Nonetheless IRAS photometry is valuable for DustPedia, particularly in the 60\,\micron\ band; fluxes in this part of the spectrum are important for constraining dust emission and properties (see \citealp{Jones2016A,Jones2017A}, and references therein).

In each of the 4 IRAS bands, a calibration uncertainty of 20\%\ \citep{Miville-Deschenes2005A,Sauvage2011A} was added in quadrature to each photometric uncertainty. IRAS fluxes are calibrated assuming a $\nu^{-1}$ reference spectrum\footnote{\label{Footnote:IRAS_Explanatory_Supplement}IRAS Explanatory Supplement: \url{https://lambda.gsfc.nasa.gov/product/iras/docs/exp.sup/}}. In total, IRAS provides an additional 2533 fluxes for the DustPedia \hersc\ sample. 

SCANPI operates by fitting the IRAS timelines. Indeed, it was not originally intended that IRAS observations would be used to be used to produce imagery; whilst IRAS maps do of course exist, the PSF asymmetry necessities dramatic down-sampling of the data, and the pointing of the resulting maps suffers from some variability\footnoteref{Footnote:IRAS_Explanatory_Supplement}. For these reasons, and in order to avoid presenting maps that are at odds with our photometry, we opt not to include IRAS imagery in the DustPedia database.

\subsubsection{IRAS SCANPI Flagging} \label{Subsubsection:Supplementary_IRAS_Flagging}

Because SCANPI works by fitting to the IRAS timelines, the resulting fluxes cannot necessarily be directly compared to the aperture-matched photometry. In the case of sources with large angular sizes, there is a risk that SCANPI will miss faint extended emission, especially if the emission distribution of the target source is not smooth; in this scenario, SCANPI might fit only the bright central region of a source, essentially shredding it. A good example of this is M\,81 (NGC\,3031), where the SCANPI 60\,\micron\ flux represents only the flux associated with the bright central bulge. 

We consider any source more extended than twice the resolution of IRAS in a given band to be at risk of being shredded by SCANPI in this way. Given the highly asymmetrical resolving power of IRAS, we assume a worst-case scenario by taking the best possible resolving power in each band\footnote{Band resolving powers taken from the Survey Array table in the Instrument Summary: \url{http://irsa.ipac.caltech.edu/IRASdocs/iras_mission.html}}. IRAS has a maximum resolution of 0.75\arcmin\ at 12 and 25\,\micron, 1.5\arcmin\ at 60\,\micron, and 3\arcmin\ at 100\,\micron. Flux densities identified as being vulnerable to shredding were flagged with an `e' flag, indicating that some emission may be extended beyond the region being measured.

Given that the DustPedia sample consists only of sources larger than 1\arcmin, all 12 and 25\,\micron\ SCANPI fluxes receive `e' flags. However, since we have WISE aperture-matched photometry covering both the 12 and 25\,\micron\ bands, at greater resolution and sensitivity than provided by IRAS, we envisage no scenario where the SCANPI photometry in these bands would be preferred; however, we include it for completeness. Note, also, that SCANPI photometry does not benefit from the foreground star removal used for our \caapr\ aperture-matched photometry.

To determine which 60 and 100\,\micron\ fluxes are at risk of shredding, we consider the FIR/submm extent of each DustPedia target, determined in the same manner as in Section~\ref{Subsubsection:Suplementary_Planck_Flagging}. Any galaxy where the major axis of the FIR/submm emission ellipse is larger than two IRAS resolving elements is deemed to be vulnerable to being shredded. Of the 675 DustPedia targets with SCANPI 60\,\micron\ fluxes, 254 (36\%) exceed this threshold and hence get flagged; however, 81 of them have reliable (ie, no major flags) \spitz\ 70\,\micron\ and/or PACS 70\,\micron\ fluxes. As a result, 502 (57\%) of our sources have good photometry at 60--70\,\micron. Of the 682 DustPedia targets with SCANPI 100\,\micron\ fluxes, 97 (14\%) are extended enough in the FIR/submm to require flagging; of these, 68 have reliable PACS 100\,\micron\ fluxes. Therefore 713 (81\%) of our galaxies have good 100\,\micron\ photometry.

Contamination flags from our aperture-matched photometry were propagated to the corresponding SCANPI fluxes. For SCANPI 12\,\micron\ photometry, if a given source had a `C' or `c' flag associated with its WISE 12\,\micron\ flux, then that flag was applied to the SCANPI flux also. Similarly, contamination flags associated with our WISE\,22\,\micron\ photometry were applied to the corresponding SCANPI 25\,\micron\ fluxes. For SCANPI 60\,\micron\ fluxes, we prefered to take contamination flags from PACS 70\,\micron; if PACS 70\,\micron\ photometry was not available, we instead used flags from, in order of preference: PACS 100\,\micron, PACS 160\,\micron, or SPIRE 250\,\micron. For SCANPI 100\,\micron\ photometry, we used the same order of preference, except that PACS 100\,\micron\ flags were preferred over PACS 70\,\micron\ flags.

\section{Photometry Validation} \label{Section:Validation}

\begin{figure*}
\begin{center}
\includegraphics[width=0.975\textwidth]{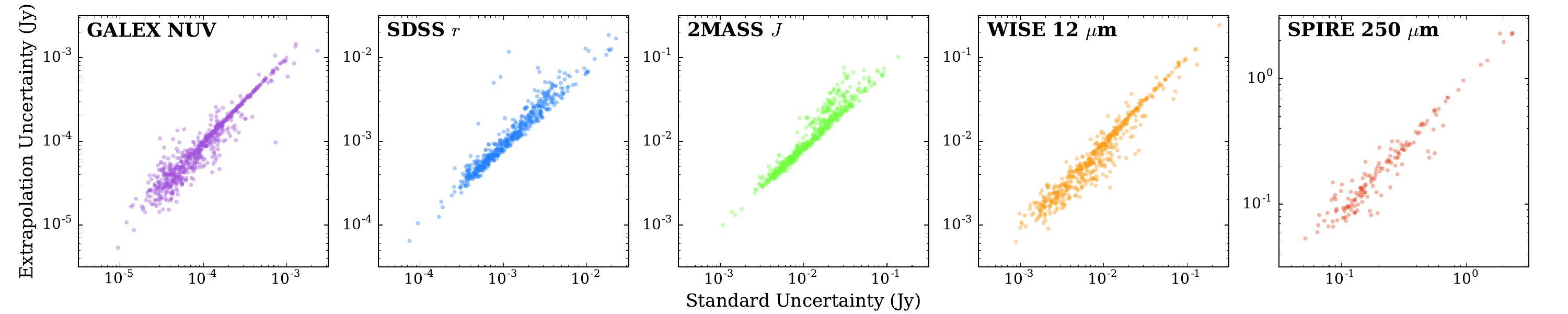}
\caption{Plots showing how uncertainty estimates determined the standard way (with full-size randomly generated sky apertures; see Section~\ref{Subsection:Uncertainty_Estimation}) compare to uncertainty estimates determined using the aperture noise extrapolation technique described in Section~\ref{Subsection:Aperture_Noise_Extrapolation}, for various bands.}
\label{Fig:CAAPR_Aperture_Noise_Extrapolation}
\end{center}
\end{figure*}

\subsection{Foreground Star Removal Validation} \label{Section:Astromagic_Validation}

We tested the foreground star removal method we described in Section~\ref{Subsection:AstroMagic} by using it to remove the stars from observations of 16 patches of sky at a range of wavelengths. 

We produced cutout maps for 16 patches of sky in SDSS $u$, SDSS $r$, 2MASS $J$, and WISE 3.4\,\micron, following the procedures laid out in Section~\ref{Subsection:Ancillary_Imagery}; this selection of data should provide a fair sampling of the range of wavelengths and noise regimes in which we perform star subtraction in our photometry. In all bands, each of the 16 maps had dimensions of 0.5\textdegree\,$\times$\,0.5\textdegree, the same size as our standard cutout images. These test cutouts were all located within 3\textdegree\ of coordinates $\alpha = 136.6^{\circ}$ $\delta = 0.45^{\circ}$ (J2000); this area of space is known to be extremely under-dense at low-redshifts \citep{Driver2011}, and contains no large galaxies in the volume sampled by DustPedia. Each of the maps was also visually inspected to confirm the absence of any nearby galaxies. As such, the vast majority of the emission seen in each of these maps should come from foreground stars, with only a small fraction coming from distant background galaxies, etc.

To remove any large-scale emission from the test cutouts, we conducted a polynomial sky subtraction, as per Section~\ref{Subsection:Polynomial_Sky_Flattening}. Having done this, we assumed that the background level associated with `empty' sky was represented by the peak of the pixel value distribution; for each test cutout, we found this level and subtracted it. Having performed this background subtraction, the sum total of the pixel values in each cutout should represent the combined flux of all the sources it contains -- primarily stars.

In SDSS $u$, SDSS $r$, 2MASS $J$, and WISE 3.4\,\micron, the median amount of flux removed from the 16 test cutouts in each band was 88\%, 90\%, 93\%, and 86\%\ respectively. This demonstrates that our star removal process is able to consistently remove the vast majority of the stellar flux from maps, across a wide range of wavelengths. And even in instances where our star removal is less effective than usual, a large majority of the stellar flux is still successfully removed -- for 90\%\ of the test cutouts, flux removal levels better than 81\%, 71\%, 84\%, and 68\%\ were achieved in each band. Moreover, note that these values are only lower limits on the fraction of the stellar flux that is removed -- even if all stellar emission was successfully removed, a low level of background flux due to distant galaxies would remain in the test cutouts. This is the reason why we omitted GALEX data from this test -- the contribution of distant galaxies to the background flux level in GALEX bands is large enough to no longer be negligible compared to that of foreground stars, preventing this method of testing from returning valid results.

Additionally, we wished to evaluate whether our foreground star removal method was prone to removing flux associated with compact sources within the target galaxies -- H{\sc ii} regions being the most obvious examples. Section~\ref{Subsection:External_Validation} compares our photometry to that of a number of independent external sources. Therein, we find no evidence of our photometry being systematically fainter than that produced by previous authors who performed by-hand identification and masking of foreground stars. We note that, as described in Section~\ref{Subsection:AstroMagic}, we repeated our photometry with star removal disabled for maps where visual inspection revealed that the process had done more harm than good by erroneously removing flux associated with the target galaxy.

\subsection{Aperture Noise Extrapolation Validation} \label{Section:Extrapolation_Validation}

As described in Section~\ref{Subsection:Aperture_Noise_Extrapolation}, in cases where a map has insufficient coverage area around the target source to allow determination of aperture noise by placing full-size sky apertures in its vicinity, \caapr\ employs a novel alternate approach. In these circumstances, \caapr\ uses `mini-apertures' of various sizes to find the relation between aperture area and aperture noise in the map, and hence extrapolate the aperture noise associated with the aperture area of the master aperture.

To evaluate the effectiveness of this approach, we use five bands where a large proportion of the DustPedia imagery comes from wide-area surveys; GALEX NUV, SDSS {\it r}-band, 2MASS {\it J}-band, WISE 12\,\micron, and SPIRE 250\,\micron. In each of these bands, we isolated the instances where the observations provide coverage of the entire cutout. For photometry of all of these maps, the aperture noise would have proceeded according to the standard uncertainty estimation technique described in Section~\ref{Subsection:Uncertainty_Estimation}, without having to resort to mini-aperture extrapolation. 

We then repeated the photometry in each of these five bands -- but did so on smaller cutout maps, with diameters only 3 times larger than the major axis of the master aperture. These smaller cutouts force the aperture noise estimation process to resort to the mini-aperture extrapolation approach. Having applied the mini-aperture noise extrapolation technique to the smaller cutouts, we compared the resulting uncertainty estimates to those from our actual photometry (which were determined with full-sized randomly generated sky apertures on the standard maps).

The results of this comparison are shown in Figure~\ref{Fig:CAAPR_Aperture_Noise_Extrapolation}. As can be seen, the uncertainty estimates produced with the mini-aperture extrapolation method agree well with the standard uncertainty estimates. The average ratios between the extrapolated and standard uncertainties are $0.94 \pm 0.18$, $0.99 \pm 0.19$, $0.94 \pm 0.19$, $0.87 \pm 0.19$, and $0.90 \pm 0.16$, for GALEX NUV, SDSS {\it r}-band, 2MASS {\it J}-band, WISE 12\,\micron, and SPIRE 250\,\micron\ respectively (the $\pm$ value indicating the RMS deviation from a ratio of 1.0 in each band). It seems that the mini-aperture extrapolation method shows a tendency to result in slightly smaller uncertainty estimates, however for all bands the scatter encompasses the 1:1 relation. As such, we deem the mini-aperture extrapolation method to be suitably accurate -- and certainly far superior than the alternative, of not attempting to estimate aperture noise when performing photometry on small maps

\subsection{External Validation} \label{Subsection:External_Validation}

In order to provide an objective means of validating our photometry, particularly with regards to identifying any systematic methodological issues, we compare our measurements to those from several independent external sources. Given that the scope of the DustPedia photometry is much larger than typical nearby galaxy surveys (both in terms of number of galaxies and number of bands; see Section~\ref{Section:Pipeline}), and given that our methodology therefore has to be much more automated, we aim to be thorough in establishing that our photometry is reliable.

When comparing DustPedia photometry to the corresponding external photometry in a given band, we consider a number of figures of merit to quantify the consistency of the two sets of fluxes. The relative scale of any offset between the sets of fluxes is quantified by the median flux ratio $\widetilde{R}$, being the median value of $S_{\rm DP} / S_{\rm ext}$, where $S_{\rm DP}$ is the DustPedia flux, and $S_{\rm ext}$ is the external measurement. The scatter in the relationship between two set of fluxes is described by $\Delta\widetilde{R}$, the median absolute deviation from $\widetilde{R}$. For two sets of fluxes in good agreement with one another, we would expect $\widetilde{R}$ to be smaller than $\Delta\widetilde{R}$, and also smaller than the calibration uncertainty of the band in question.

The other main figure of merit we consider when comparing photometry is $\chi^{[-1;1]}$, which is the fraction of galaxies for which $\chi$ lies in the range $-1 < \chi < 1$; to clarify the significance of this, we must first explain precisely what is meant by $\chi$. When comparing two sets of fluxes, we calculate $\chi$ for each galaxy, defined as:

\begin{equation}
\chi = \frac{ S_{\rm DP} - S_{\rm ext} }{ \sqrt{ \Delta S_{\rm DP}^2 + \Delta S_{\rm ext}^2 } }
\label{Equation:Chi}
\end{equation}

where $\Delta S_{\rm DP}$ and $\Delta S_{\rm ext}$ are the uncertainties on the DustPedia and external fluxes respectively. The quadrature sum term $\sqrt{ \Delta S_{\rm DP}^2 + \Delta S_{\rm ext}^2 }$ represents the mutual uncertainty between a pair of measurements. Assuming well-behaved Gaussian uncertainties, then 68\%\ of values of $\chi$ should lie in the range $-1 < \chi < 1$ -- ie, the $\chi$ distribution should have a mean of 0 and a standard deviation of 1. The figure of merit $\chi^{[-1;1]}$ describes what fraction {\it actually} lie in the range $-1 < \chi < 1$ for a given set of fluxes. $\chi^{[-1;1]}$ is a useful value; merely knowing that the median flux ratio for a set of fluxes is $\widetilde{R} = 1.05$, for example, tells us very little if we do not also know the uncertainties on those fluxes. If the typical flux uncertainty is 20\%, then $\widetilde{R} = 1.05$ indicates good agreement for that set of fluxes -- whereas if the typical flux uncertainty is 1\%, then that same median flux ratio of $\widetilde{R} = 1.05$ indicates rather poor agreement. By using $\chi^{[-1;1]}$, we are able to quantify how good the agreement is for a set of fluxes with consideration given to the uncertainties on the measurements.

Throughout this section, we only consider fluxes with significance \textgreater\,5\,$\sigma$, and which have no major flags associated with them, nor `e' flags (in the case of \planck\ CCS2 and IRAS SCANPI).

\subsubsection{Herschel Reference Survey} \label{Subsubsection:HRS_Validation}

The \hersc\ Reference Survey (HRS; \citealp{Boselli2010}) is a volume- and magnitude-limited survey of 323 nearby galaxies conducted by \hersc, with published multiwavelength photometry in 13 UV--submm bands. The HRS is ideally suited to serving as a reference for comparison with DustPedia; we have 288 galaxies in common, and our photometry includes all 13 bands for which they have published measurements. The HRS SPIRE photometry is presented in \citet{Ciesla2012B}, PACS photometry in \citet{Cortese2014A}, WISE photometry in \citet{Ciesla2014A}, and SDSS and GALEX photometry in \citet{Cortese2012C}.

The published GALEX UV--MIR photometry of the HRS is not corrected for Galactic extinction. However, 46\%\ of the galaxies of the HRS are located in the Virgo Cluster (as defined by objects included in the Virgo Cluster Catalogue of \citealp{Binggeli1985A}), which lies in a region of sky noted for having conspicuously large amounts of Galactic cirrus, despite its high galactic latitude \citep{Auld2013A,Bianchi2017A}. As such, to allow for a fair comparison, we extinction-correct all HRS fluxes at wavelengths \textless\,10\,\micron\ in the same manner as for own fluxes (as per Section~\ref{Subsection:Aperture_Photometry}). 

\begin{figure*}
\begin{center}
\includegraphics[width=0.4375\textwidth]{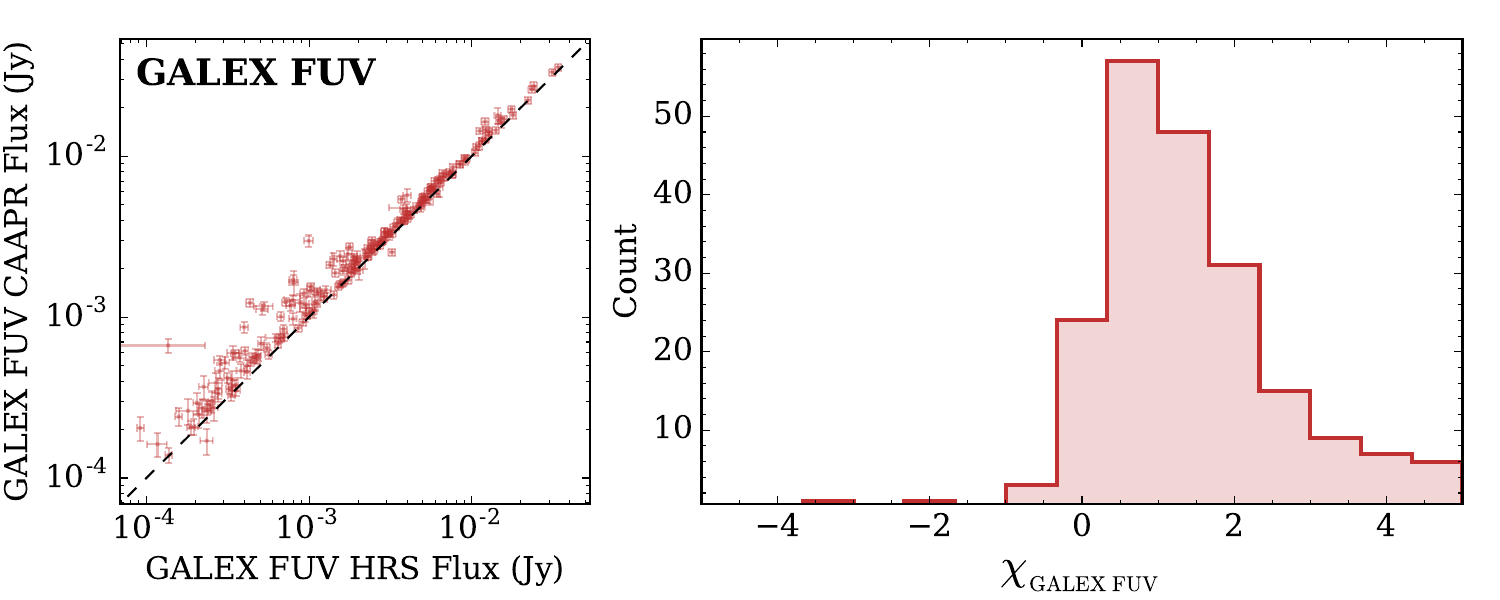}
\includegraphics[width=0.4375\textwidth]{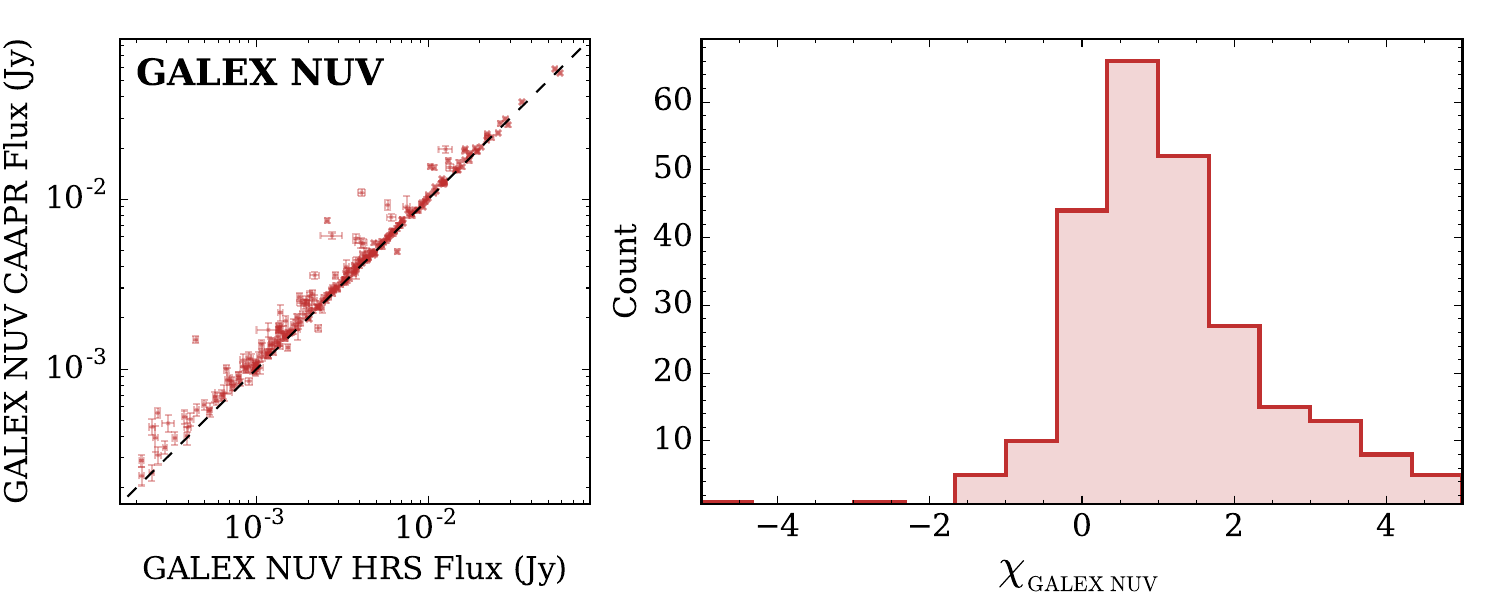}
\includegraphics[width=0.4375\textwidth]{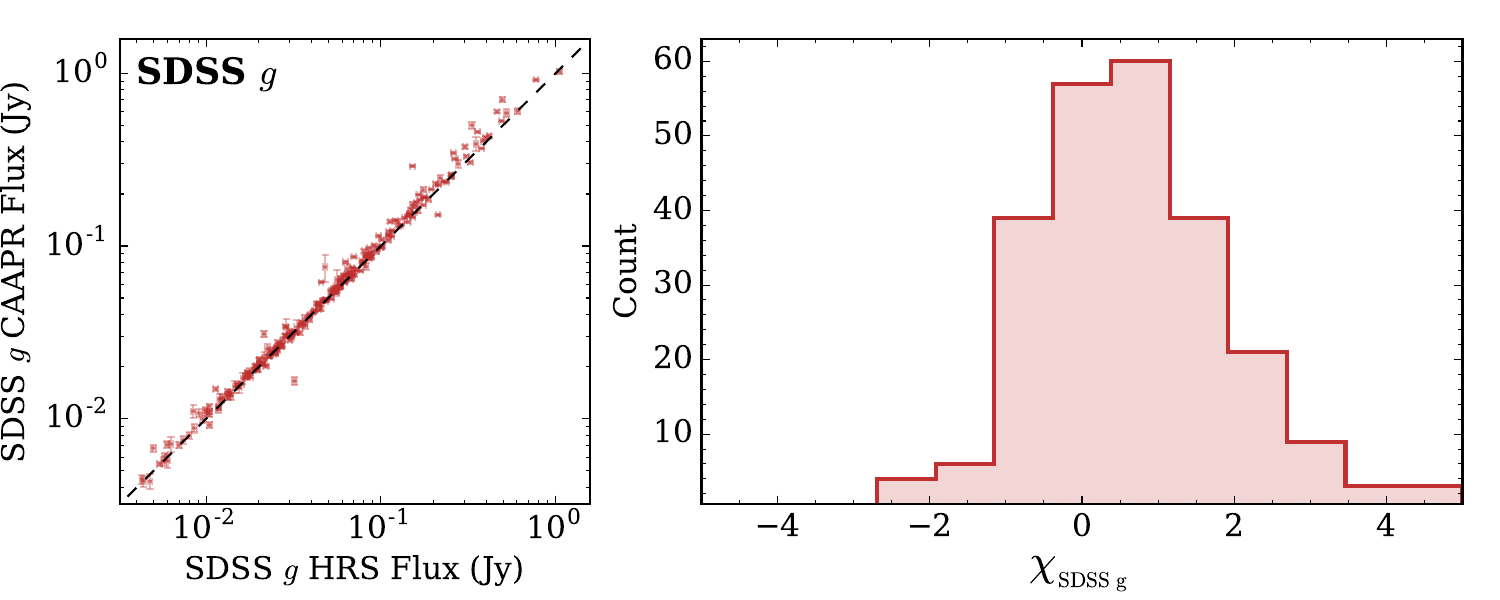}
\includegraphics[width=0.4375\textwidth]{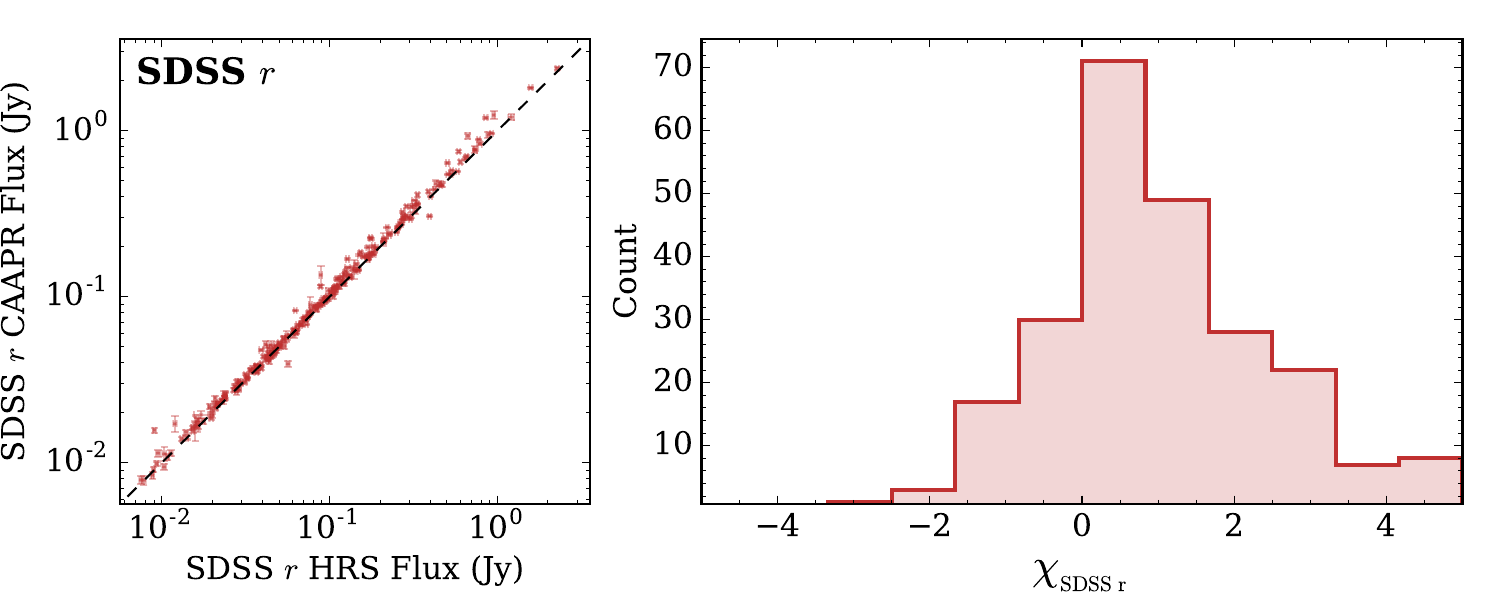}
\includegraphics[width=0.4375\textwidth]{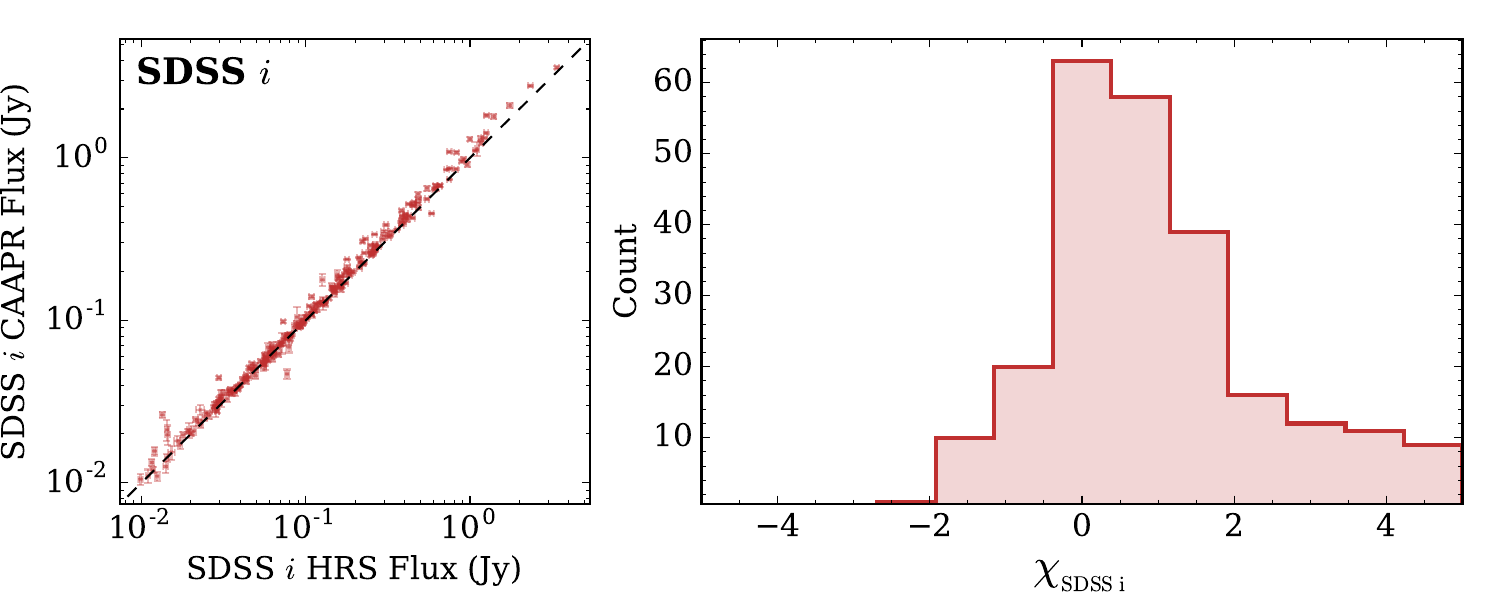}
\includegraphics[width=0.4375\textwidth]{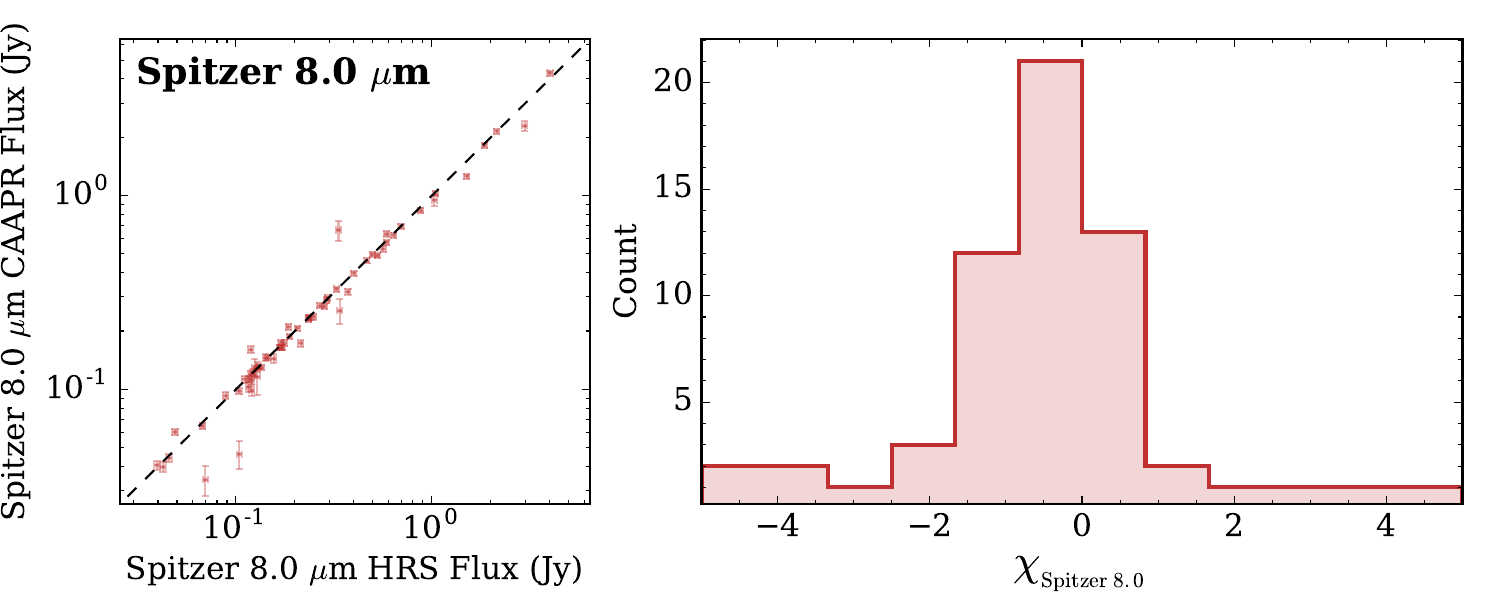}
\includegraphics[width=0.4375\textwidth]{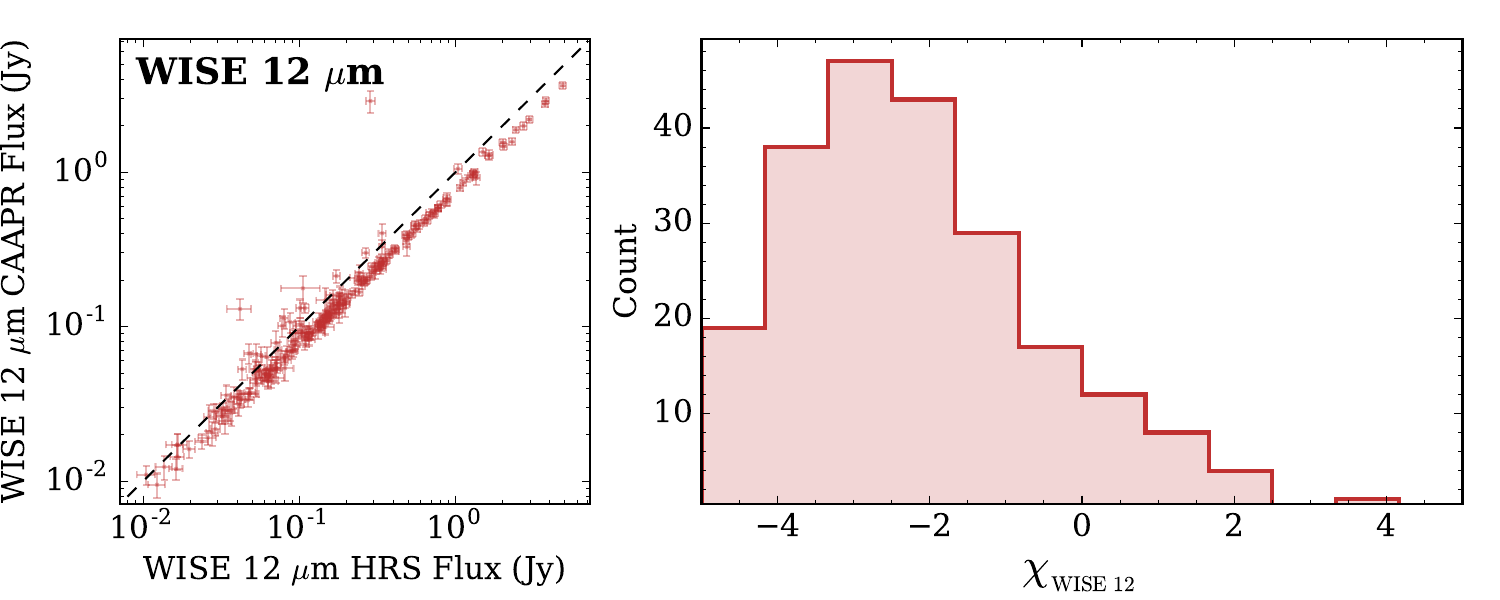}
\includegraphics[width=0.4375\textwidth]{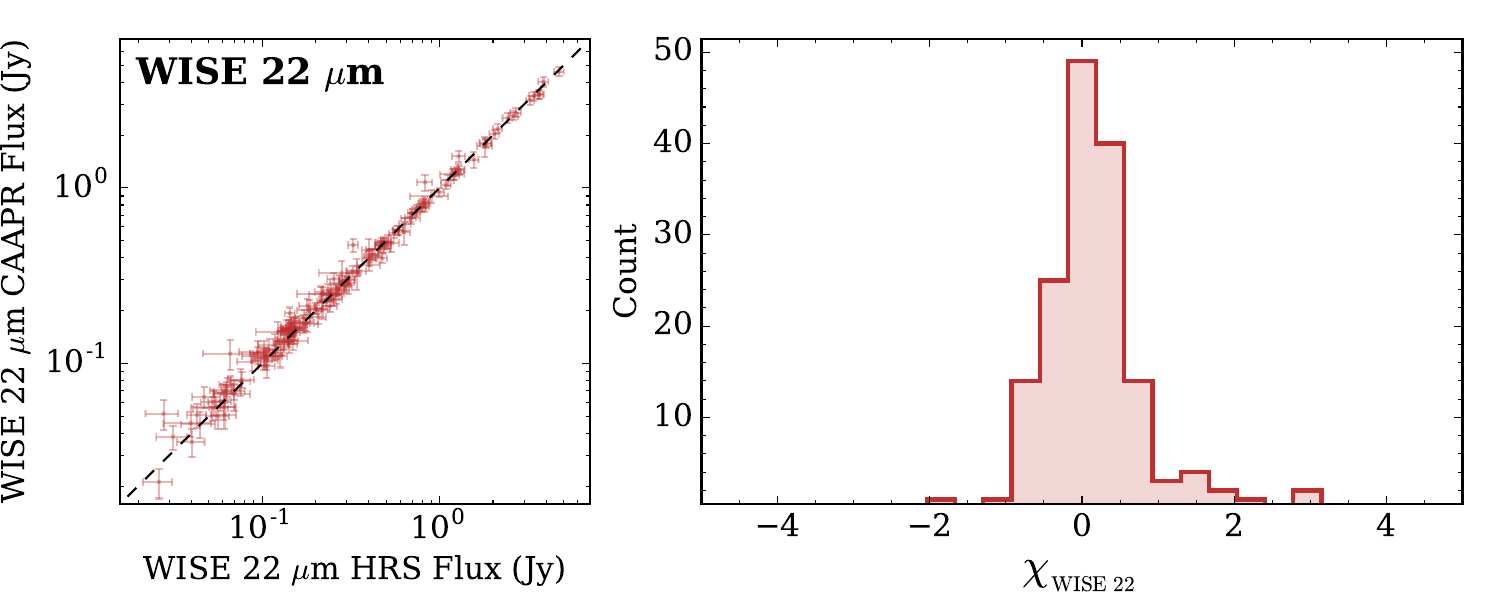}
\includegraphics[width=0.4375\textwidth]{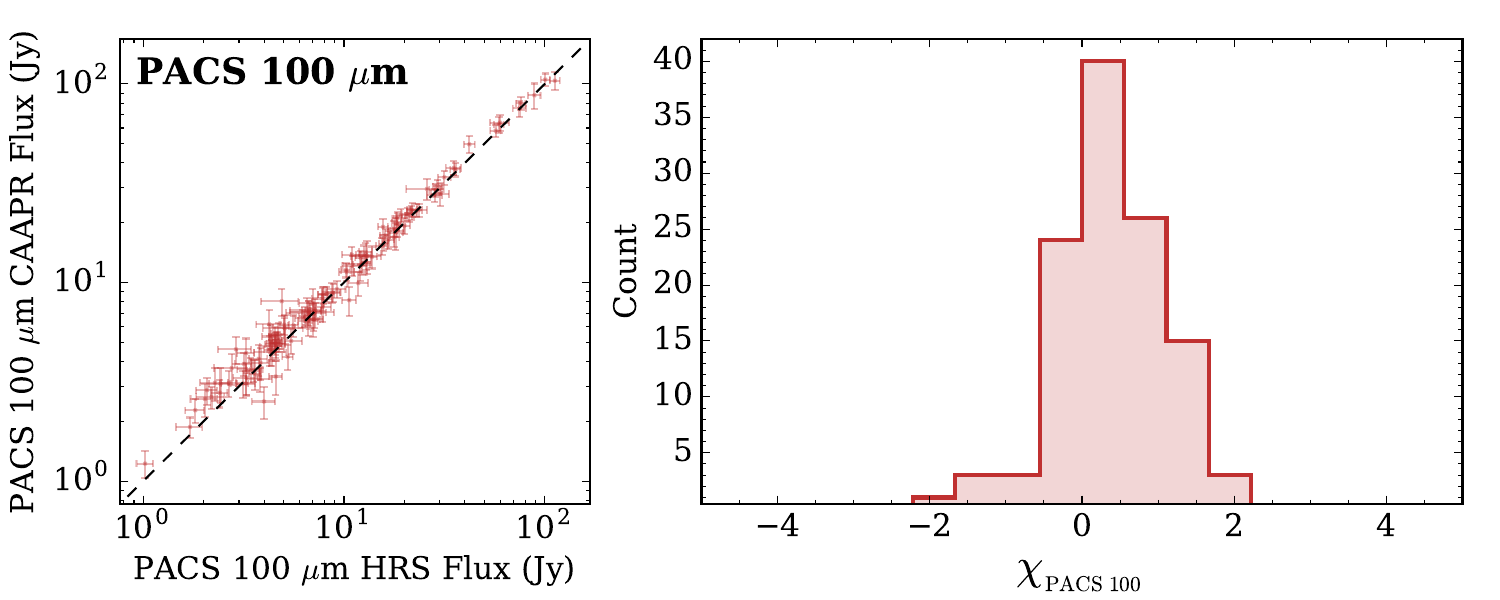}
\includegraphics[width=0.4375\textwidth]{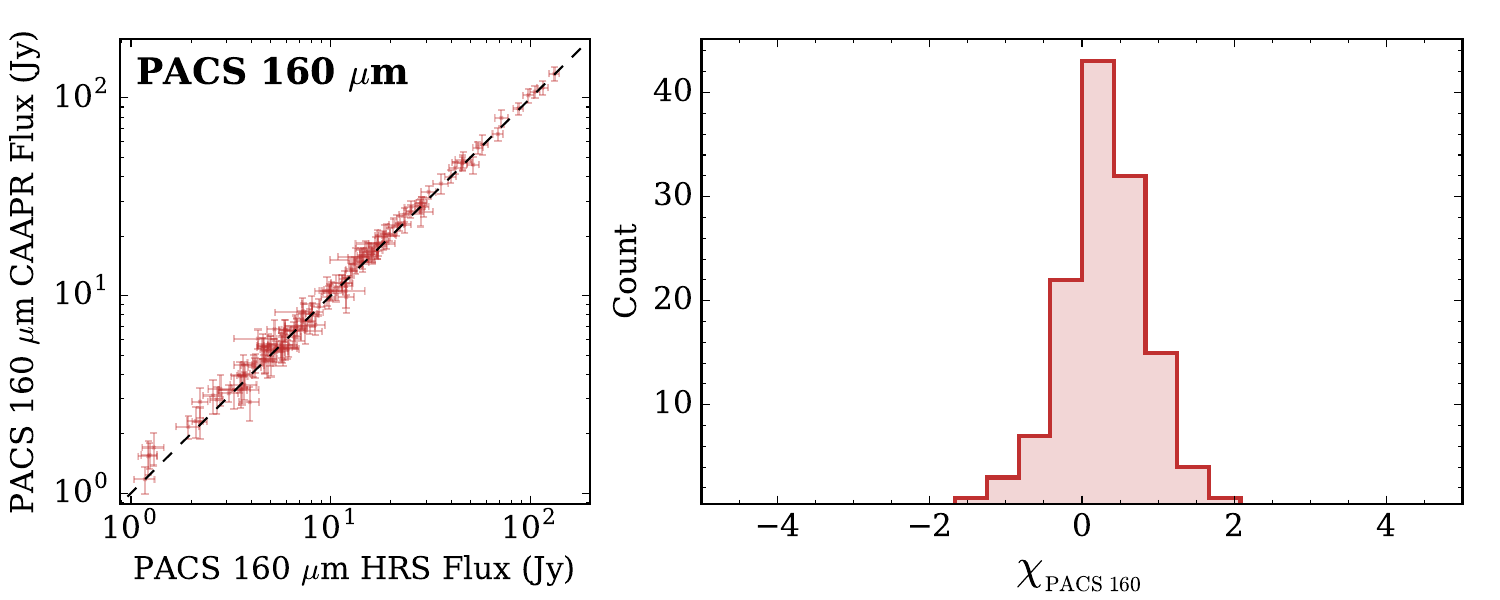}
\includegraphics[width=0.4375\textwidth]{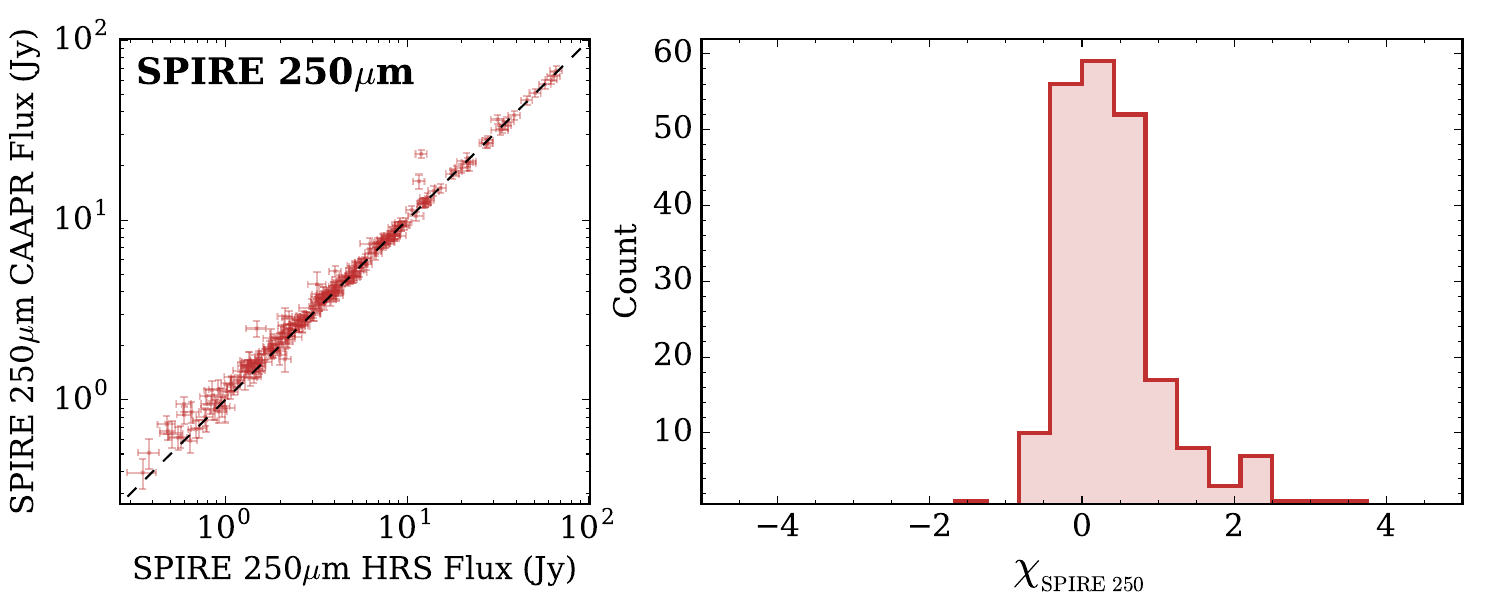}
\includegraphics[width=0.4375\textwidth]{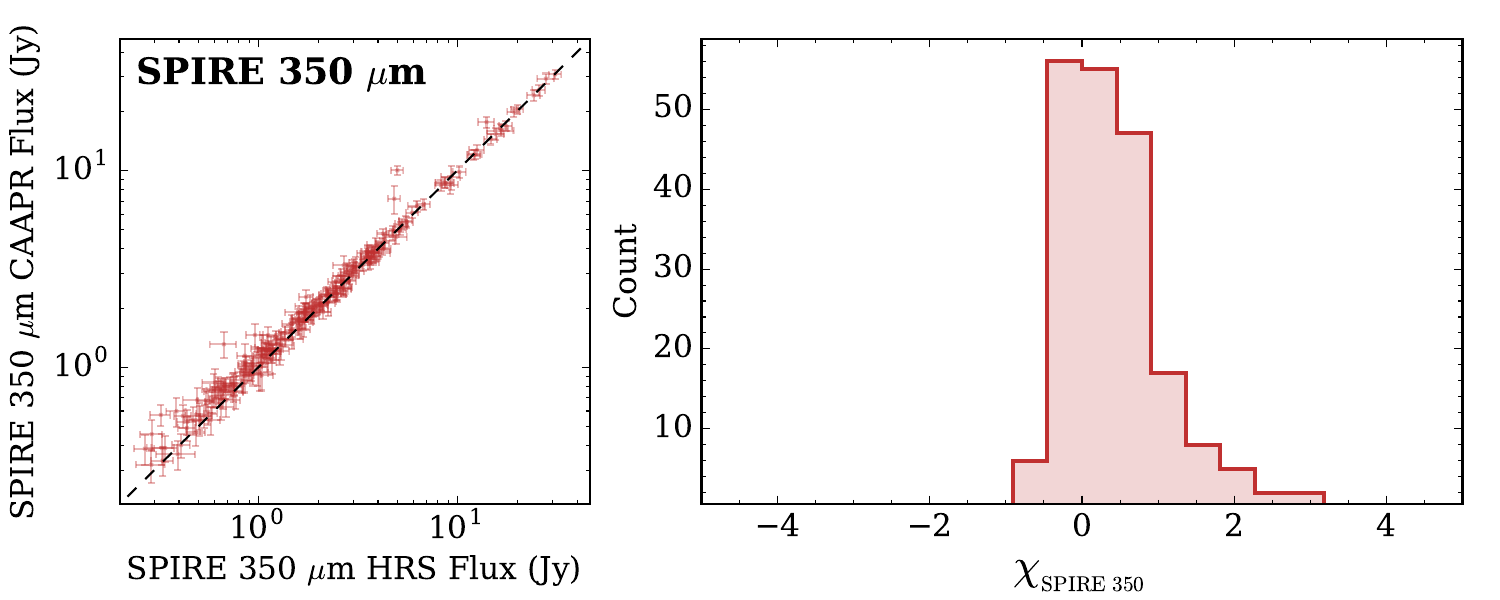}
\includegraphics[width=0.4375\textwidth]{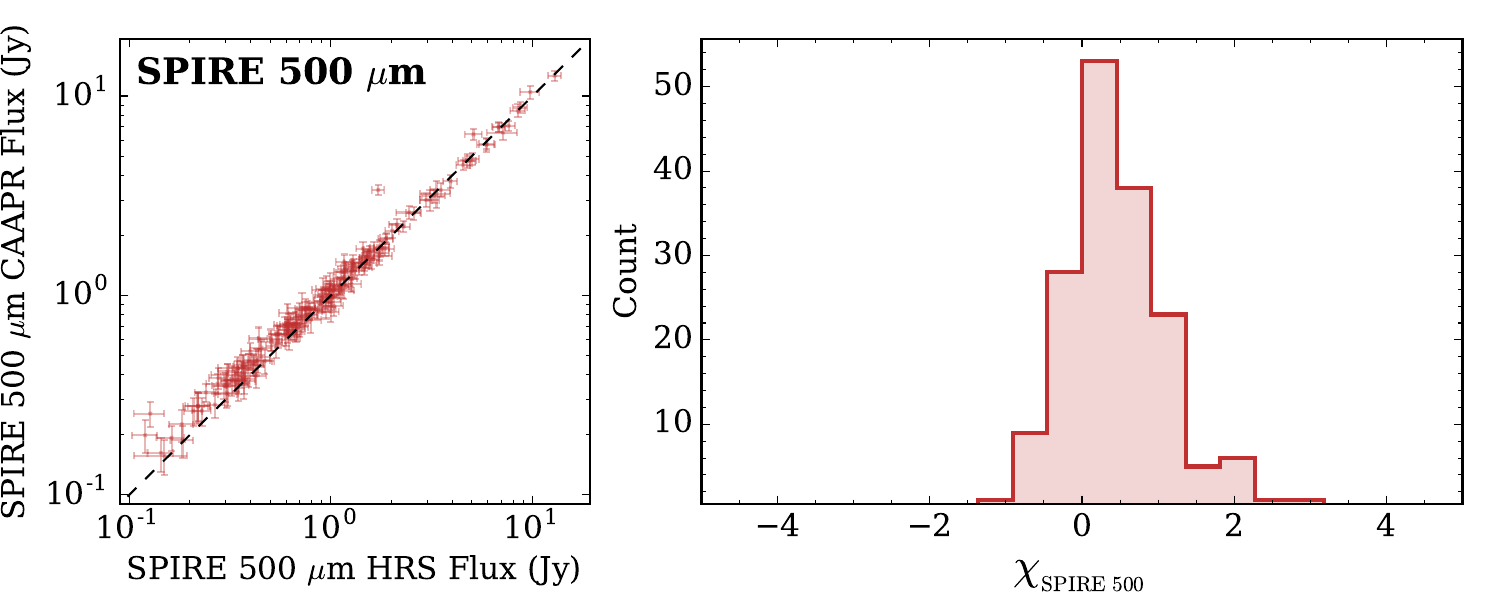}
\includegraphics[width=0.4375\textwidth]{Validation_Placeholder.pdf}
\caption{Comparison of DustPedia \caapr\ and HRS photometry. Left-hand plot for each band directly plots \caapr\ and HRS photometry against each other, with 1:1 relationship indicated by dashed black line. Right-hand plot for each band shows $\chi$ (as per Equation~\ref{Equation:Chi}) distribution for that set of fluxes (binning dictated by the Freedman-Diaconis rule; \citealp{Freedman1981A}). Noticeable deviations for GALEX FUV, GALEX NUV, and WISE 12\,\micron\ are addressed in Sections~\ref{Subsubsubsection:HRS_GALEX_Validation} and \ref{Subsubsubsection:HRS_WISE_Validation}.}
\label{Fig:CAAPR_HRS_Validation}
\end{center}
\end{figure*}

 Figure~\ref{Fig:CAAPR_HRS_Validation} illustrates how our photometry compares to that of the HRS -- both in terms of $\chi$, and the direct relationships between the reported fluxes. The various figures of merit we consider are provided for each band in Table~\ref{Table:CAAPR_HRS_Validation}.

\begin{table}
\begin{center}
\footnotesize
\caption{Figures of merit for comparison of DustPedia \caapr\ and HRS photometry in the 13 bands shared between the samples. }
\label{Table:CAAPR_HRS_Validation}
\begin{tabular}{lSSS}
\toprule \toprule
\multicolumn{1}{c}{Band} &
\multicolumn{1}{c}{$\widetilde{R}$} &
\multicolumn{1}{c}{$\Delta\widetilde{R}$} &
\multicolumn{1}{c}{$\chi^{[-1;1]}$}  \\
\midrule
GALEX FUV & 1.105 & 0.075 & 0.390 \\
GALEX NUV & 1.053 & 0.047 & 0.460 \\
SDSS $g$ & 1.030 & 0.049 & 0.488 \\
SDSS $r$ & 1.033 & 0.032 & 0.421 \\
SDSS $i$ & 1.037 & 0.040 & 0.492 \\
\spitz\ 8.0\,\micron\ & 0.985 & 0.032 & 0.587 \\
WISE 12\,\micron\ & 0.776 & 0.035 & 0.159 \\
WISE 22\,\micron\ & 1.006 & 0.043 & 0.917 \\
PACS 100\,\micron\ & 1.036 & 0.053 & 0.800 \\
PACS 160\,\micron\ & 1.030 & 0.047 & 0.891 \\
SPIRE 250\,\micron\ & 1.031 & 0.046 & 0.848 \\
SPIRE 350\,\micron\ & 1.044 & 0.059 & 0.844 \\
SPIRE 500\,\micron\ & 1.061 & 0.067 & 0.801 \\
\bottomrule
\end{tabular}
\end{center}
\tablefoot{Given figures are $\widetilde{R}$ (median flux ratio), $\Delta\widetilde{R}$ (median absolute deviation in flux ratio), and $\chi^{[-1;1]}$ (fraction of values for which $-1 < \chi < 1$ ); all of these terms are explained in detail at the start of Section~\ref{Subsection:External_Validation}.}
\end{table}

\paragraph{SPIRE} \label{Subsubsubsection:HRS_SPIRE_Validation}

The \caapr\ and HRS photometry in the SPIRE bands compares favourably, with $\widetilde{R} < \Delta\widetilde{R}$ and $ \chi^{[-1,1]} > 0.8$ in all three bands. We note that $\widetilde{R}$ gets progressively larger as wavelength increases -- which is to be expected given that we applied aperture corrections to our SPIRE fluxes to account for the effects of beam spread flux outside the master aperture, whereas the HRS did not \citep{Ciesla2012B}. Our average aperture corrections are 2.7\%, 3.2\%, and 5.8\%\ at 250, 350, and 500\,\micron\ respectively (for sources detected at \textgreater\,3\,$\sigma$ in all three bands), mirroring the increase in $\widetilde{R}$. If we subtract these average corrections from the median flux offsets, the remaining differences would be almost zero, at $\widetilde{R}_{250} = 1.009$, $\widetilde{R}_{350} = 1.012$, and $\widetilde{R}_{500} = 1.002$.

\paragraph{PACS} \label{Subsubsubsection:HRS_PACS_Validation}

Our PACS photometry is in excellent agreement with HRS\footnote{Note that before performing a comparison with the HRS PACS photometry, we corrected their fluxes to account for an issue in {\sc Scanamorphos} \citep{Roussel2013A}, the pipeline used to reduce the HRS PACS maps, relating to the weightings of the relative areas of the reference pixels on the focal plane. These weightings were not fully implemented in {\sc Scanamorphos} until after the HRS published their PACS data. We accounted for this by multiplying the published HRS photometry by 1.01 at 100\,\micron\ and 0.93 at 160\,\micron; these factors represent the average change in fluxes measured from extended-source maps reduced using updated versions of {\sc Scanamorphos}.}. The offsets in both bands are considerably smaller than the scatter -- whilst the scatter in both bands is less than half the 7\% PACS calibration uncertainty. Moreover, both bands enjoy tight $\chi$ distributions. 

\paragraph{WISE} \label{Subsubsubsection:HRS_WISE_Validation}

Our WISE 22\,\micron\ fluxes are in superb agreement with those of the HRS\footnote{Note that we have removed the factor 0.93 colour correction that \citet{Ciesla2014A} apply to all of their WISE 22\,\micron\ fluxes, to allow valid comparison to our fluxes, which are not colour-corrected.}, with a median flux ratio of $\widetilde{R}_{22} = 1.006$, and scatter of $\Delta\widetilde{R}_{22} = 0.043$ -- comfortably within the 5.6\%\ calibration uncertainty of the band. Similarly, the $\chi$ distribution at 22\,\micron\ is very tight, with $\chi^{[-1,1]}_{22} = 0.917$ (which if anything suggests a possible over-estimation of the photometric uncertainties). 

In contrast to the concurrence at WISE 22\,\micron, the WISE 12\,\micron\ fluxes measured by \caapr\ are seriously offset from those quoted by the HRS, with $\widetilde{R}_{12} = 0.776$ -- far exceeding anything that could be ascribed to simple scatter or normal methodological differences. The details of our investigation into the source of this discrepancy are given in Appendix~\ref{AppendixSection:WISE_HRS}. But to summarise, the final result of our investigation was that we found {\it three} reasons why the WISE 12\,\micron\ fluxes measured by the HRS will differ from those measured by \caapr: a factor 1.585 difference due to their incorrect use of the Preliminary Data Release unit calibration when they were in fact using All-Sky Data Release maps; a factor 0.929 difference due to the fact they used the DN-to-Jy given by the WISE documentation, as opposed to the zero-point magnitudes from the map headers (an apparent contradiction in the WISE data); and a factor 0.909 difference due to their smaller apertures and lack of foreground star removal. Combined, these three effects should make the HRS fluxes brighter than our own by a factor of 1.338 -- ie, an expected median offset of $\widetilde{R} = 0.747$, which is very similar to the actual median offset of $\widetilde{R}_{12} = 0.776$. The remaining 2.9\%\ difference is much smaller than the WISE 12\,\micron\ calibration uncertainty of 4.6\%, and well within the $\Delta\widetilde{R}_{22} = 0.035$ scatter -- suggesting we have successfully isolated all of the significant causes of disagreement between our respective photometry.

\paragraph{{\it Spitzer}} \label{Subsubsubsection:HRS_Spitzer_Validation}

Our \spitz\ 8.0\,\micron\ photometry is in good agreement with that of the HRS; the median flux ratio of $\widetilde{R}_{8.0} = 0.985$  is well within both the 3\%\ calibration uncertainty of the band and the $\Delta\widetilde{R}_{8.0} = 0.032$ scatter. Although the $\chi^{[-1,1]}_{8.0} = 0.587$ is somewhat smaller than optimal, DustPedia and the HRS only share 65 galaxies with high-quality \spitz\ 8.0\,\micron\ \caapr\ fluxes, so small-number statistics prevent us from inferring too much from the $\chi$ distribution. 

\begin{figure*}
\begin{center}
\includegraphics[width=0.4375\textwidth]{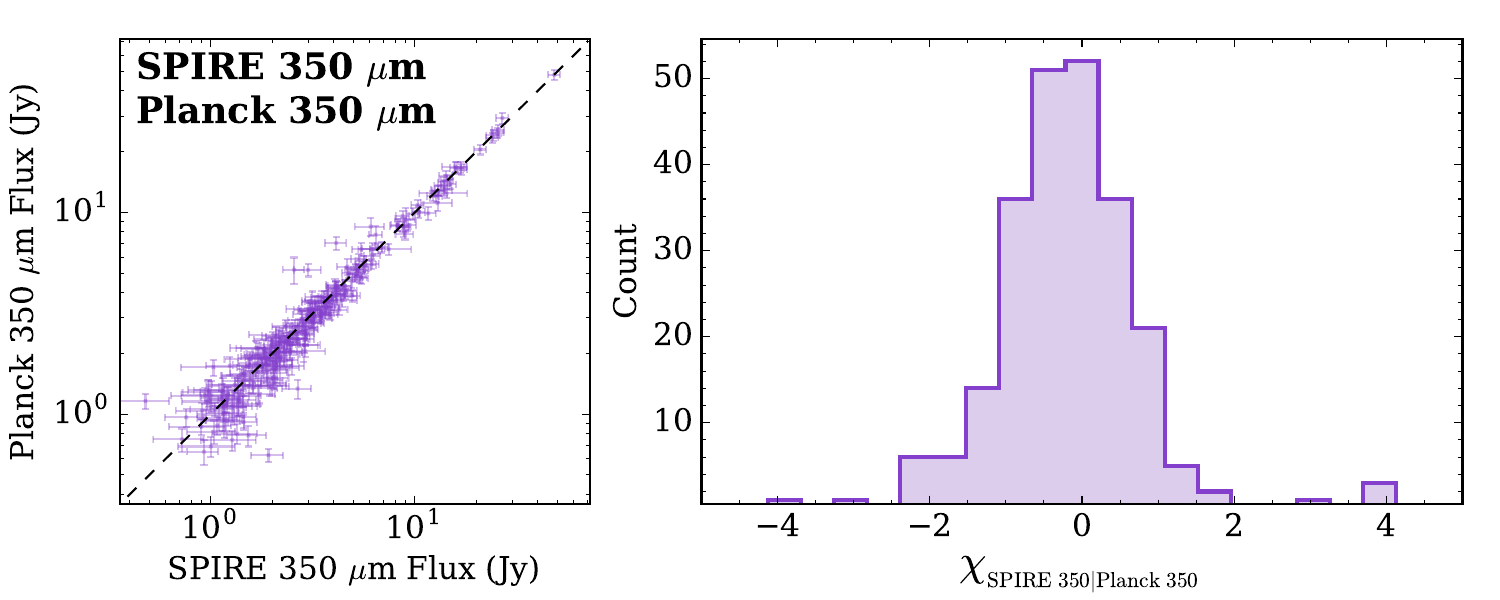}
\includegraphics[width=0.4375\textwidth]{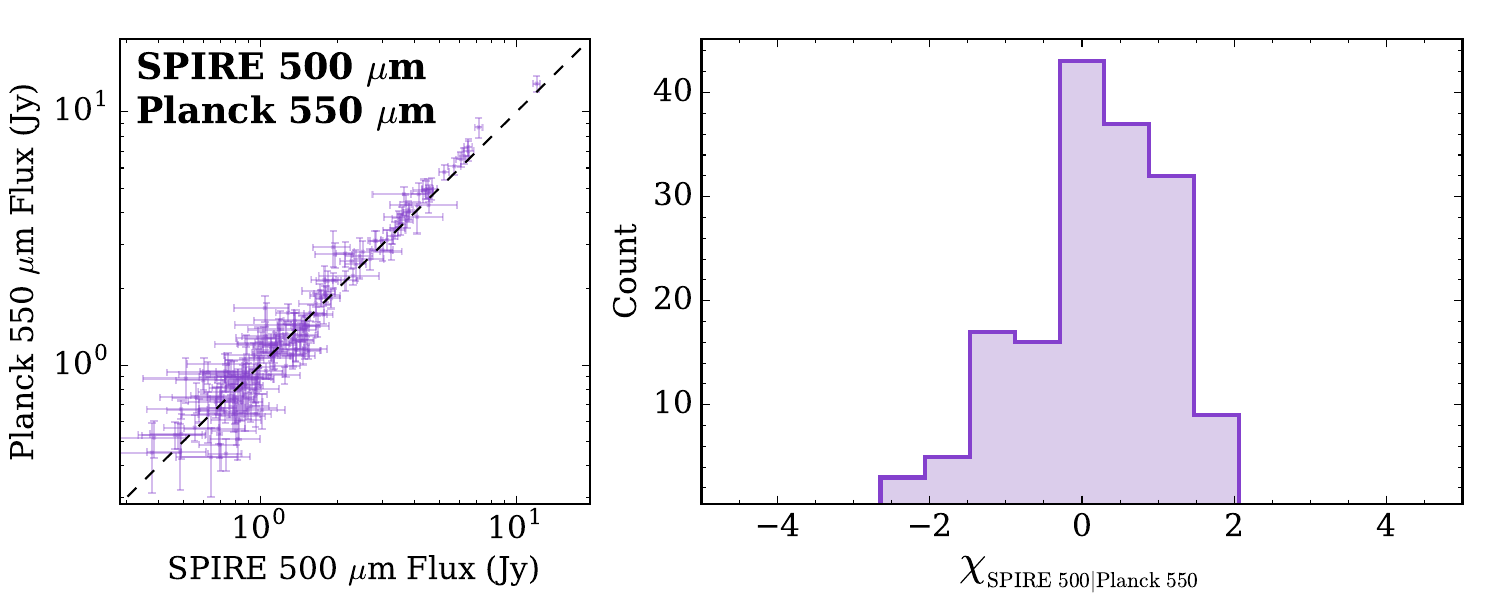}
\caption{Comparison of DustPedia \caapr\ and \planck\ CCS2 photometry in similar bands. Plots as per Figure~\ref{Fig:CAAPR_HRS_Validation}.}
\label{Fig:Planck_Internal_Validation}
\end{center}
\end{figure*}

\paragraph{SDSS} \label{Subsubsubsection:HRS_SDSS_Validation}

Our SDSS photometry shows only a small offset from the HRS values, with median flux ratios $\widetilde{R} < 1.037$ in all three bands (the HRS only report SDSS photometry for{\it gri}-bands). Nonetheless, this is larger than both $\Delta\widetilde{R}$ and the calibration uncertainties for each band, so it is probably a real effect. However a discrepancy this small is within the realm of what can be ascribed to unremarkable methodological differences. \citet{Cortese2012C} do not describe their foreground star removal process in detail, which in itself could account for an offset of this size through minor methodological differences (eg, differences in distinguishing between bright H{\sc ii} regions in the target galaxies from true foreground stars). The SDSS $\chi$ distributions in Figure~\ref{Fig:CAAPR_HRS_Validation} are well-behaved, showing few outliers, although they are a little broad (with $\chi^{[-1;1]} < 0.49$ for all bands).

\paragraph{GALEX} \label{Subsubsubsection:HRS_GALEX_Validation}

Our GALEX photometry shows a systematic offset from that of the HRS, with $\widetilde{R}_{\rm FUV} = 1.105$ and $\widetilde{R}_{\rm NUV} = 1.053$; both of these offsets are significant, given the bands' respective calibration uncertainties of 4.5\%\ and 2.7\%, and scatters of $\Delta\widetilde{R}_{\rm FUV} = 0.075$ and $\Delta\widetilde{R}_{\rm NUV} = 0.047$. The cause of this offset is not immediately obvious. As with the HRS SDSS photometry, no detailed description is given of their foreground star removal process, so this is a potential methodological source of some of the discrepancy. Also, the HRS fluxes are asymptotic measurements; as this method assumes a smooth curve-of-growth, it is possible that our apertures (designed to ensure all practically-recoverable flux is recorded) encompass additional flux excluded by their technique. \citet{Cortese2012C} find that their fluxes are on average $\sim$\,9\%\ fainter than those reported by \citet{GilDePaz2007A} for the 62 galaxies they have in common (though this may in part be due to the fact that \citealp{GilDePaz2007A} use the older GR2/GR3 GALEX data release) -- similar to the discrepancy we find in FUV, and even larger than we find in NUV. On the other hand, we note that the HRS GALEX fluxes are systematically {\it brighter} than those reported by \citeauthor{Bai2015B} (\citeyear{Bai2015B}; who use the current GR6/GR7 GALEX data release, and have 90 galaxies in common with the HRS), by an average of 4.0\%\ in FUV and 2.4\%\ in NUV. It therefore seems that the disagreement between our GALEX photometry and that of the HRS is in line with the typical variation between authors.

\subsubsection{\textit{Planck} CCS2} \label{Subsubsection:Planck_CCS2_Validation}

The wavelength coverage of \planck\ and \hersc\ overlap by design, to allow them to conduct complimentary observations; they have similar spectral response functions at 500--550\,\micron, and nearly identical response functions at 350\,\micron\ \citep{Bertincourt2016A}. As such, our supplementary \planck\ CCS2 photometry is well suited to being cross-validated with our SPIRE \caapr\ photometry. Plots comparing the two sets of fluxes are shown in Figure~\ref{Fig:Planck_Internal_Validation}, whilst figures of merit for the cross-validation are given in the upper block of Table~\ref{Table:CAAPR_Supplementary_Validation}. For the CCS2 photometry, we excluded any fluxes with a `e' flag (in addition to the standard exclusion of major flags).

At 350\,\micron\ the agreement between the two sets of fluxes is very good; there is no meaningful offset, whilst the $\chi$ distribution is tight and Gaussian.

For 500--550\,\micron\, we first colour-corrected the SPIRE 500\,\micron\ fluxes to allow for a valid comparison; we followed \citet{Baes2014A} in multiplying the SPIRE 500\,\micron\ fluxes by a correction factor of 0.83, which assumes a 20\,K modified black body spectrum with an emissivity slope of $\beta = 2.0$. The two sets of photometry seem reasonably consistent. Whilst it appears that the CCS2 fluxes might be slightly brighter (with $\widetilde{R}_{500|550} = 1.05$), this is not only within the calibration uncertainty of both bands, but also within the range of deviation that could be ascribed to our colour correction. For example, were we to assume a $\beta = 1.5$ emissivity slope, the colour correction factor applied to the SPIRE 500\,\micron\ fluxes would instead be 0.87, which would almost entirely remove the observed offset.

\begin{figure*}
\begin{center}
\includegraphics[width=0.4375\textwidth]{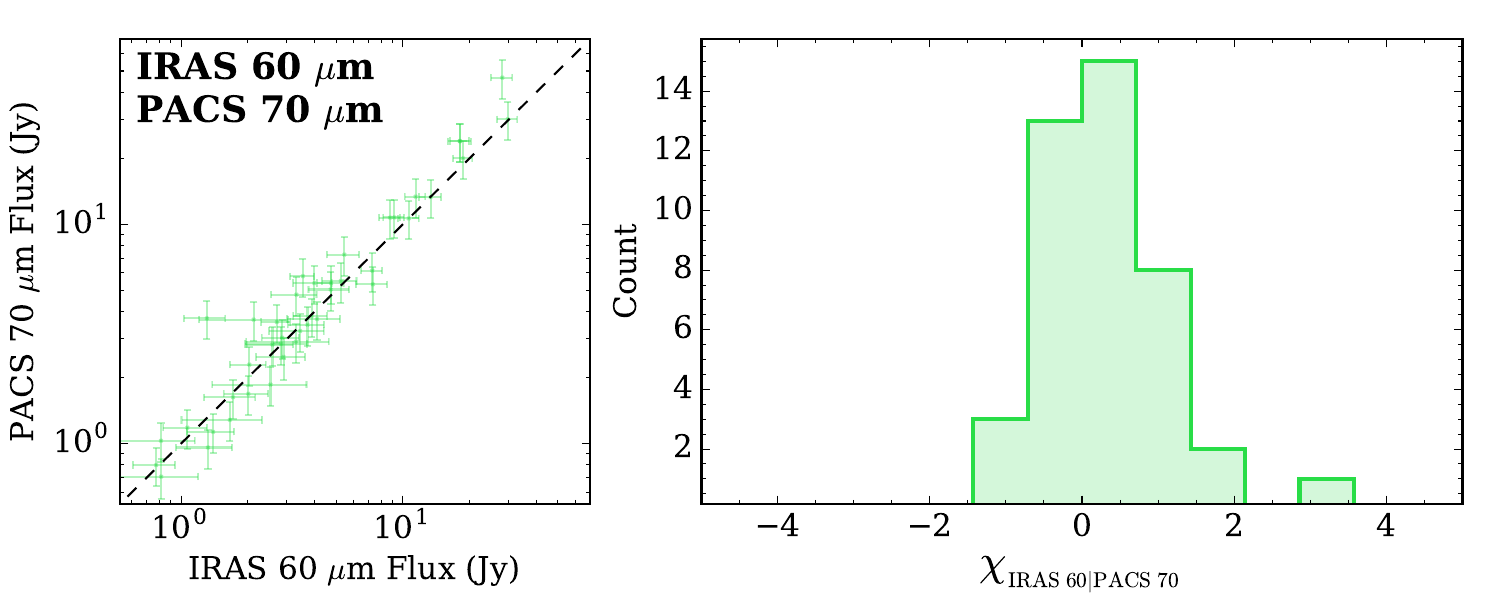}
\includegraphics[width=0.4375\textwidth]{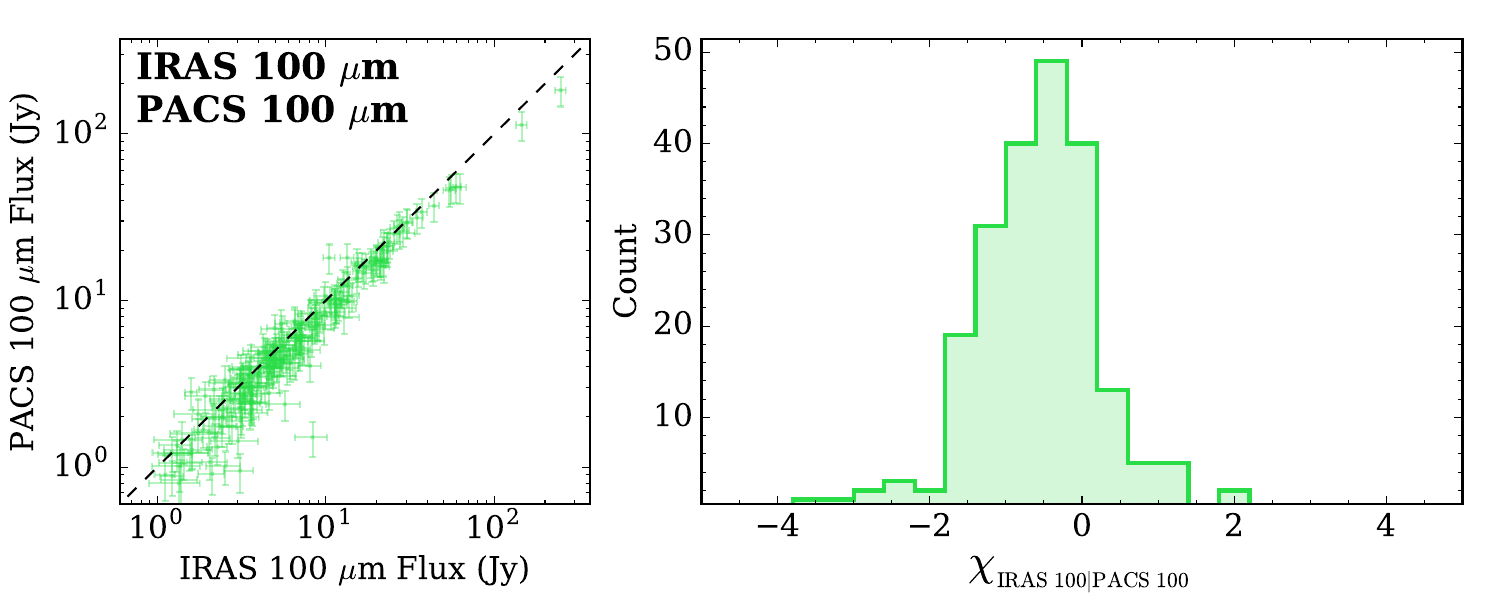}
\caption{Comparison of DustPedia \caapr\ and IRAS SCANPI photometry in similar bands. Plots as per Figure~\ref{Fig:CAAPR_HRS_Validation}.}
\label{Fig:IRAS_Internal_Validation}
\end{center}
\end{figure*}

\begin{table}
\begin{center}
\footnotesize
\caption{Figures of merit for comparison of DustPedia \caapr\ photometry with \planck\ CCS2 (upper block of values) and IRAS SCANPI (lower block of values) supplementary photometry. }
\label{Table:CAAPR_Supplementary_Validation}
\begin{tabular}{llrrr}
\toprule \toprule
\multicolumn{1}{c}{$S_{1}$} &
\multicolumn{1}{c}{$S_{2}$} &
\multicolumn{1}{c}{$\widetilde{R}$} &
\multicolumn{1}{c}{$\Delta\widetilde{R}$} &
\multicolumn{1}{c}{$\chi^{[-1;1]}$}  \\
\midrule
SPIRE 350\,\micron\ & \planck\ 350\,\micron\ & 0.970 & 0.089 & 0.787 \\
SPIRE 500\,\micron\ & \planck\ 550\,\micron\ & 1.050 & 0.103 & 0.698 \\
\midrule
IRAS 60\,\micron\ & PACS 70\,\micron\ & 1.057 & 0.170 & 0.738 \\
IRAS 100\,\micron\ & PACS 100\,\micron\ & 0.875 & 0.096 & 0.690 \\
\bottomrule
\end{tabular}
\end{center}
\tablefoot{The bands being compared are denoted by $S_{1}$ and $S_{2}$ (such that flux ratios are $S_{1}/S_{2}$); for consistency, $S_{1}$ represents the shortest wavelength band of each pair. Other column definitions the same as for Table~\ref{Table:CAAPR_HRS_Validation}.}
\end{table}

\subsubsection{IRAS SCANPI} \label{Subsubsection:IRAS_SCANPI_Validation}

\begin{figure*}
\begin{center}
\includegraphics[width=0.4375\textwidth]{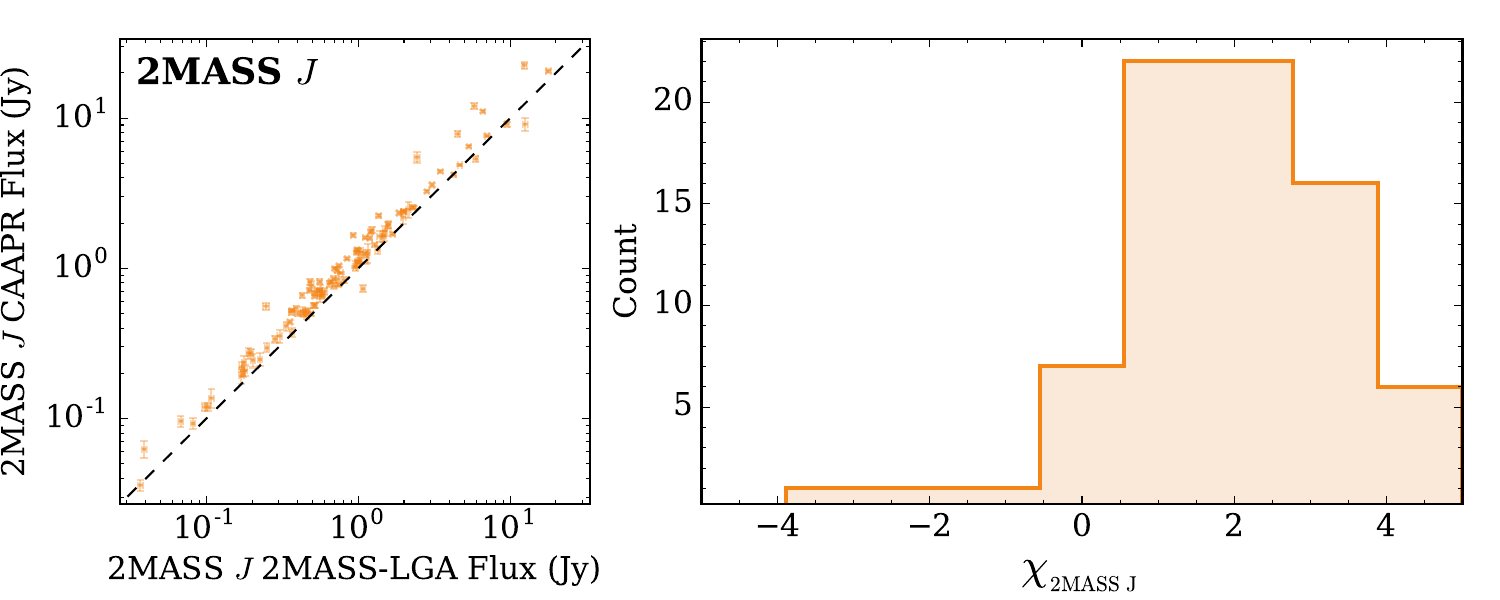}
\includegraphics[width=0.4375\textwidth]{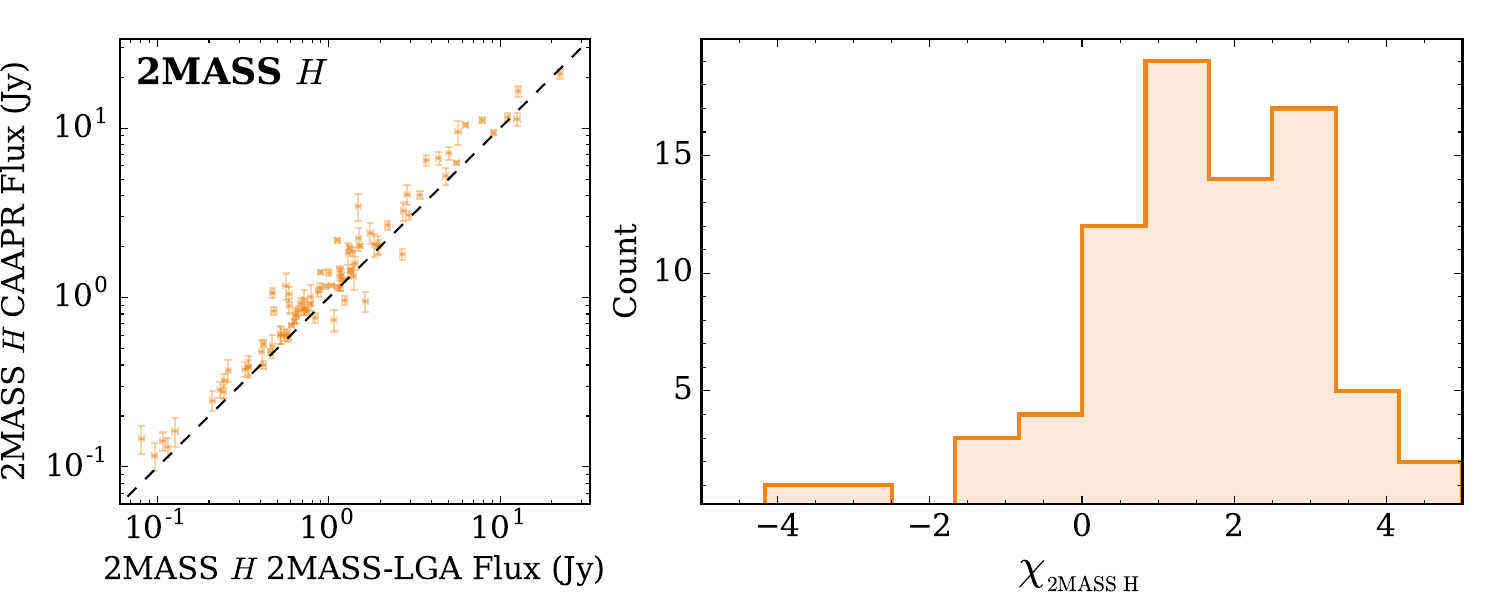}
\includegraphics[width=0.4375\textwidth]{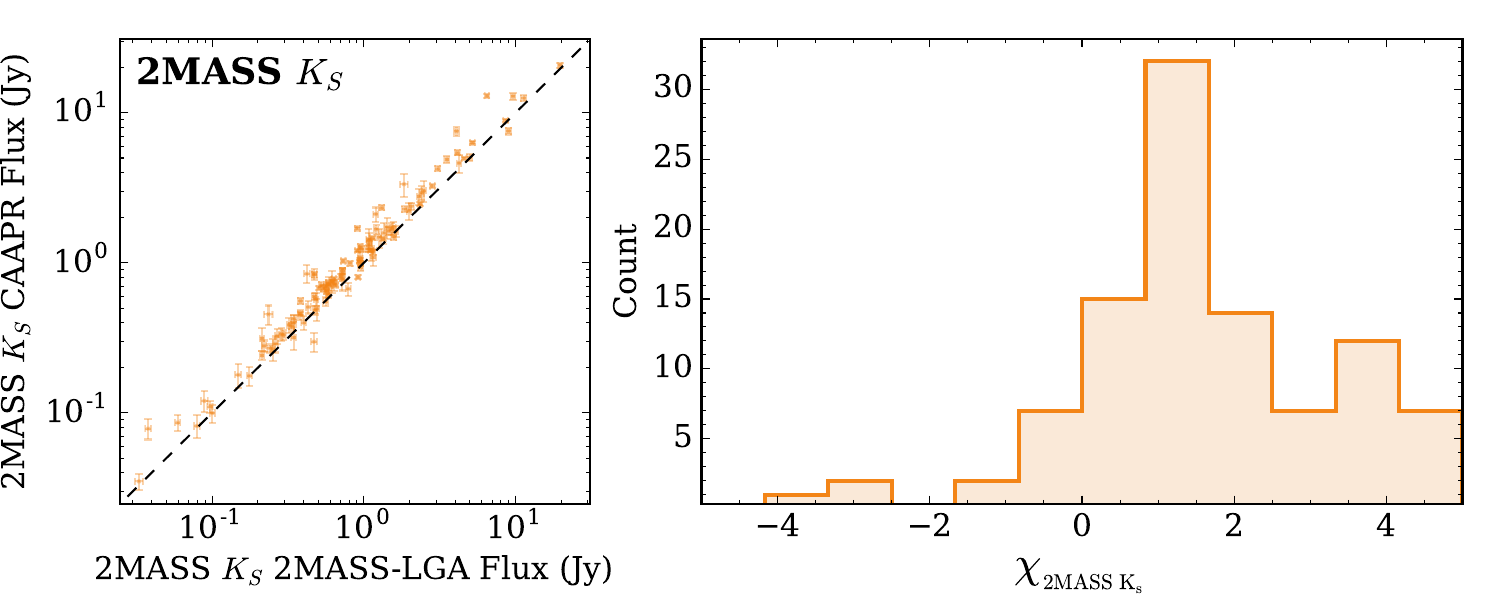}
\includegraphics[width=0.4375\textwidth]{Validation_Placeholder.pdf}
\caption{Comparison of DustPedia \caapr\ photometry to 2MASS-LGA photometry of the galaxies in both samples. Plots as per Figure~\ref{Fig:CAAPR_HRS_Validation}. Note that Figure~\ref{AppendixFig:CAAPR_2MASS_Repeat_Validation} shows a version of this plot where the \caapr\ photometry used the 2MASS-LGA apertures, instead of our own.}
\label{Fig:CAAPR_2MASS_Validation}
\end{center}
\end{figure*}

It is important to ensure that our supplementary IRAS SCANPI photometry is consistent with our aperture-matched \caapr\ photometry, especially at 60\,\micron\ (an important part of the spectrum for SED modelling, as previously discussed) where IRAS provides the only available photometry for \textgreater\,40\%\ of the DustPedia galaxies. Plots comparing PACS 70 and 100\,\micron\ fluxes to SCANPI 60 and 100\,\micron\ fluxes are shown in Figure~\ref{Fig:IRAS_Internal_Validation}, whilst  for the cross-validation figures of merit are given in the lower block of Table~\ref{Table:CAAPR_Supplementary_Validation}.

To allow for a valid comparison between instruments, colour corrections for IRAS were taken from Section~VI.C, Table~VI.C.6 of the IRAS Explanatory Supplement\footnoteref{Footnote:IRAS_Explanatory_Supplement}, whilst colour corrections for PACS were taken from \citet{Muller2011A}. Emission from galaxies in the IRAS--PACS wavelength regime tends to be due to emission from a mixture of sources (cold dust, warm dust, and small-grain emission). The full SED modelling required to describe the relative contributions of these components is beyond the scope of this work; for the purposes of colour-correction, here we simply assume a reference spectrum described by an appropriate choice of power law, of the form $S_{\nu} \propto \nu^{\alpha}$.

To compare SCANPI 60\,\micron\ and PACS 70\,\micron\ photometry, we colour-corrected both sets of fluxes assuming a power law index of $\alpha = -2.0$ as the reference spectrum\footnote{\label{Footnote:IRAS_Colour_Correction} A power law slope of $\alpha = -2.0$ at 60\,\micron\ and $\alpha = -1.0$ at 100\,\micron\ would approximate a fairly unremarkable galaxy SED at those wavelengths -- for example, that of a two-component modified blackbody SED with a cold dust temperature of 20\,K, a warm dust temperature of 45\,K, a cold-to-warm mass ratio of 100:1, and an emissivity slope of $\beta = 2$. See \citet{MWLSmith2012A}, \citet{Ciesla2014A}, and \citet{CJRClark2015A} for examples of typical FIR-submm SEDs of nearby galaxies. See also \citet{Casey2012E} for discussion of $\alpha = -2.0$ being a typical power law in the $\sim$\,60\,\micron\ range.}; this colour-correction included translating the PACS 70\,\micron\ fluxes to an effective central wavelength of 60\,\micron\ as per the prescription of \citet{Muller2011A}. The result suggests SCANPI and PACS fluxes are compatible, with the median offset of $\widetilde{R}_{60|70} = 1.057$ -- which is not only less than the sizeable calibration uncertainty of IRAS, but also less than the much more modest PACS calibration uncertainty.

To compare SCANPI and PACS photometry at 100\,\micron, we assumed a power law slope of $\alpha = -1.0$ for the reference spectrum\footnoteref{Footnote:IRAS_Explanatory_Supplement}; conveniently, this is the standard reference spectrum for both instruments, meaning no colour-correction was necessary. The $\widetilde{R}_{100|100} = 0.875$ median offset is less than the 20\%\ SCANPI 100\,\micron\ calibration uncertainty, and the $\chi$ distribution is tight and Gaussian enough to yield a nearly-ideal $\chi^{[-1,1]}_{100|100} = 0.690$. And given the fact that our colour-corrections are only approximate in this instance, the agreement between SCANPI and PACS at 100\,\micron\ seems satisfactory.

\begin{table}
\begin{center}
\footnotesize
\caption{Figures of merit for comparison of DustPedia \caapr\ and 2MASS-LGA photometry.}
\label{Table:CAAPR_2MASS-LGA_Validation}
\begin{tabular}{lSSS}
\toprule \toprule
\multicolumn{1}{c}{Band} &
\multicolumn{1}{c}{$\widetilde{R}$} &
\multicolumn{1}{c}{$\Delta\widetilde{R}$} &
\multicolumn{1}{c}{$\chi^{[-1;1]}$}  \\
\midrule
2MASS $J$ & 1.205 & 0.105 & 0.124 \\
2MASS $H$ & 1.182 & 0.134 & 0.244 \\
2MASS $K_{S}$ & 1.172 & 0.092 & 0.237 \\
\midrule
{\it 2MASS-LGA $J$} & {\it 1.008} & {\it 0.028} & {\it 0.701} \\
{\it 2MASS-LGA $H$} & {\it 0.996} & {\it  0.025} & {\it 0.713} \\
{\it 2MASS-LGA $K_{S}$} & {\it 1.001} & {\it 0.021} & {\it 0.776} \\
\bottomrule
\end{tabular}
\end{center}
\tablefoot{ The upper block of figures are for the direct comparison between the DustPedia \caapr\ and 2MASS-LGA photometry; the lower block of figures (in italic) compare the 2MASS-LGA photometry to \caapr\ photometry repeated using the 2MASS-LGA apertures. Column definitions the same as for Table~\ref{Table:CAAPR_HRS_Validation}.}
\end{table}

\begin{figure*}
\begin{center}
\includegraphics[width=0.4375\textwidth]{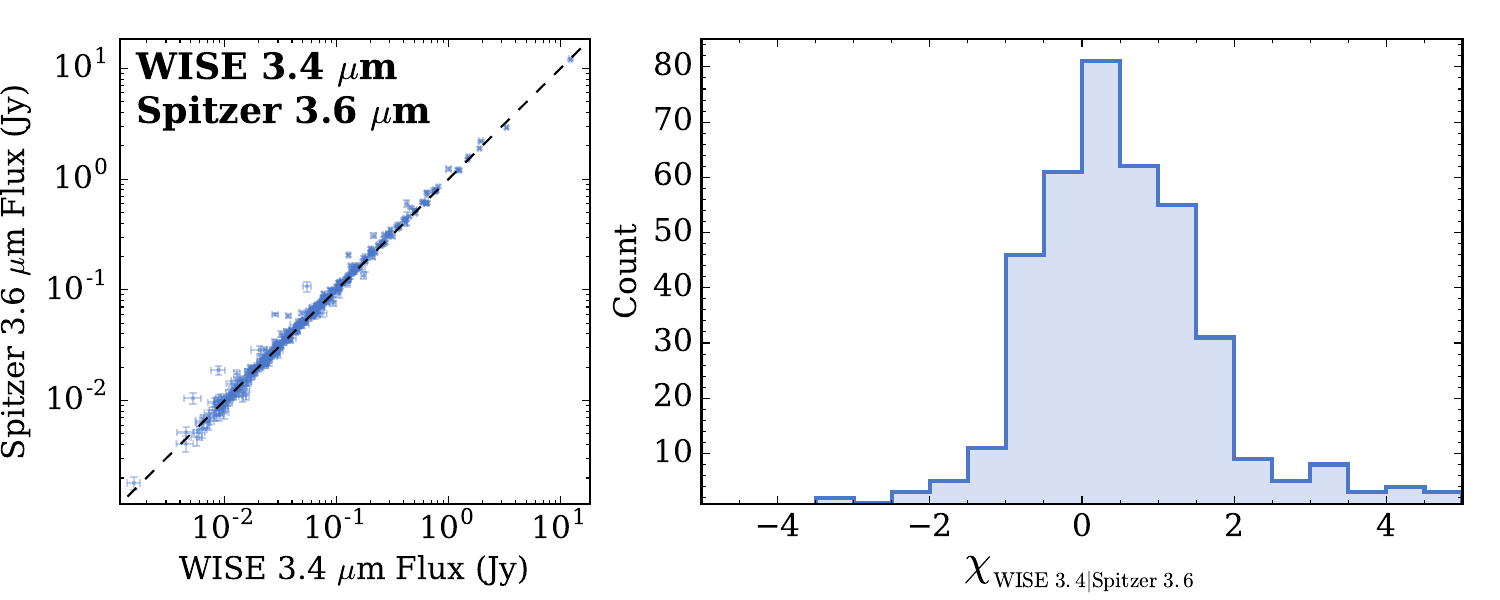}
\includegraphics[width=0.4375\textwidth]{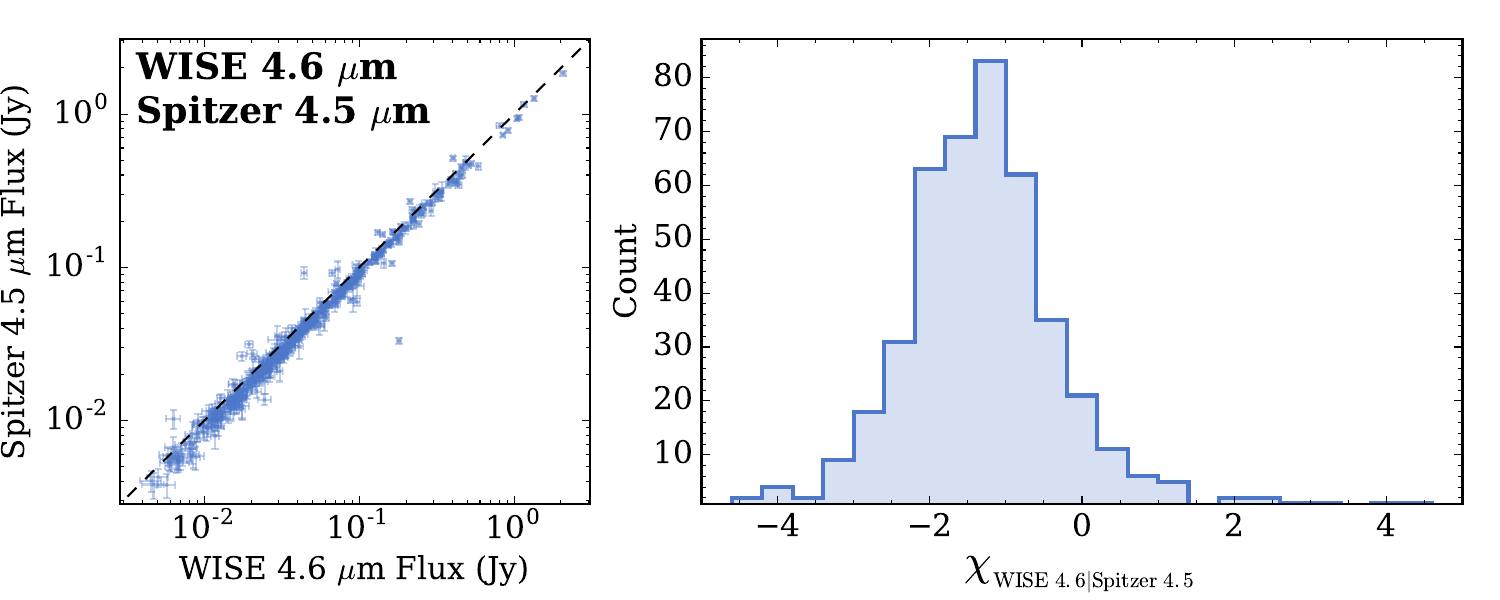}
\includegraphics[width=0.4375\textwidth]{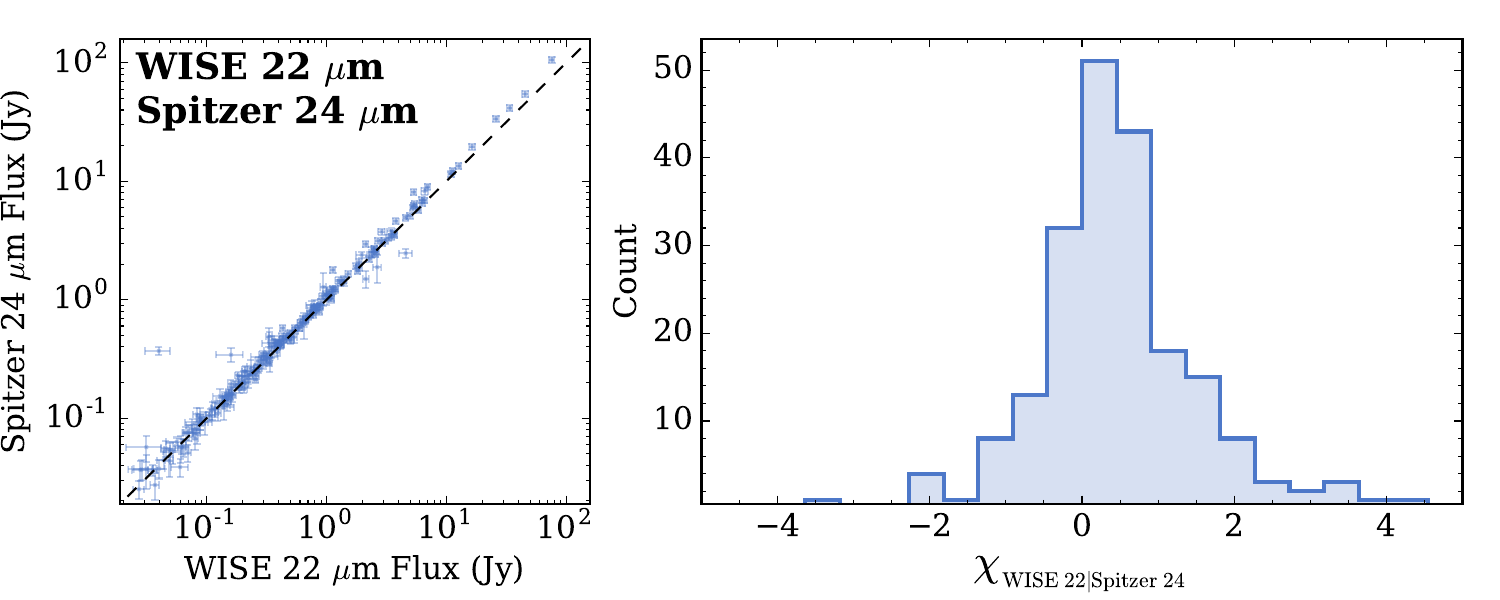}
\includegraphics[width=0.4375\textwidth]{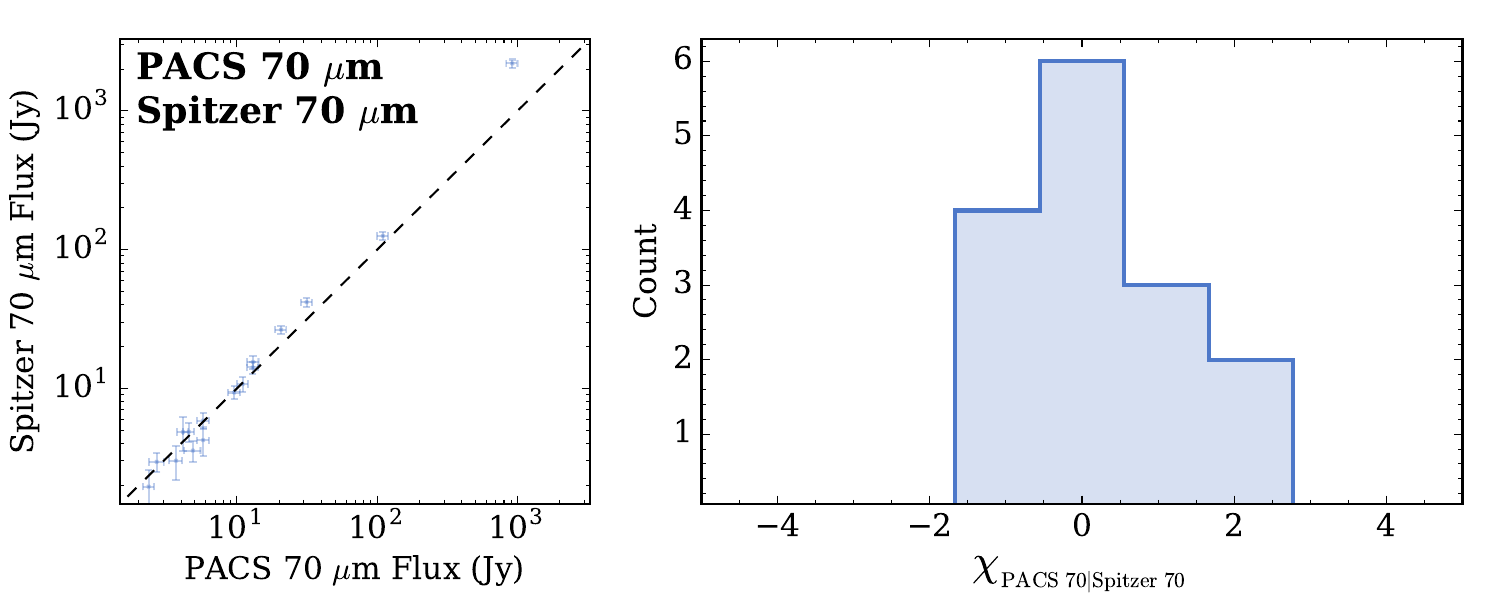}
\includegraphics[width=0.4375\textwidth]{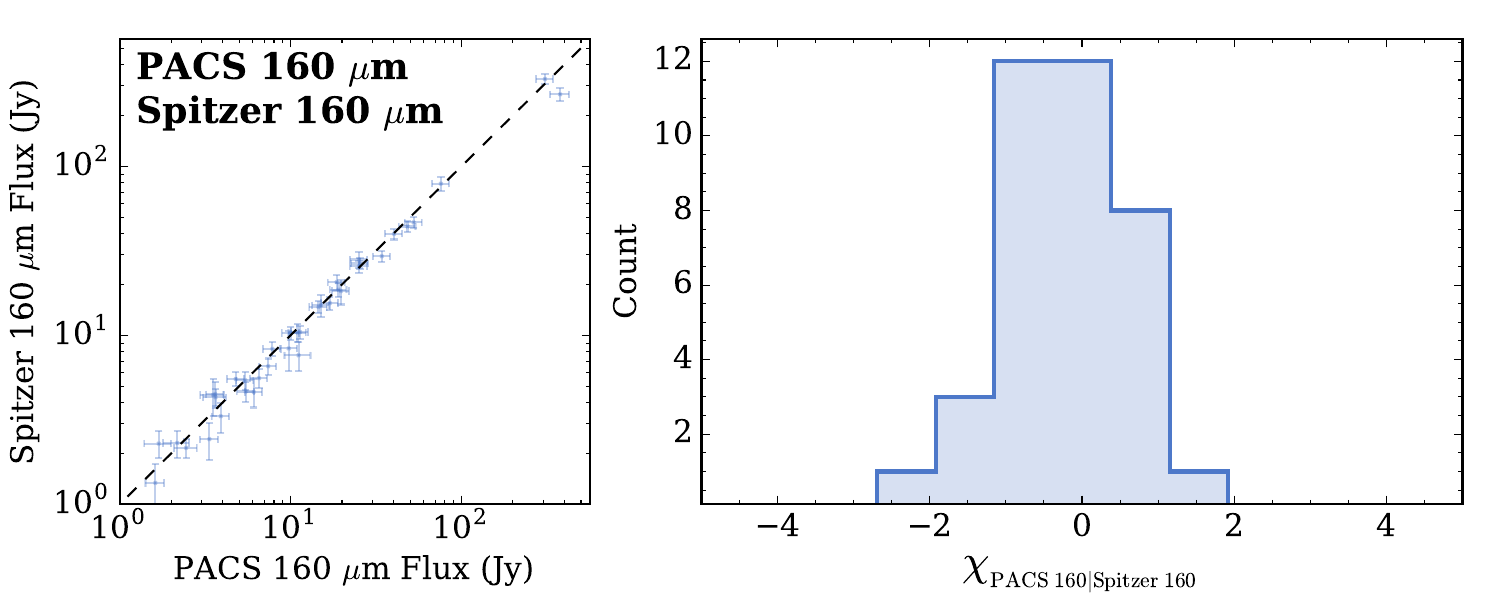}
\includegraphics[width=0.4375\textwidth]{Validation_Placeholder.pdf}
\caption{Comparison of DustPedia \caapr\ photometry in similar bands. Plots as per Figure~\ref{Fig:CAAPR_HRS_Validation}.}
\label{Fig:CAAPR_Internal_Validation}
\end{center}
\end{figure*}

\subsubsection{2MASS Large Galaxy Atlas} \label{Subsubsection:2MASS-LGA_Validation}

The only instrument for which HRS, \planck\ CCS2 and IRAS SCANPI photometry provide no independent external validation for our aperture-matched \caapr\ photometry is 2MASS. To remedy this, we refer to the 2MASS Large Galaxy Atlas (2MASS-LGA; \citealp{Jarrett2003A}). The 2MASS-LGA has 129 galaxies in common with DustPedia, spread across the sky (including in the plane of the Milky Way). We specifically compare to the 2MASS-LGA large-aperture Kron photometry, as it is the most similar of their photometric measures to our own. Plots comparing the \caapr\ and 2MASS-LGA photometry for our common sources are shown in Figure~\ref{Fig:CAAPR_2MASS_Validation}, and the validation figures of merit are given in the upper block of Table~\ref{Table:CAAPR_2MASS-LGA_Validation}.

In all three {\it JHK$_{S}$}-bands, the 2MASS-LGA photometry compares poorly with our own. Our fluxes are systematically brighter by an average of 19\%, and the $\chi$ distributions are broad, non-Gaussian, and asymmetrical in all bands. However, the photometric apertures used by the 2MASS-LGA are significantly smaller than our own; on average, our apertures encompasses 8 times more area. Visual inspection of the 2MASS-LGA apertures reveals that they often do not encompass all of the stellar emission of the target galaxies -- especially when compared to the deeper WISE 3.4\,\micron\ data, which shows an appreciable fraction of some galaxies' NIR flux can extend beyond the 2MASS-LGA apertures. This strongly suggests that the reason we record systematically brighter {\it JHK$_{S}$}-band photometry was that we recovered flux excluded by the much smaller apertures of the 2MASS-LGA. 

To test this hypothesis, we repeated our photometry, but instead used the apertures employed by the 2MASS-LGA; the resulting comparison plots are shown in Appendix~\ref{AppendixSection:2MASS_LGA_Repeat}, with validation figures of merit in the lower block of Table~\ref{Table:CAAPR_2MASS-LGA_Validation}. As can be seen, this photometry is in near-perfect agreement with that of the 2MASS-LGA. The offsets have been eliminated (to the \textless\,1\%\ level), and the $\chi$ distributions are Gaussian and narrow (with $\chi^{[-1,1]} > 0.7$) in all bands. We can therefore confidently state that difference in aperture sizes is the cause of the disagreement between our fluxes. Given that our larger apertures contain flux missed by the 2MASS-LGA (as we as being matched to the rest of our photometry), we deem our own photometry to be preferable. The high degree of scatter seen in Figure~\ref{Fig:CAAPR_2MASS_Validation} is an unavoidable consequence of our larger apertures, driving up the aperture noise associated with our measurements (an effect to which 2MASS is particularly vulnerable to due to sky brightness; see Section~\ref{Subsubsection:2MASS_Imagery}).

For a number of our galaxies, visual inspection of the multiwavelength SEDs indicated that the {\it H}-band flux can appear conspicuously offset above or below the $J$-band and $K_{S}$-band fluxes\footnote{NB, here we are once again referring to our standard aperture-matched \caapr\ photometry, not the repeated 2MASS photometry employed in the previous paragraph}. Our investigation of this is detailed in Appendix~\ref{AppendixSection:2MASS_LGA_Bump}; to summarise, we find that our {\it H}-band photometry does not exhibit any systematic offset from that of the 2MASS-LGA, but there is evidence of non-Gaussianity in our {\it H}-band photometric uncertainties.

\begin{table}
\begin{center}
\footnotesize
\caption{Figures of merit for internal comparison of DustPedia photometry in similar bands. }
\label{Table:Internal_Validation}
\begin{tabular}{llSSS}
\toprule \toprule
\multicolumn{1}{c}{$S_{1}$} &
\multicolumn{1}{c}{$S_{2}$} &
\multicolumn{1}{c}{$\widetilde{R}$} &
\multicolumn{1}{c}{$\Delta\widetilde{R}$} &
\multicolumn{1}{c}{$\chi^{[-1;1]}$}  \\
\midrule
WISE 3.4\,\micron\ & \spitz\ 3.6\,\micron\ & 1.027 & 0.044 & 0.636 \\
\spitz\ 4.5\,\micron\ & WISE 4.6\,\micron\ & 0.901 & 0.041 & 0.308 \\
WISE 22\,\micron\ & \spitz\ 24\,\micron\ & 1.040 & 0.064 & 0.686 \\
PACS 70\,\micron\ & \spitz\ 70\,\micron\ & 1.083 & 0.106 & 0.563 \\
PACS 160\,\micron\ & \spitz\ 160\,\micron\ & 0.976 & 0.091 & 0.838 \\
\bottomrule
\end{tabular}
\end{center}
\tablefoot{Column definitions the same as for Tables~\ref{Table:CAAPR_HRS_Validation} and \ref{Table:CAAPR_Supplementary_Validation}. For consistency, $S_{1}$ represents the shortest wavelength band of each pair.}
\end{table}

\subsection{Internal Validation} \label{Subsection:Internal_Validation}

Thanks to wavelength overlap between the various instruments used for our photometry, there is good scope for internal validation, comparing \caapr\ photometry in bands that are similar to one another. Figure~\ref{Fig:CAAPR_Internal_Validation} shows plots comparing the overlapping bands; the cross-validation figures of merit are listed in Table~\ref{Table:Internal_Validation}. We employ the same figures of merit we did for external validation; for instance, when comparing two bands, $S_{1}$ and $S_{2}$, the median flux ratio between the two is denoted by $\widetilde{R}_{S_{1}|S_{2}}$. 

\subsubsection{WISE--\textit{Spitzer}} \label{Subsubsection:Internal_WISE-Spitzer_Validation}

The WISE 3.4\,\micron\ and \spitz\ 3.6\,\micron\ bands have extremely similar relative system responses, as do WISE 4.6\,\micron\ and \spitz\ 4.5\,\micron\ (see Figure~1 of \citealp{Jarrett2011B}). To allow comparison between both pairs of bands, we colour correct all fluxes assuming a Rayleigh-Jeans reference spectrum; colour corrections for WISE were taken from Table~1 of \citet{Wright2010F}, whilst corrections for \spitz\ were taken from Table~4.3, Section~4.4 of the IRAC Instrument Handbook. These corrections are small, being \textless\,1.2\%\ in all cases. 

We find our WISE 3.4\,\micron\ and \spitz\ 3.6\,\micron\ fluxes to be in excellent agreement, with the median offset smaller than either bands' calibration uncertainty, and a good $\chi^{[-1,1]}_{3.4|3.6}$ value of 0.636. This is in keeping with the findings of \citet{Jarrett2012A} and \citet{Papovich2016A}, who report that the two bands exhibit no noticeable offset. 

However, we do find a definite offset between our WISE 4.6\,\micron\ and \spitz\ 4.5\,\micron\ photometry, with $\widetilde{R}_{4.6|4.5} = 0.901$. This is twice the size of the 4.5\%\ mutual calibration uncertainty between the bands (ie, $\sqrt{3\%^{2}+3.4\%^{2}}$). Previous authors have reported varying degrees of disagreement between these bands, with \citet{Jarrett2012A} finding offsets in the range $0.89 < \widetilde{R}_{4.6|4.5} < 1.02$ for a small sample of extended nearby galaxies, whilst \citet{Papovich2016A} find $\widetilde{R}_{4.6|4.5} = 0.98$ for a large sample of high redshift galaxies. \citet{Jarrett2012A} point out that the expected flux offset depends on the nature of the target galaxy, from 0.94 for ellipticals, to 0.93--0.99 for standard spirals, to 1.09 for starbursts; given that it is beyond the scope of this work to perform the full SED-fitting necessary to provide source-specific reference spectra, such differences could account for a large proportion of the offset we find. As the matter stands, users of this photometry should be aware of the potential disagreement here; we are hopeful that the full SED-fitting to be performed in future DustPedia papers will allow for source-specific colour corrections that negate the issue. And notwithstanding the offset, the $\chi$ distribution for these bands is excellent.

Given that the part of the spectrum sampled by the WISE 22\,\micron\ and \spitz\ 24\,\micron\ bands has contributions from stellar emission, hot dust emission, and PAH-like emission, in proportions that vary greatly between sources, we are unable to apply any colour-corrections in the absence of full SED-fitting. As per the corrections given in Section~4.3.5 of the MIPS Instrument Handbook, adjusting \spitz\ 24\,\micron\ fluxes to the reference spectrum employed by WISE requires a correction of only 0.1\%\ (thanks to the close similarity at 24\,\micron\ between the 10$^{4}$\,K blackbody reference spectrum used by \spitz-MIPS and perfect Rayleigh-Jeans reference spectrum used by WISE). The two sets of fluxes are entirely compatible; the median flux ratio of  $\widetilde{R}_{22|24} = 1.04$ is smaller than the WISE 22\,\micron\ calibration uncertainty, and the minimal scatter yields an almost ideal $\chi^{[-1,1]}_{22|24} = 0.686$. This agreement between the two sets of fluxes is good enough that it essentially rules out the risk of egregious differences arising when source-by-source colour corrections are applied during full SED-fitting.

\subsubsection{PACS--\textit{Spitzer}} \label{Subsubsection:Internal_PACS-Spitzer_Validation}

 PACS and \spitz\ have two almost-identical bands in common, at 70 and 160\,\micron. To colour-correct the 70\,\micron\ fluxes for direct comparison, we follow the same procedure as in Section~\ref{Subsubsection:IRAS_SCANPI_Validation}, assuming an $S_{\nu} \propto \nu^{-1}$ reference spectrum. At 160\,\micron\ we assume a 20\,K blackbody reference spectrum. In both cases we adjust the PACS fluxes as per \citet{Muller2011A} (including the translation factor to account for the difference in central wavelengths between the PACS and \spitz\ bands), and adjust the \spitz\ fluxes as per Section~4.3.5 of the MIPS Instrument Handbook.

At both 70 and 160\,\micron, the PACS and \spitz\ fluxes are in good agreement, with average offsets smaller than the \spitz\ calibration uncertainties in both cases (and also smaller than the PACS calibration uncertainty at 160\,\micron). And both $\chi$ distributions appear acceptable, given the small number of sources involved.

Whilst it is satisfying to find such good agreement between PACS and \spitz\ at 160\,\micron, it is worth noting that there are only 5 sources without major flags where \spitz\ 160\,\micron\ coverage is available when PACS 160\,\micron\ coverage is not (and only 3 of those are \textgreater\,3\,$\sigma$). As such, there are almost no instances at 160\,\micron\ where users will be unable to use superior PACS fluxes instead of those from \spitz.

\begin{figure*}
\begin{center}
\includegraphics[width=0.975\textwidth]{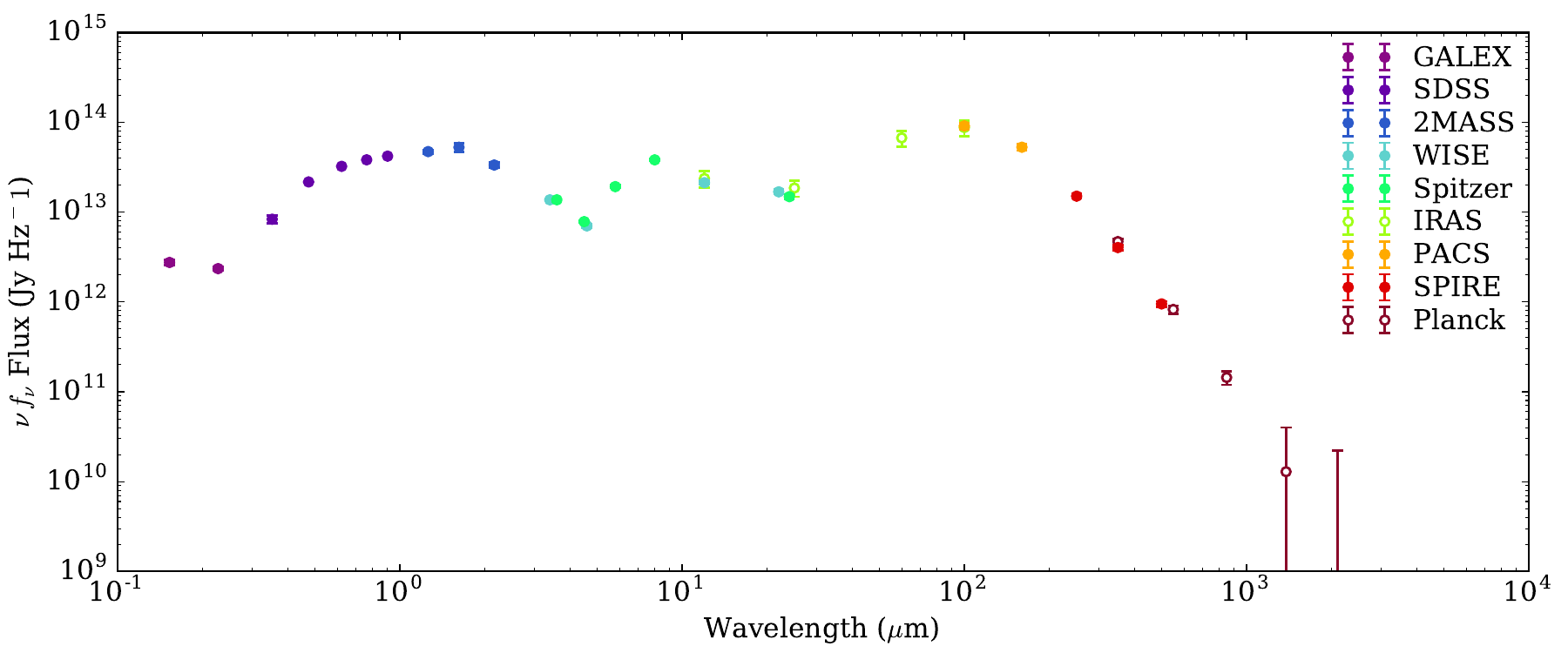}
\caption{UV--mm SED of NGC\,3683, demonstrating the rich photometric coverage possible for the DustPedia galaxies. Fluxes from our aperture-matched photometry are marked with solid circles, whilst fluxes from our supplementary \planck\ CCS2 and IRAS SCANPI photometry are marked with hollow circles.}
\label{Fig:Example_SED}
\end{center}
\end{figure*}

\section{Data Products} \label{Section:Data_Products}

\subsection{Photometry} \label{Subsection:Photometry}

The DustPedia photometry tables can be accessed via the InfraRed Science Archive (IRSA)\footnote{\url{https://irsa.ipac.caltech.edu/data/Herschel/DustPedia/overview.html}, assigned DOI \url{https://www.ipac.caltech.edu/doi/irsa/10.26131/IRSA660}.}, and from the VizieR catalogue service\footnote{Available at the CDS (see title footnote).}. Separate tables are provided for our aperture-matched \caapr\ photometry, the supplementary IRAS SCANPI photometry, and the supplementary \planck\ CCS2 photometry. Each table has a row for every DustPedia galaxy. Each band has 3 columns; one giving the flux, one giving the uncertainty on the flux (with calibration uncertainty included, as per Section~\ref{Subsection:Uncertainty_Estimation}), and one stating any flags associated with that flux. If the flux and error are blank for a given source in a particular band, this indicates that there is no data for the target in that band, or that there was insufficient data to make a measurement (ie, that the map in question was so small that no pixels were located in the background annulus). Whenever a measurement could be made, the resulting value is given, regardless of whether or not the measurement represents a detection. For maps where a flux could be measured, but where too little sky area was available to determine the aperture noise (even for \caapr's mini-aperture extrapolation), then we quote a negative uncertainty; this indicates the uncertainty only incorporates the contribution of the instrument's calibration uncertainty.

Extended-source corrections have been applied to the WISE and \spitz-IRAC photometry, as per Sections~\ref{Subsubsection:WISE_Imagery} and \ref{Subsubsubsection:Spitzer_IRAC_Imagery}.

In total, we present 21,857 photometric measurements; 18,254 fluxes from our aperture-matched \caapr\ photometry, along with 1,079 \planck\ CCS2 fluxes, and 2,533 IRAS SCANPI fluxes. A typical DustPedia source has photometry in 25 bands; Figure~\ref{Fig:Example_SED} shows the example UV--mm SED of NGC\,3683, a DustPedia galaxy with rich photometric coverage, having fluxes in 32 bands.

Our photometry demonstrates that the DustPedia sample spans a wide range of luminosities; amongst the DustPedia sources with \textgreater\,5\,$\sigma$ detections, SPIRE 250\,\micron\ luminosity ranges from 6.1--10.6\,${\rm log_{10} L_{\odot}}$, whilst WISE 3.4\,\micron\ luminosity (a standard proxy for stellar mass) ranges from 5.0--10.4\,${\rm log_{10} L_{\odot}}$. We span a similarly wide range of colour, with NUV-$r$ colour ranging from 0.8--6.3\,mag, and FUV-$K_{S}$ colour (a good proxy for morphology, see \citealp{GilDePaz2007A}) ranging from 0.6--8.8\,mag.

\subsection{Sample Catalogue}  \label{Subsection:Sample_Catalogue}

Also provided is the sample catalogue, providing the key properties of each DustPedia galaxy, such as position, morphology, inclination, optical D25, etc. This catalogue provides a number of distance measures; because of the difficulty in establishing reliable distances to nearby galaxies, we have established an order of preference.

Our preferred distance estimates are the redshift-independent distances provided by the HyperLEDA database, as these have been homogenised to account for methodological differences between references. For galaxies without redshift-independent HyperLEDA distances, our next preferred values are the redshift-independent distances provided by the {\sc Nasa/ipac} Extragalactic Database (NED\footnote{\url{https://ned.ipac.caltech.edu/ui/}}); whilst these are not homogenised like the HyperLEDA values, we take the quoted average value for each galaxy to minimise bias. If neither HyperLEDA nor NED redshift-independent distances are available for a source, we use the flow-corrected redshift-derived values provided by NED; these distances were calculated assuming a Hubble constant of ${\rm H_{0} = 73.24\,km\,s^{-1}\,Mpc^{-1}}$ \citep{Riess2016B}, and have been corrected for bulk deviation from Hubble flow arising from the influence of the Virgo Cluster, the Great Attractor, and the Shapley Supercluster \citep{Mould2000B}. For each galaxy in the DustPedia database we list each distance measure available, along with the preferred value as per the hierarchy described here.

\subsection{FITS Images} \label{Subsection:FITS Images}

The standardised DustPedia imagery (as described in Section~\ref{Section:Imagery}) can be accessed from IRSA\footnote{\url{https://irsa.ipac.caltech.edu/data/Herschel/DustPedia/overview.html}, assigned DOI \url{https://www.ipac.caltech.edu/doi/irsa/10.26131/IRSA660}.}, in the form of FITS images. The database may be queried according to sky coordinates. Data for individual galaxies may also be searched for by name . The maps may also be retrieved programmatically via URL. For example, the SPIRE 250\,\micron\ map of NGC\,0891 can be directly downloaded with the following URL: \url{https://irsa.ipac.caltech.edu/data/Herschel/DustPedia/Cutouts/SPIRE/NGC0891_SPIRE_250.fits}.  Other maps can be retrieved in a similar manner, by changing the galaxy, instrument, and band names given in the URL; the cutout directory index can be explored here: \url{https://irsa.ipac.caltech.edu/data/Herschel/DustPedia/Cutouts/}

In total, the imagery presented in this work represents 21,724 FITS images (not counting error maps), including the 3,297 non-photometric DSS maps. A typical DustPedia source has maps providing coverage of 25 bands.

\section{Summary} \label{Section:Summary}

We have presented the imagery and photometry of the DustPedia sample, covering  875 extended nearby galaxies in 42 UV--radio bands -- every nearby extended galaxy that was observed by the \hersc\ Space Observatory. The centrepiece of the dataset we present is our consistent multiwavelength aperture-matched photometry, encompassing 27 bands (from GALEX, SDSS, 2MASS, WISE, \spitz, and \hersc). In combination with supplementary IRAS and \planck\ photometry. This represents 21,857 fluxes; on average, each DustPedia galaxy possesses photometry in 25 bands. Additionally, we have produced imagery spanning 38 bands, including custom reductions of \hersc\ observations, along with standardised preparations of archival UV--radio data.

To perform our aperture-matched photometry, we developed the Comprehensive \& Adaptable Aperture Photometry Routine (\caapr), which is designed to determine fluxes and uncertainties that can can be directly compared across a wide range of wavelengths and data types. \caapr\ features a novel technique of determining aperture noise, even in maps with very little background coverage around the target source; this allows us to determine uncertainties consistently across the wide range of observations we employ.

We perform extensive validation of our photometry, comparing overlapping bands internally, and comparing to independent external sources. We find that our photometry is consistently reliable, with no excessive discrepancies.

The resulting database is being made publicly available, for the benefit of the astronomical community.

\section*{Acknowledgements} \label{Section:Acknowledgements}

The authors kindly thank the anonymous referee. This work was improved by the referee's attention to detail and constructive comments.

The DustPedia project is funded by the European Union, as a European Research Council (ERC) 7\textsuperscript{th} Framework Program (FP7) project (PI Jon Davies, proposal 606824).

JF acknowledges the financial support from UNAM-DGAPA-PAPIIT IA104015 grant, M\'exico.

This research has made use of Astropy, a community-developed core Python package for Astronomy (\url{http://www.astropy.org/}; \citealp{Astropy2013}). This research has made use of TOPCAT (\url{http://www.star.bris.ac.uk/~mbt/topcat/}; \citealp{Taylor2005A}), which was initially developed under the UK Starlink project, and has since been supported by PPARC, the VOTech project, the AstroGrid project, the AIDA project, the STFC, the GAVO project, the European Space Agency, and the GENIUS project. This research has made use of NumPy (\url{http://www.numpy.org/}; \citealp{VanDerWalt2011B}), SciPy (\url{http://www.scipy.org/}), and MatPlotLib (\url{http://matplotlib.org/}; \citealp{Hunter2007A}). This research made use of APLpy, an open-source plotting package for Python (\url{https://aplpy.github.io/}; \citealp{Robitaille2012B}). This research has made use of the scikit-learn machine learning library (\url{http://scikit-learn.org}; \citealp{Scikit-Learn2011}) and the scikit-image computer vision library (\url{http://scikit-image.org/}; \citealp{Scikit-Image2014}). This research has made use of code written by Adam Ginsburg (\url{https://github.com/keflavich}), kindly made available under the GNU General Public License (\url{http://www.gnu.org/copyleft/gpl.html}). 

This research made use of \montage\ (\url{http://montage.ipac.caltech.edu/ }), which is funded by the National Science Foundation under Grant Number ACI-1440620, and was previously funded by the National Aeronautics and Space Administration's Earth Science Technology Office, Computation Technologies Project, under Cooperative Agreement Number NCC5-626 between NASA and the California Institute of Technology.

This research has made use of the SIMBAD database (\url{http://simbad.u-strasbg.fr/simbad/}; \citealp{Wenger2000D}) and the VizieR catalogue access tool (\url{http://vizier.u-strasbg.fr/viz-bin/VizieR}; \citealp{Ochsenbein2000B}), both operated at CDS, Strasbourg, France. This research has made use of the NASA/IPAC Infrared Science Archive (IRSA; \url{http://irsa.ipac.caltech.edu/frontpage/}), and the NASA/IPAC Extragalactic Database (NED; \url{https://ned.ipac.caltech.edu/}), both of which are operated by the Jet Propulsion Laboratory, California Institute of Technology, under contract with the National Aeronautics and Space Administration. This research has made use of the HyperLEDA database (\url{http://leda.univ-lyon1.fr/}; \citealp{Makarov2014A}).

This research has made use of GALEX data obtained from the Mikulski Archive for Space Telescopes (MAST); support for MAST for non-HST data is provided by the NASA Office of Space Science via grant NNX09AF08G and by other grants and contracts (MAST is maintained by STScI, which is operated by the Association of Universities for Research in Astronomy, Inc., under NASA contract NAS5-26555). 

This research has made use of data from the 3\textsuperscript{rd} phase of the Sloan Digital Sky Survey (SDSS-III). Funding for SDSS-III has been provided by the Alfred P. Sloan Foundation, the Participating Institutions, the National Science Foundation, and the U.S. Department of Energy Office of Science. The SDSS-III web site is \url{http://www.sdss3.org/}. SDSS-III is managed by the Astrophysical Research Consortium for the Participating Institutions of the SDSS-III Collaboration including the University of Arizona, the Brazilian Participation Group, Brookhaven National Laboratory, Carnegie Mellon University, University of Florida, the French Participation Group, the German Participation Group, Harvard University, the Instituto de Astrofisica de Canarias, the Michigan State/Notre Dame/JINA Participation Group, Johns Hopkins University, Lawrence Berkeley National Laboratory, Max Planck Institute for Astrophysics, Max Planck Institute for Extraterrestrial Physics, New Mexico State University, New York University, Ohio State University, Pennsylvania State University, University of Portsmouth, Princeton University, the Spanish Participation Group, University of Tokyo, University of Utah, Vanderbilt University, University of Virginia, University of Washington, and Yale University. 

This research has made use of the NASA SkyView service (\url{http://skyview.gsfc.nasa.gov/current/cgi/query.pl}). SkyView has been developed with generous support from the NASA AISR and ADP programs (P.I. Thomas A. McGlynn) under the auspices of the High Energy Astrophysics Science Archive Research Center (HEASARC) at the NASA/ GSFC Astrophysics Science Division. 

This research has made use of the Digitized Sky Survey (DSS). The DSS was produced at the Space Telescope Science Institute under U.S. Government grant NAG W-2166. The images of these surveys are based on photographic data obtained using the Oschin Schmidt Telescope on Palomar Mountain and the UK Schmidt Telescope. The plates were processed into the present compressed digital form with the permission of these institutions. The National Geographic Society - Palomar Observatory Sky Atlas (POSS-I) was made by the California Institute of Technology with grants from the National Geographic Society. The Second Palomar Observatory Sky Survey (POSS-II) was made by the California Institute of Technology with funds from the National Science Foundation, the National Aeronautics and Space Administration, the National Geographic Society, the Sloan Foundation, the Samuel Oschin Foundation, and the Eastman Kodak Corporation. The Oschin Schmidt Telescope is operated by the California Institute of Technology and Palomar Observatory. The UK Schmidt Telescope was operated by the Royal Observatory Edinburgh, with funding from the UK Science and Engineering Research Council (later the UK Particle Physics and Astronomy Research Council), until 1988 June, and thereafter by the Anglo-Australian Observatory. The blue plates of the southern Sky Atlas and its Equatorial Extension (together known as the SERC-J), the near-IR plates (SERC-I), as well as the Equatorial Red (ER), and the Second Epoch [red] Survey (SES) were all taken with the UK Schmidt telescope at the AAO. All material not subject to one of the above copyright provisions is copyright 1995 by the Association of Universities for Research in Astronomy, Inc. Supplemental funding for sky-survey work at the ST ScI is provided by the European Southern Observatory. 

This research makes use of data products from the Two Micron All Sky Survey, which is a joint project of the University of Massachusetts and the Infrared Processing and Analysis Center/California Institute of Technology, funded by the National Aeronautics and Space Administration and the National Science Foundation.

This research makes use of data products from the Wide-field Infrared Survey Explorer, which is a joint project of the University of California, Los Angeles, and the Jet Propulsion Laboratory/California Institute of Technology, and NEOWISE, which is a project of the Jet Propulsion Laboratory/California Institute of Technology. WISE and NEOWISE are funded by the National Aeronautics and Space Administration.

This work is based in part on observations made with the {\it Spitzer} Space Telescope, which is operated by the Jet Propulsion Laboratory, California Institute of Technology under a contract with NASA.

This research makes use of data from \planck, a project of the European Space Agency, which received support from: ESA; CNES and CNRS/INSU- IN2P3-INP (France); ASI, CNR, and INAF (Italy); NASA and DoE (USA); STFC and UKSA (UK); CSIC, MINECO, JA, and RES (Spain); Tekes, AoF, and CSC (Finland); DLR and MPG (Germany); CSA (Canada); DTU Space (Denmark); SER/SSO (Switzerland); RCN (Norway); SFI (Ireland); FCT/MCTES (Portugal); ERC and PRACE (EU).

\hersc\ is an ESA space observatory with science instruments provided by European-led Principal Investigator consortia and with important participation from NASA. The \hersc\ spacecraft was designed, built, tested, and launched under a contract to ESA managed by the \hersc/\planck\ Project team by an industrial consortium under the overall responsibility of the prime contractor Thales Alenia Space (Cannes), and including Astrium (Friedrichshafen) responsible for the payload module and for system testing at spacecraft level, Thales Alenia Space (Turin) responsible for the service module, and Astrium (Toulouse) responsible for the telescope, with in excess of a hundred subcontractors.

\bibliographystyle{aa}
\bibliography{ChrisBib}

\begin{appendix}

\section{FITS Image Data Unit Conversions} \label{AppendixSection:Map_Unit_Conversions}

\begin{table}
\begin{center}
\footnotesize
\caption{Description of map unit conversions for various UV--MIR ancillary data for which pixel units of input archival maps were not already expressed in terms of Jy.}
\label{AppendixTable:Map_Unit_Conversions}
\begin{tabular}{lllSS}
\toprule \toprule
\multicolumn{1}{c}{Facility} &
\multicolumn{1}{c}{Band} &
\multicolumn{1}{c}{Native Units} &
\multicolumn{1}{c}{$m_{\it ZP}$} &
\multicolumn{1}{c}{$\Delta m_{\it AB}$}  \\
\cmidrule(lr){4-5}
\multicolumn{1}{c}{} &
\multicolumn{1}{c}{} &
\multicolumn{1}{c}{} &
\multicolumn{2}{c}{(mag)} \\
\midrule
GALEX & FUV & counts s$^{-1}$ & 18.82 & \textendash \\
GALEX & NUV & counts s$^{-1}$ & 20.08 & \textendash \\
SDSS & {\it u} & nanomaggies & 22.50 & -0.04 \\
SDSS & {\it g} & nanomaggies & 22.50 & +0.00 \\
SDSS & {\it r} & nanomaggies& 22.50 & +0.00 \\
SDSS & {\it i} & nanomaggies& 22.50 & +0.00 \\
SDSS & {\it z} & nanomaggies& 22.50 & +0.02 \\
2MASS & {\it J} & DN & \textendash & +0.91 \\
2MASS & {\it H} & DN & \textendash & +1.39 \\
2MASS & {\it K$_{S}$} & DN & \textendash & +1.85 \\
WISE & 3.4\,\micron\ & DN & 20.5 & +2.669 \\
WISE & 4.6\,\micron\ & DN & 19.5 & +3.339 \\
WISE & 12\,\micron\ & DN & 18.0 & +5.174 \\
WISE & 22\,\micron\ & DN & 30.0 & +6.620 \\
\bottomrule
\end{tabular}
\end{center}
\end{table}

Table~\ref{AppendixTable:Map_Unit_Conversions} provides information regarding the pixel units of the original maps for our various ancillary data, as taken from the official archive for each facility, and the relevant quantaties we employed to convert these pixel units to our consistent Jy\,pix$^{-1}$ units. See Section~\ref{Subsection:Ancillary_Imagery} for the full details of each ancillary data facility.

The $m_{\it ZP}$ column of Table~\ref{AppendixTable:Map_Unit_Conversions} provides the zero-point magnitude used to convert the native map units $S$ to Pogson magnitudes $m$, as per $m =  m_{\it ZP} - 2.5 \log_{10}(S)$. In the case of 2MASS, each tile has its own independently-calibrated zero-point magnitude, provided in the FITS header; as such no $m_{\it ZP}$ values can be listed for the 2MASS bands.

The $m_{\it ZP}$ values listed are the values provided by each facility, and as such give magnitudes in whatever native magnitude system each facility employs. Both 2MASS and WISE use Vega magnitudes, whilst the SDSS uses SDSS magnitudes (which are similar to AB magnitudes, but with small offsets in $u$ and $z$). The $\Delta m_{\it AB}$ column of Table~\ref{AppendixTable:Map_Unit_Conversions} gives the offset between these native magnitudes for each band, and the corresponding AB magnitudes, such that $m_{\it AB} = m + \Delta m_{\it AB}$. GALEX employs AB magnitudes by default, so there is no $\Delta m_{\it AB}$ offset.

\section{External Validation -- Investigating Severe Disagreement with HRS WISE 12\,\micron\ Photometry} \label{AppendixSection:WISE_HRS}

As described in Section~\ref{Subsubsubsection:HRS_WISE_Validation}, our photometry significantly differs from that of the HRS at 12\,\micron, with our fluxes being fainter by an average factor of 0.776. Here we discuss in detail how we determined the sources of this disagreement.

After investigation, it transpired that the HRS converted their WISE measurements from `DN' (the WISE term for instrumental pixel units) to Jy using the conversion provided in the WISE Preliminary Data Release Explanatory Supplement\footnote{\url{http://wise2.ipac.caltech.edu/docs/release/prelim/expsup/wise_prelrel_toc.html}} \citep{Ciesla2014A}, despite the fact the maps they used came from the WISE All-Sky Data Release (L.\,Ciesla, {\it priv. comm.}). The 12\,\micron\ maps of the All-Sky Data Release have a very different zero-point from that of the Preliminary Data Release, causing the HRS to overestimate their 12\,\micron\ fluxes by a factor of 1.585 (the 12\,\micron\ zero-point of the AllWISE Data Release, used by us, is the same as that of the All-Sky Data Release). 

Our investigation also revealed an apparent inconsistency in the WISE All-Sky and AllWISE data releases. As described in Section~\ref{Subsubsection:WISE_Imagery}, we converted our WISE imagery from DN to Vega magnitudes using the zero-point magnitudes provided in the FITS headers, then to AB magnitudes using the offsets listed in the Table~1, Section~IV.3.a AllWISE Explanatory Supplement, and thereby to Jy (we use this conversion method, as our preference is for using the photometric calibration information provided with the data itself). However, if we instead use the DN-to-Jy conversion factor listed in Table~1, Section~IV.3.a of the AllWISE Explanatory Supplement (mirroring the approach of the HRS), the resulting final map units are not the same -- map units produced by using the DN-to-Jy conversion from the documentation differ by a factor of 0.929 from the map units produced by using the zero-point magnitudes in the FITS headers. 

Another potential source of discrepancy between the \caapr\ and HRS 12\,\micron\ fluxes is the fact that the HRS do not perform any foreground star removal, and they use much smaller photometric apertures than our own (theirs being designed to contain only the visible MIR emission). To test the impact of these differences, we repeated our photometry but using the same apertures as \citet{Ciesla2014A}, with foreground star removal disabled -- the resulting fluxes differed from our actual fluxes by a factor of 0.909 (for \textgreater\,5\,$\sigma$ measurements).

The combined effect of these three differences is that we should expect our WISE 12\,\micron\ photometry to be fainter than the WISE 12\,\micron\ photometry of the HRS by a factor of 0.747 -- very close to the actual factor of 0.776 that we encounter.

\section{External Validation -- Repeating the 2MASS LGA Photometry} \label{AppendixSection:2MASS_LGA_Repeat}

\begin{figure*}
\begin{center}
\includegraphics[width=0.4375\textwidth]{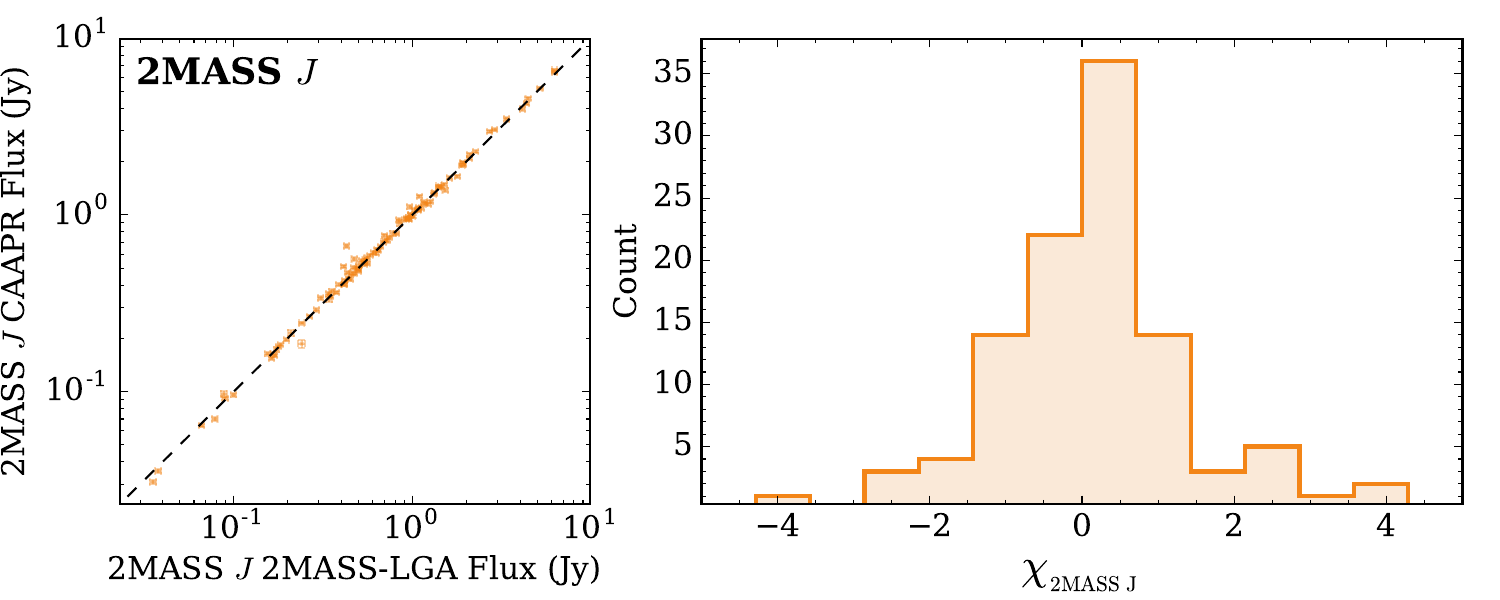}
\includegraphics[width=0.4375\textwidth]{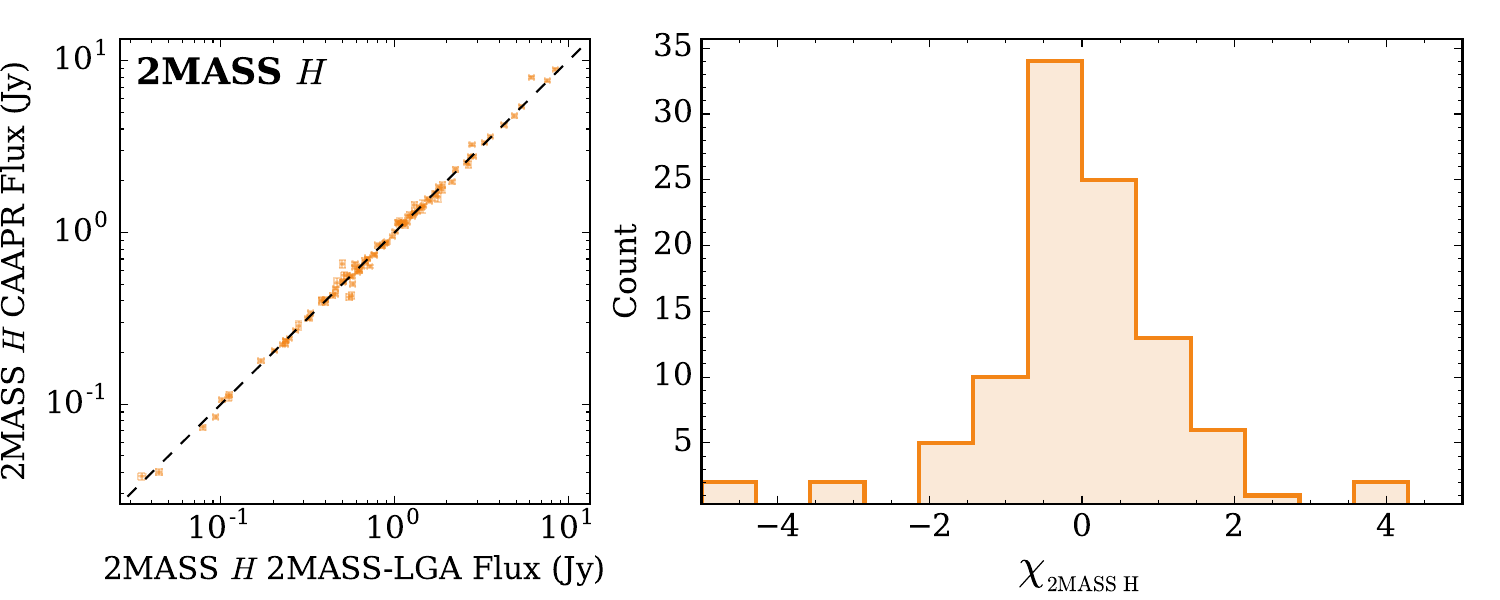}
\includegraphics[width=0.4375\textwidth]{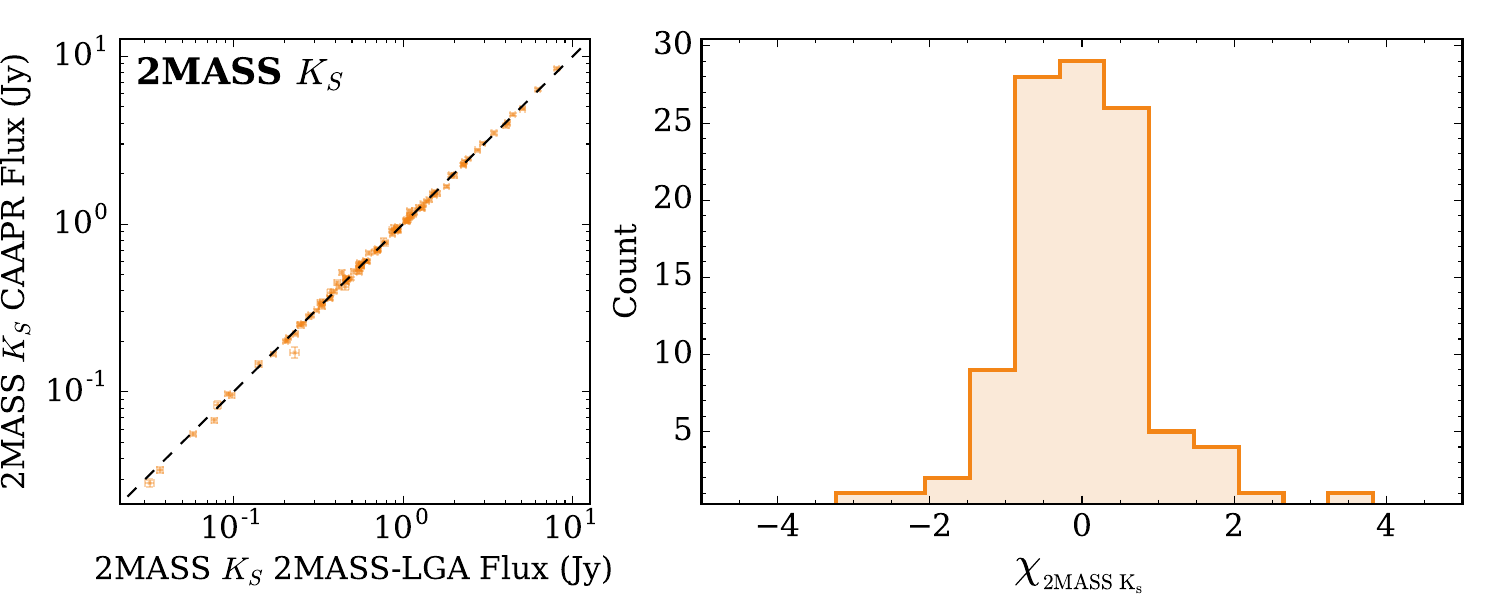}
\includegraphics[width=0.4375\textwidth]{Validation_Placeholder.pdf}
\caption{Comparison of \caapr\ photometry performed using the apertures employed by the 2MASS-LGA, to photometry the 2MASS-LGA themselves report. Plots as per Figure~\ref{Fig:CAAPR_HRS_Validation}.}
\label{AppendixFig:CAAPR_2MASS_Repeat_Validation}
\end{center}
\end{figure*}

In Section~\ref{Subsubsection:2MASS-LGA_Validation}, we repeat our 2MASS photometry, in order to compare it with the 2MASS-LGA. We use \caapr\ to perform this repeat photometry, but instead of our own apertures we employ the apertures used by the 2MASS-LGA. This is to establish whether the fact that our apertures are consistently larger than those of the 2MASS-LGA is the reason why our photometry is systematically brighter. The plots of the results of this comparison are shown in Figure~\ref{AppendixFig:CAAPR_2MASS_Repeat_Validation}. The findings from this comparison are discussed in Section~\ref{Subsubsection:2MASS-LGA_Validation}.

\section{External Validation -- Assessing the reliability of 2MASS $H$-Band Photometry} \label{AppendixSection:2MASS_LGA_Bump}

\begin{figure}
\begin{center}
\includegraphics[width=0.475\textwidth]{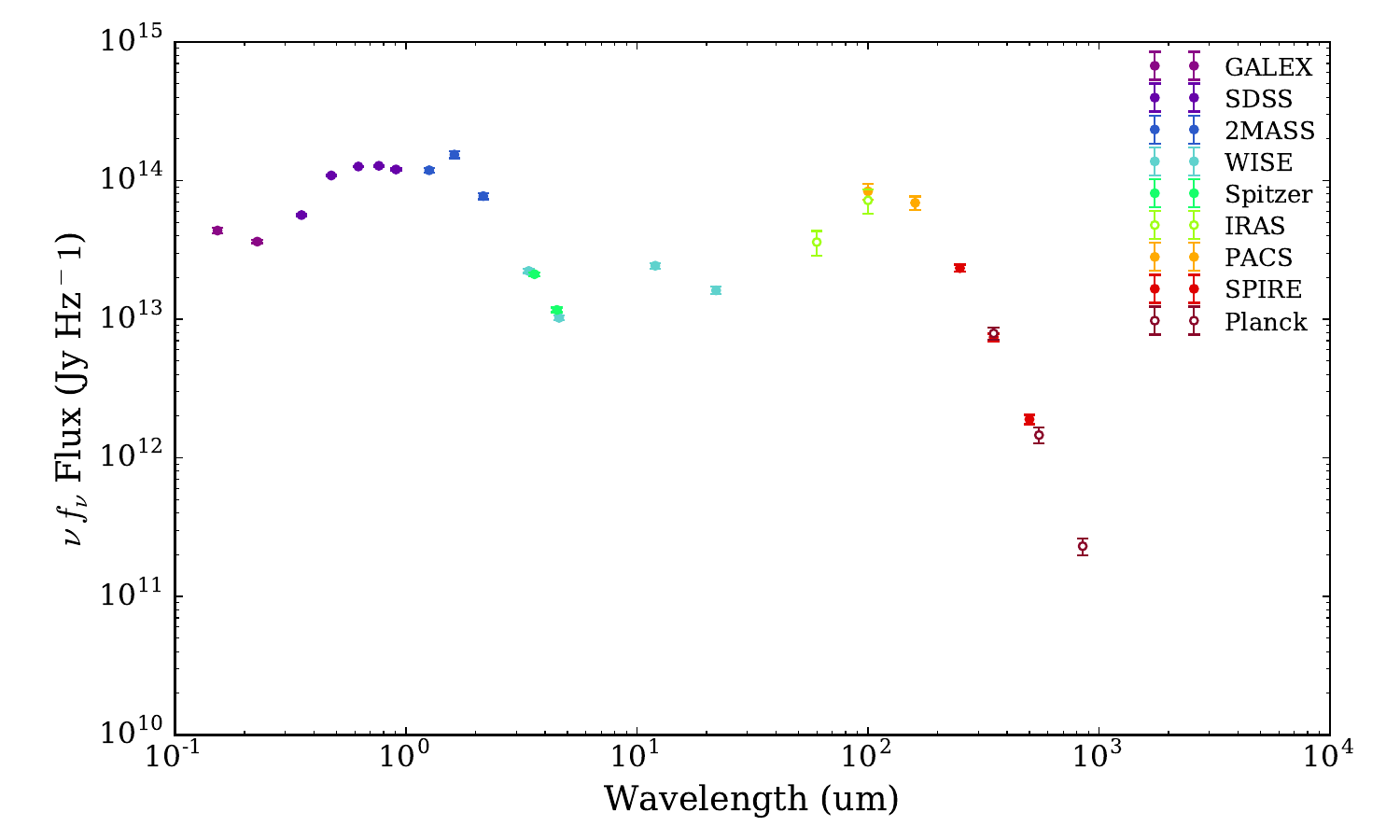}
\caption{UV--mm SED of NGC\,3631, showing an especially pronounced example of a `bump' in our 2MASS {\it H}-band photometry. 2MASS photometry is plotted in dark blue, with {\it H}-band hence being the central of the three 2MASS fluxes plotted (at 1.65\,\micron).}
\label{AppendixFig:Example_2MASS_Bump}
\end{center}
\end{figure}

When visually inspecting the multiwavelength SEDs of our galaxies, we found that our {\it H}-band photometry sometimes appears conspicuously offset above or below the $J$-band and $K_{S}$-band, manifesting as a `bump' or `dip' in the SED at {\it H}-band; see Figure~\ref{AppendixFig:Example_2MASS_Bump} for an example. 

To establish whether this was indicative of some sort of systematic issue, we calculated the $[J/H]$ and $[H/K_{S}]$ colours of our galaxies using our aperture-matched \caapr\ photometry. If the {\it H}-band fluxes were suffering from a systematic artefact, we would expect the colours calculated with our \caapr\ photometry to be systematically offset from the equivalent colours calculated with the 2MASS LGA photometry. However, we find that the \caapr\ colours show no systematic difference from the 2MASS LGA colours; an offset can be ruled out to the \textless\,1.4\%\ level. What we do find is that the \caapr\ photometry exhibits a much larger {\it scatter} in colour than the 2MASS LGA photometry. Whilst we should expect \caapr\ colours to exhibit more scatter (due to the larger uncertainties of our aperture-matched photometry), the {\it degree} to which the scatter is larger for the \caapr\ colours is greater than we would predict assuming the photometric uncertainties to be Gaussian -- specifically, 65\%\ greater for $[J/H]$ and 26\%\ greater for $[H/K_{S}]$. 

For comparison, we also calculated the $[J/K_{S}]$ colours for both the \caapr\ and 2MASS-LGA photometry. Once again, we found greater scatter with the \caapr\ photometry than with that of the 2MASS-LGA -- however, the increase was within the bounds of what would be expected given the larger \caapr\ uncertainties (assuming they behave in a Gaussian manner).

We can therefore infer that the conspicuous {\it H}-band offsets found for some sources are due to non-Gaussianity in our {\it H}-band uncertainties. This results in there being more outliers amongst our {\it H}-band photometry than we would otherwise expect (such as that shown in Figure~\ref{AppendixFig:Example_2MASS_Bump}). Users who are concerned by the this non-Gaussianity have the option of not using the {\it H}-band photometry -- the $J$-band $K_{S}$-band fluxes do not seem to suffer from a similar problem, given that $[J/K_{S}]$ colours behave as expected.

The non-Gaussianity of the {\it H}-band uncertainties could arise from the complex nature of the sky brightness suffered by 2MASS {\it H}-band observations. The arcminute scales of the {\it H}-band sky brightness make it particularly problematic for nearby-galaxy photometry.

\end{appendix}

\end{document}